\documentclass[a4paper,11pt]{article}

\usepackage{jheppub} 
\usepackage{amsmath}

\usepackage[T1]{fontenc} 
\usepackage{latexsym,amsmath,amsfonts,amssymb,amsthm,mathrsfs, mathtools}

\usepackage[dvipsnames]{xcolor}
\usepackage{setspace}
\newcommand{\floor}[1]{\left\lfloor #1 \right\rfloor}

\title{Lagrangian Schur index and Bethe ansatz type formula}
\allowdisplaybreaks

\author{Yutong Li,}
\author{Yiwen Pan,}

\affiliation{School of Physics, Sun Yat-sen University,\\No. 135 Xingangxi Road, Guangzhou, Guangdong, China}

\emailAdd{liyt83@mail2.sysu.edu.cn}

\emailAdd{panyw5@mail.sysu.edu.cn}

\abstract{We propose a surprisingly elementary method to compute the Schur index in closed-form for general $\mathcal{N} = 2$ Lagrangian theories. The method is inspired by the Bethe-ansatz-type formula for $\mathcal{N} = 1$ superconformal index. We identify issues underlying the original derivation: the loss of periodicity property upon integration and the omitted poles outside of the annulus region. We circumvent the problems and transform integration into solving a simple difference equation. The final result is expressed as a finite sum of quasi-Jacobi forms involving twisted Eisenstein series. We test our method on different types of theories, including BCD-type $\mathcal{N} = 4$ theories and various $\mathcal{N} = 2$ quiver gauge theories.}

\makeatletter
\gdef\@fpheader{}
\makeatother

\begin{document} 
\maketitle


\section{Introduction}

Supersymmetric field theories contain a rich set of protected quantities that can be computed exactly and non-perturbatively. These quantities often serve as bridges into different areas of physics and mathematics, such as quantum gravity, geometry of complex manifolds, or representation theory of different types of algebras. One such quantity is the superconformal index \cite{Witten:1986bf,Benini:2013xpa,Imamura:2011su,Kapustin:2011jm,Romelsberger:2005eg,Kinney:2005ej,Gadde:2011ik,Gadde:2011uv,Kim:2012ava}, which counts the number (decorated by signs and other symmetry data) of certain local operators in a superconformal field theory (SCFT). The superconformal index is invariant under continuous deformations of the theory preserving the superconformal symmetry, and therefore it is a powerful tool to study strongly coupled SCFTs and their dualities \cite{Seiberg:1994pq,Dolan:2008qi,Spiridonov:2008zr,Rastelli:2014jja}.

In this paper we focus on the Schur limit of four dimensional (4D) $\mathcal{N} = 2$ superconformal index \cite{Gadde:2011ik,Gadde:2011uv,Beem:2013sza}. It is defined as a supertrace $\mathcal{I} = \operatorname{str}q^{E - R}b^f$ over the Hilbert space of the theory quantized on $S^3$, where $E$ is the scaling dimension, $R$ is the Cartan of the $SU(2)_R$ R-symmetry, and $f$ is a flavor charge. It is identified with the vacuum character of the associated chiral algebra \cite{Beem:2013sza,Beem:2014rza,Lemos:2014lua,Xie:2016evu,Choi:2017nur}, and has significant applications in the study of chiral algebra and its representation theory \cite{Arakawa:2016hkg,Creutzig:2017qyf,Cordova:2017mhb,Nishinaka:2018zwq,Bianchi:2019sxz,Creutzig:2017uxh,Creutzig:2018lbc,Arakawa:2018egx,Arakawa:2023cki,Xie:2019zlb}. The (defect) Schur index is also closely related to the Coulomb branch physics of the $\mathcal{N} = 2$ theory \cite{Cordova:2016uwk,2017arXiv170906142F,Fredrickson:2017yka,Dedushenko:2018bpp}. The Schur index of $\mathcal{N} = 4$ theories is also related to the giant graviton partition function in the dual AdS$_5$ gravity \cite{Arai:2020qaj,Imamura:2021ytr,Gaiotto:2021xce,Liu:2022olj,Eniceicu:2023uvd,Beccaria:2023zjw,Hatsuda:2023imp,Eleftheriou:2023jxr,Deddo:2024liu,Beccaria:2024vfx,Beccaria:2024szi,Imamura:2024zvw,Ezroura:2024wmp,Deddo:2025jrg}. It is therefore of particular interest to compute the Schur index analytically and in closed-form.

There have been a series of efforts in computing Lagrangian Schur index in closed form, which is given by a multivariate contour integral of an elliptic integrand,
\begin{equation}
  \mathcal{I} = \oint_{|a_i| = 1} \bigg[\frac{da}{2\pi i a}\bigg] \mathcal{Z}(a) \ , \qquad
  \mathcal{Z}(..., a_i q , ...) = \mathcal{Z}(..., a_i, ...) \ .
\end{equation}
The early work \cite{Bourdier:2015wda} applies fermi-gas formalism to compute the unflavored Schur index of $\mathcal{N} = 4$ $SU(N)$ super Yang-Mills (SYM) theory, which was generalized to $\mathcal{N} = 2$ circular quiver gauge theories \cite{Bourdier:2015sga}. The flavored Schur (Wilson line) index in closed-form of $\mathcal{N} = 4$ $SU(N)$ SYM was recently computed by generalizing the fermi-gas formalism  \cite{Hatsuda:2022xdv,Hatsuda:2023iwi,Hatsuda:2025mvj}. The unflavored Schur index for $\mathcal{N} = 4$ SYMs with gauge group $SO(N)$ and $USp(N)$ was also computed \cite{Du:2023kfu}. For more general $\mathcal{N} = 2$ theories, other methods are needed. In \cite{Pan:2021mrw,Guo:2023mkn}, several integration formulas are proposed to carry out exact contour integrals for relatively simple $\mathcal{N} = 2$ theories. In particular, the Schur index of all $A_1$ class-$\mathcal{S}$ theories can be computed exactly and written in a fairly compact form. Modularity also helps constrain the unflavored Schur index, which can be expanded in terms of (quasi-)modular forms \cite{Beemetal,Huang:2022bry,Du:2023kfu}. More recently, the work \cite{vanLeuven:2025gwr} replaces the Haar measure with a reduced measure in the contour integral, and is able to carry out the $SU(2)$ integration for Schur (and more generally the superconformal index) using the standard Cauchy residue theorem without the problemetic contribution from the non-isolated singularity. There is also the  $q$-deformed Yang-Mills formalism applicable to the class-$\mathcal{S}$ theories \cite{Gadde:2011ik,Gadde:2011uv}. With all the remarkable progress, it is still challenging to compute the Schur index of general $\mathcal{N} = 2$ theories in closed form. Since the integrand of the Schur index has no simple universal functional structure, it is unlikely to find a generalization of fermi-gas formalism to other $\mathcal{N} = 2$ theories. The only universal feature to exploit is ellipticity.

As a special limit of the superconformal index, Schur index would benefit from any scheme of exact computation of the superconformal index. Luckily, there has been substantial effort along this line. In particular, \cite{Benini:2018mlo} proposes a Bethe ansatz type formula to compute the $\mathcal{N} = 1$ superconformal index (inspired by \cite{Closset:2017bse} on generalized supersymmetric index on $\mathcal{M}_{g, p}\times S^1$). The formula makes use of the periodicity of the integrand of the $\mathcal{N} = 1$ superconformal index which leads to a set of functions $\mathcal{Q}_i$ with elliptic property. With a series of intriguing tricks, the paper claims that the superconformal index can be written as a sum over Bethe ansatz solutions $\mathcal{Q}_i = 1$ within certain region, where each solution contributes its residue. The formula plays an important role in analyzing $\mathcal{N} = 4$ superconformal index and its relation with blackhole entropy \cite{Benini:2018ywd,GonzalezLezcano:2019nca,Lanir:2019abx,Benini:2020gjh,GonzalezLezcano:2020yeb,Lezcano:2021qbj,Benini:2021ano,Aharony:2021zkr,David:2021qaa,Colombo:2021kbb,Amariti:2024bsr,Amariti:2025vjd}. It is therefore natural to view any $\mathcal{N} = 2$ superconformal index as some special $\mathcal{N} = 1$ superconformal index, and anticipate a Bethe ansatz type closed-form formula for the Schur index.

Unfortunately, the above idea quickly runs into a series of problems. Ultimately we identify some key issues that underlie the derivation of the Bethe ansatz type formula in \cite{Benini:2018mlo}: the multivariate periodic behavior the derivation relies on is actually broken by the multivariate integration, and the higher dimensional residue theorem in general demands poles outside of the listed annulus region. The difficulties in classifying poles and computing residues, including the presence of infinite and even continuous singularities were discussed in \cite{ArabiArdehali:2019orz,Lezcano:2021qbj,Benini:2021ano}, and recently addressed in detail \cite{Cabo-Bizet:2024kfe}.

Nonetheless, the series of tricks in \cite{Benini:2018mlo} can be reorganized to carry out exact computation of almost any Lagrangian Schur index, and cast the result in terms of $\eta(\tau)$, $\vartheta_i$ and the Eisenstein series. Our approach exploits the ellipticity of the Schur index integrand, and reduces the multivariate integration to solving a series of difference equations. These difference equations take a fairly simple form
\begin{equation}
  n(aq) - n(a) = P(a),
\end{equation}
where $P(a)$ denotes any polynomial of the Eisenstein series, and it turns out that $n(a)$ can be solved with a simple closed-form formula,
\begin{equation}
  n(a) = \sum_{r = 0}^{\infty} \frac{B_r}{r!} \partial^{r - 1}P(a) \ , 
\end{equation}
where $\partial$ denotes the weight-lowering operator that maps $E_k \Big[\substack{\pm 1 \\ a b}\Big] \to E_{k - 1}\Big[\substack{\pm 1 \\ ab}\Big]$. With $n(a)$, the integration of any elliptic function $\mathcal{Z}(a)$ times $P(a)$ can be carried out easily by residue theorem,
\begin{equation}
  \oint_{|a| = 1} \bigg[\frac{da}{2\pi i a}\bigg] \mathcal{Z}(a) P(a)
  = \sum_{J} \operatorname{Res}_{a = a_J} \frac{1}{a} \mathcal{Z}(a) n(a) \ ,
\end{equation}
where $a_J$ are poles of $\frac{1}{a}\mathcal{Z}(a)n(a)$ within the annulus region $|q| \le |a| < 1$. It is crucial to carry out the multivariate integration one variable at a time to avoid missing poles in a higher dimensional space, where some of the poles may be outside of the annulus region.

The paper is organized as follows. In section \ref{sec:BAE}, we review the Bethe ansatz type formula in \cite{Benini:2018mlo} and point out the problems in the derivation. In section \ref{sec:integration-formula}, we reshape the Bethe ansatz formula into a new approach to compute Lagrangian Schur index in closed form. In section \ref{sec:examples}, we apply the method to several examples of $\mathcal{N} = 2$ theories.

\section{Bethe ansatz type formula}\label{sec:BAE}

The application of integrable system techniques and in particular Bethe ansatz to the study of exact supersymmetric partition functions has been a fruitful direction \cite{Nekrasov:2009uh,Nekrasov:2009ui,Orlando:2010uu}. In particular, four dimensional generalized $\mathcal{N} = 1$ superconformal index can be computed as a sum of fiberings and handle-gluing operators over Bethe vacua \cite{Closset:2017bse}. The subsequent work \cite{Benini:2018mlo}, which inspired this paper, proposes a similar method computing the $\mathcal{N} = 1$ superconformal index exactly as a sum over Bethe ansatz solutions. It is tempting to adopt the formula and specialize it to the Schur limit to find a Bethe ansatz type formula for the Schur index. However, as we will see, there are several problems in this approach. In this section we will first review the Bethe ansatz formula for superconformal index, and then discuss the problems hidden in the method.

\subsection{The formula}
Consider the $\mathcal{N} = 2$ superconformal index $\mathcal{I}(p,q,t)$. It is defined as a supertrace over the Hilbert space of the theory quantized on $S^3$, and hence it is intrisically a series in $p,q,t$, where the coefficients capture the representation-theoretic data under the flavor symmetry. The index is independent of exactly marginal coupling, and hence can be computed in the free limit of the theory. For a Lagrangian theory with a gauge group, the index can be written as a multivariate contour integral,
\begin{align}
  \mathcal{I}(p,q,t) = \oint_{|a| = 1} \left[{\frac{da}{2\pi i a}}\right] \mathcal{Z}(a; p,q,t) \ ,
\end{align}
where $a$ collectively denotes the gauge fugacities $a_1, ..., a_r$, and the integration measure incorporates the Haar measure, implying counting only gauge invariant operators in the theory. In principle one can include flavor fugacities in the index. For simplicity we often omit the dependence on $p, q, t$ in the notation. More precisely, the integral should be understood as the result of sequential integration along the unit circles, or equivalently, by taking the constant term of the integrand $\mathcal{Z}$. In practice, we can expand the integrand $\mathcal{Z}(a;p,q,t)$ as a $p,q,t$-series with coefficients being functions of $a$, and extract the $a$-constant term of the coefficients to recover the index.

The integrand $\mathcal{Z}(a)$ is usually some ratio of elliptic Gamma functions $\Gamma(z; p, q)$ which capture contrubitions from vecthr and hypermultiplets. It enjoys the following periodic property,
\begin{equation}
  \Gamma(pz; p, q) = \theta(z; q) \Gamma(z; p, q) \ ,
\end{equation}
where $\theta(z; q) = \prod_{n=0}^{\infty} (1 - z q^n)(1 - z^{-1} q^{n+1})$ is the $q$-theta function. The $q$-theta function is related to the Jacobi theta functions $\vartheta_1(\mathfrak{z})$ by $\vartheta_1(\mathfrak{z}) = i q^{\frac{1}{8}}z^{-1/2}(q;q)\theta(z;q)$.

The periodic property of the elliptic Gamma function leads to the following behavior of $\mathcal{Z}(a)$ under large gauge transformation. Focusing on the special case $p = e^{2\pi i \sigma} = h^a$, $q = e^{2\pi i \tau} = h^b$ where $a,b$ are positive coprime integers (and therefore $\tau = b\omega$ for $h = e^{2\pi i \omega}$), then
\begin{equation}
  \mathcal{Z}(a_1, ..., a_i h^{-ab}, ..., a_n) = \mathcal{Q}_i(a_1, ..., a_n) \mathcal{Z}(a_1, ..., a_i, ..., a_n) \ ,
\end{equation}
Here the $\mathcal{Q}$ functions are simple product of theta functions (referred to as the quasi-periodicity in \cite{Cabo-Bizet:2024kfe}). The crucial observation is that each $\mathcal{Q}$ is actually doubly periodic in all $a_j$ (or, more precisely, in $\mathfrak{a}_j$) due to the cancellation of the flavor and $U(1)_R$ ABJ anomaly, gauge, and gravity anomalies,
\begin{align}
  \mathcal{Q}_i(\mathfrak{a}_1, ..., \mathfrak{a}_j + 1, ..., \mathfrak{a}_n) = & \ \mathcal{Q}_i(\mathfrak{a}_1, ..., \mathfrak{a}_j + \omega, ..., \mathfrak{a}_n) = \mathcal{Q}_i(\mathfrak{a}_1, ..., \mathfrak{a}_n) \ , \quad h = e^{2\pi i \omega} \ .
\end{align}

Exploiting the double periodicity of $\mathcal{Q}_i$, the multivariate integral can be manipulated,
\begin{align}
  \mathcal{I} = \oint \left[{\frac{da}{2\pi i a}}\right]
  \prod_i \frac{1 - \mathcal{Q}_i(a)}{1 - \mathcal{Q}_i(a)} \mathcal{Z}(a) \ .
\end{align}
The numerator $1 - \mathcal{Q}_i(a)$ can be absorbed into $\mathcal{Z}$ by performing a shift of the $a$ variable in $\mathcal{Z}$,
\begin{align}
  \prod_i (1 - \mathcal{Q}_i(a)) \mathcal{Z}(a) = \sum_{\ell = 0}^{r} \frac{(-1)^\ell}{\ell!} \sum_{i_1 \ne ... \ne i_\ell}^{r} \mathcal{Z}(a_1, ..., a_{i_1} h^{-ab}, ..., a_{i_\ell} h^{-ab}, ..., a_n) \ ,
\end{align}
where the shift of the $a$ variables can be further absorbed into a \emph{simultaneous} shift of integration contour,
\begin{align}
  \mathcal{I} = \sum_{\ell = 0}^{r} \sum_{i_1 \ne ... \ne i_\ell}^{r} \frac{(-1)^\ell}{\ell!}  \oint_{\mathcal{C}_{i_1, ..., i_\ell}} \left[{\frac{da}{2\pi i a}}\right] \frac{\mathcal{Z}(a)}{\prod_i (1 - \mathcal{Q}_i(a))} \ .
\end{align}
The contours $\mathcal{C}_{i_1, ..., i_\ell}$ form the boundary of some sort of higher dimensional annulus which suggests the use of residue theorem, collecting the residues of poles ``encircled'' by the contour. The poles may come from the zero of the denominator $1 - \mathcal{Q}_i(a)$, or from the poles of $\mathcal{Z}(a)$. The paper carefully argued the absence of poles from the numerator $\mathcal{Z}$ inside the annulus bounded by the contour, and the integral receives only contributions from the zeros of $1 - \mathcal{Q}_i(a)$.

Finally, the application of ``residue theorem'' leads to the Bethe ansatz formula for the index which sums over solutions $\hat a$ to the following Bethe ansatz equations (BAE),
\begin{align}
  \mathcal{I} = \sum_{\hat{a} \in \text{BAE}} \frac{\mathcal{Z}(\hat{a})}{\det \Big(\frac{\partial \mathcal{Q}_i}{\partial \log a_j}\Big)\big|_{a = \hat{a}}} \ ,
\end{align}
where $\hat a$ solve the Bethe ansatz equations
\begin{align}
  \mathcal{Q}_i(\hat{a}) = 1 \ , \qquad i = 1, ..., r \ , \qquad 1 < |a_i| < |h|^{-ab} \ .
\end{align}
Although the Bethe ansatz formula is only explicitly valid for the special case $p = h^a$, $q = h^b$, it is expected to determine the generic $p,q$ case as well via analytic continuation. The residue used in the formula is often referred to as the \emph{Grothendieck residue}.

\subsection{The problems}
We'd like to explore such a Bethe ansatz type formula for computing Schur index. After all, the Schur index is a special limit of the superconformal index. However, one immediately runs into several problems if we were to apply the above manipulations and formula.

The first problem is fairly obvious. When one tries to apply the formula in an $\mathcal{N} = 2$ theory and simply takes the Schur limit $t \to q$, the resulting integrand $\mathcal{Z}$ is automatically doubly periodic for $\mathfrak{a}_i$: 
\begin{equation}
  \mathcal{Z}(\mathfrak{a}_1, \cdots, \mathfrak{a}_i + 1 \cdots, \mathfrak{a}_n)
  = \mathcal{Z}(\mathfrak{a}_1, \cdots, \mathfrak{a}_i + \tau, \cdots, \mathfrak{a}_n) = \mathcal{Z}(\mathfrak{a}_1, \cdots, \mathfrak{a}_n) \ .
\end{equation}
Therefore the key functions $\mathcal{Q}$'s are simply trivial.

The second problem is more severe: it is the intrinsic flaw within the original Bethe ansatz type formula.  We can phrase the problem in two ways.  First, the Bethe ansatz formula uses the simple periodicity $\mathcal{Z}(a_1,..., a_i q ..., a_n) = \mathcal{Q}_i(a_1, ..., a_n)\mathcal{Z}(a_1,  ..., a_n)$ to absorb all $\mathcal{Q}_i$ into $\mathcal{Z}$, and subsequently shift the contour \emph{simultaneously} for all integration variables $a_i$. However, viewing the integral as a sequential integration, after the first, say, the $a_1$ integration, the remaining integrand with respect to $a_2, a_3, \cdots$ no longer enjoys the simple periodic property. We will see a simple example where the integrand is elliptic, that after the first integral the remaining integrand is no longer elliptic due to the presence of Eisenstein series.

The second way to phrase the problem is the incorrect application of residue theorem with multiple complex variables. In \cite{Benini:2018mlo}, after absorbing the $\mathcal{Q}$ into $\mathcal{Z}$ and shifting the contours, one formally encounters multivariate (for simplicity, consider a rank-two case) contour integral of the form
\begin{equation}
  \mathcal{I} = \left({\oint_{|z_2| = 1} - \oint_{|z_2| = |q|}}\right)
  \left({\oint_{|z_1| = 1} - \oint_{|z_1| = |q|}}\right) \frac{dz_1}{2\pi i} f(z_1, z_2) \ .
\end{equation}
The result in \cite{Benini:2018mlo} follows from picking up poles within the annulus region bounded by the tori, namely those satisfying simultaneously
\begin{equation}
  |q| < |z_1| < 1, \qquad |q| < |z_2| < 1\ .
\end{equation}
However, this is not true in general. The toroidal contour actually inevitably receives contributions from poles outside of the annulus region, which can be seen by a more careful deformation of the contour avoiding any singularities.

Let us present two simple examples. Consider the simple rational integrand,
\begin{equation}
  f(z_1, z_2) = \frac{1}{(z_1 - a)(z_2 - b)(z_1 - z_2 c)} \ , \qquad |a| \sim |q|^{\frac{1}{3}}, \quad |b|, |c| \sim |q|^{\frac{2}{3}} \ , \quad |q| < 1 \ .
\end{equation}
We will compute the integral using two different methods. First we carry out the sequential integration. Consider the $z_2$ integration along $|z_2| = 1$, the inner $z_1$ integration picks up residues from $z_1 = a$ and $z_1 = z_2 c$ which sit in between the $|z_1| = 1$ and $|z_1| = |q|$, giving total residue zero,
\begin{equation}
  \mathop{\operatorname{Res}}_{z_1 = a} + \mathop{\operatorname{Res}}_{z_1 = z_2 c} = \frac{1}{(b - z_2)(-a + c z_2)} - \frac{1}{(b - z_2)(-a + c z_2)} = 0 \ .
\end{equation}
For the $z_2$ integration along $|z_2| = |q|$, the inner $z_1$ integration picks up only the pole $z_1 = a$ (the pole $z_1 = z_2 c$ sits outside of the contour since $|z_2 c| \sim |q|^{\frac{5}{3}} < |q|$), and the left over $z_2$ integration also vanishes (since $|b| \sim |q|^{\frac{2}{3}} > |q|$, and $|a/c| \sim |q|^{-\frac{1}{3}} > |q|$),
\begin{equation}
  - \oint_{|z_2| = |q|}\frac{dz_2}{2\pi i} \frac{1}{(b - z_2)(-a + c z_2)}
  = 0\ ,
\end{equation}
Hence, sequential integration gives
\begin{equation}
  \mathcal{I} = 0 \ .
\end{equation}
Equivalently, computing the constant term of $z_1 z_2 f(z_1, z_2)$ gives zero,
\begin{equation}
  \operatorname{CT} \frac{z_1 z_2}{(z_1 - a)(z_2 - b)(z_1 - z_2 c)} = 0 \ .
\end{equation}

The second method is to sum over poles of the full integrand $f(z_1, z_2)$ within the region $|q| < |z_1| < 1$ and $|q| < |z_2| < 1$, given by
\begin{equation}
  z_1 = a, z_2 = b \qquad \Rightarrow \qquad \mathcal{I} = \frac{1}{a - b c} \ .
\end{equation}
Apparently, this is different from the result from sequential integration. Note that the pole at $(z_1, z_2) = (bc, b)$ is outside of the annulus region since $|bc| = |q|^{\frac{4}{3}} < |q|$, which is precisely the pole missing from the approach.

The discrepancy does not mean that the residue theorem is inapplicable in the multivariate scenario. Instead, it only means that we need to be careful with the pole to be collected: the $n$-dimensional contour does not separate an ``interior'' from an ``exterior'' in $\mathbb{C}^n$, hence the bounding contour does not necessarily mean picking up all the poles ``inside the bounded region'', in this case, the higher dimensional annulus region. The multivariate contour integral enjoys the standard invariance under continuous deformation of the contour, provided it doesn't cross any singularities. Let us denote the above contour as $T^2(1,1) - T^2(1, |q|) - T^2(|q|, 1) + T^2(|q|, |q|)$, where $T^2$ suggests the toroidal shape of the contour, and the two arguments denote the radii of the two circles inside the $z_1, z_2$ planes. The singularities sit at the following loci,
\begin{equation}
  \{z_1 = a, \forall z_2\} \cup \{\forall z_1, z_2 = b\} \cup \{z_1 = z_2 c\} \ .
\end{equation}
These planes intersect at the \emph{simultaneous poles} $(a, b)$ and $(bc, b)$. The second pole sits outside of the region $|q| < |z_1| < 1$ and $|q| < |z_2| < 1$.
\begin{itemize}
  \item The $T^2(1,1)$ can be continuously deformed to ``encircle'' $(a,b), (bc, b) \in \mathbb{C}^2$. First the $|z_2| = 1$ contour shrinks to $|z_2| \gtrsim |q|^{\frac{2}{3}} = |b|$. Along the way, the contour never intersects the singular loci: during the process $|q|^{\frac{4}{3}} \lesssim |z_2 c| < |q^{\frac{2}{3}}| < 1$, and $z_1 = z_2 c$ is never satisfied as $|z_1| =1$. Then $|z_1| = 1$ contour shrinks to $|z_1| \gtrsim |q|^{\frac{1}{3}} = |a|$, and again it never crosses the singularity $z_1 = z_2 c$ since $|z_2c| < |q|^{\frac{2}{3}} < |z_1| < 1$. Apparently, if the contour further shrinks $z_1$ or $z_2$, it will eventually encounter either $\{z_1 = a, \forall z_2\}$, $\{\forall z_1, z_2 = b\}$ or $\{z_1 = z_2 c\}$. Therefore the $z_1, z_2$ contour $T^2(1,1)$ links all three singular planes, and the residue theorem picks up the Grothendieck residues of all the intersections (simultaneous poles)
  \begin{equation}
    \oint_{T^2(1,1)} \frac{dz_1}{2\pi i} \frac{dz_2}{2\pi i} f(z_1, z_2) = \frac{1}{J(a,b)}\operatorname{Res}_{(a,b)} + \frac{1}{J(bc, b)} \operatorname{Res}_{(bc, b)} = 0\ ,
  \end{equation}
  where $J$ denotes the Jacobian at the pole. 
  \item The contour $T^2(1, |q|)$ can shrink to a point in the $z_2$ plane without crossing either of the singular loci, hence contributing zero.
  \item The contour $T^2(|q|, 1)$ can also shrink to a point $z_1 = 0$ along $z_1$-plane, since $|z_1| \le |q| < |q|^{\frac{2}{3}} = |z_2 c|$.
  \item Finally, the last contour $T^2(|q|, |q|)$ can also be shrunk to $z_1 = 0$.
\end{itemize}
Therefore, the integral sums to zero, as already predicted by the sequential integration method or constant term. To summarize, the problem with the original Bethe ansatz formula is that it chooses to discard the poles outside of the annulus region, in this case, the pole at $(z_1, z_2) = (bc, b)$.

Next we consider an elliptic example, which is more relevant to the index computation,
\begin{align}\label{eq:example-original}
  \mathcal{I} = \oint_{|a_i| = 1} \frac{da_1}{2\pi i a_1} \frac{da_2}{2\pi i a_2} 
  \underbrace{\frac{\vartheta_1(\mathfrak{a}_1 - \mathfrak{a}_2)^2 \vartheta_1(\mathfrak{a}_1 + \mathfrak{a}_2)^2}{
    \prod_\pm
    \vartheta_4(\pm(\mathfrak{a}_1 - \mathfrak{a}_2) + \mathfrak{b}_1)
    \vartheta_4(\pm(\mathfrak{a}_1 + \mathfrak{a}_2) + \mathfrak{b}_2)
  }}_{\mathcal{Z}} \ .
\end{align}
It is straightforward to see that the integrand built out of the Jacobi theta functions is elliptic with respect to both $\mathfrak{a}_1$ and $\mathfrak{a}_2$, \emph{e.g.,}
\begin{equation}
  \mathcal{Z}(\mathfrak{a}_1 + \tau, \mathfrak{a}_2)
  = \mathcal{Z}(\mathfrak{a}_1, \mathfrak{a}_2 + \tau) = \mathcal{Z}(\mathfrak{a}_1, \mathfrak{a}_2) \ .
\end{equation}

First we show that integration breaks the simple periodicity, in this case, the ellipticity. One can check the value of $a_1$-integration by first expanding the integrand in $q$ series, and then performing the $a_1$-integration term by term. The result is
\begin{equation}
  (4 + a_2^2 + a_2^{-2})q^{\frac{1}{2}} - \frac{2(1 + a_2^2)^2 (b_2 + b_1^2 b_2 + b_1(1 + b_2^2))}{a_2^2 b_1 b_2}q + \cdots \ .
\end{equation}
Alternatively, using the exact integration formula \cite{Pan:2021mrw}, the $a_1$-integration gives the closed-form
\begin{align}
= & \ \frac{\vartheta_1(\mathfrak{a}_2)^{4}}
{\prod_{\alpha=\pm 1}\prod_{i=1}^{2}
\vartheta_4\bigl(\mathfrak{a}_2+\alpha\,\mathfrak{b}_i\bigr)}
\nonumber\\
&\quad+
\sum_{\alpha=\pm 1}\sum_{i=1}^{2}
\frac{i\,
\vartheta_4\bigl(2\mathfrak{a}_2+\alpha\,\mathfrak{b}_i\bigr)^{2}\,
\vartheta_4\bigl(\mathfrak{b}_i\bigr)^{2}}
{\eta(\tau)^{3}\,
\vartheta_1\bigl(2\alpha\,\mathfrak{b}_i\bigr)
\prod_{j\neq i}\prod_{\beta=\pm 1}
\vartheta_1\bigl(
2\mathfrak{a}_2+\alpha\,\mathfrak{b}_i+\beta\,\mathfrak{b}_j
\bigr)}E_1\begin{bmatrix}-1\\ a_2 b_i^{\alpha}\end{bmatrix} \ .
\end{align}
This expands to exactly the same $q$-series. Now, the first term in the closed-form is elliptic in $\mathfrak{a}_2$,
\begin{equation}
  \frac{\vartheta_1(\mathfrak{a}_2 + \tau)^{4}}{\prod_{\alpha = \pm} \prod_{i = 1}^2 \vartheta_4(\mathfrak{a}_2 + \tau + \alpha \mathfrak{b}_i)}
  =
  \frac{\vartheta_1(\mathfrak{a}_2)^{4}}{\prod_{\alpha = \pm} \prod_{i = 1}^2 \vartheta_4(\mathfrak{a}_2 + \alpha \mathfrak{b}_i)} \ .
\end{equation}
The theta ratio in the second term is also elliptic in $\mathfrak{a}_2$, however, the Eisenstein series $E_1 \Big[\substack{-1 \\  a_2 b_i^{\alpha}} \Big]$ is not,
\begin{equation}
  E_1 \begin{bmatrix}
    -1 \\ a_2 q b_i^\alpha
  \end{bmatrix}
  = E_1 \begin{bmatrix}
    -1 \\ a_2 b_i^\alpha
  \end{bmatrix} - 1 \ .
\end{equation}
The loss of periodicity after integration means that simultaneous shift of all variables and absorption of the $\mathcal{Q}$ functions into $\mathcal{Z}$ will in general \emph{alter} the value of the integral. In fact, the difference before and after $\tau$ (or $q$) shift is
\begin{align}
\sum_{\alpha = \pm}& \ \frac{
  i\vartheta_4(2\mathfrak{a}_2+\alpha\mathfrak{b}_1)^2
  \vartheta_4(\mathfrak{b}_1)^2
}{
  \eta(\tau)^3
  \vartheta_1(2\alpha\mathfrak{b}_1)
  \vartheta_1(2\mathfrak{a}_2+\alpha\mathfrak{b}_1-\mathfrak{b}_2)
  \vartheta_1(2\mathfrak{a}_2+\alpha\mathfrak{b}_1+\mathfrak{b}_2)
}
\nonumber\\
&+\frac{
  i\vartheta_4(2\mathfrak{a}_2+\alpha\mathfrak{b}_2)^2
  \vartheta_4(\mathfrak{b}_2)^2
}{
  \eta(\tau)^3
  \vartheta_1(2\alpha\mathfrak{b}_2)
  \vartheta_1(2\mathfrak{a}_2+\alpha\mathfrak{b}_2-\mathfrak{b}_1)
  \vartheta_1(2\mathfrak{a}_2+\alpha\mathfrak{b}_2+\mathfrak{b}_1)
} \ ,
\end{align}
which is a combination of the Grothendieck residues of some of the poles inside the annulus.

Next we show again that picking up poles within the annulus does not give the correct integral. As will be shown shortly in the next section, the integral is equivalent to 
\begin{align}\label{eq:example-rewrite}
  \mathcal{I} = & \ \oint_{\mathcal{C}_2} \frac{da_2}{2\pi i a_2}\oint_{\mathcal{C}_1} \frac{da_1}{2\pi i a_1}
  \frac{\vartheta_1(\mathfrak{a}_1 - \mathfrak{a}_2)^2 \vartheta_1(\mathfrak{a}_1 + \mathfrak{a}_2)^2}{
    \prod_\pm
    \vartheta_4(\pm(\mathfrak{a}_1 - \mathfrak{a}_2) + \mathfrak{b}_1)
    \vartheta_4(\pm(\mathfrak{a}_1 + \mathfrak{a}_2) + \mathfrak{b}_2)
  }\prod_{i = 1}^2 \frac{1}{1-Q_i}  \\
  = & \ \oint_{\mathcal{C}_2} \frac{da_2}{2\pi i a_2}\oint_{\mathcal{C}_1} \frac{da_1}{2\pi i a_1}
  \frac{\vartheta_1(\mathfrak{a}_1 - \mathfrak{a}_2)^2 \vartheta_1(\mathfrak{a}_1 + \mathfrak{a}_2)^2}{
    \prod_\pm
    \vartheta_4(\pm(\mathfrak{a}_1 - \mathfrak{a}_2) + \mathfrak{b}_1)
    \vartheta_4(\pm(\mathfrak{a}_1 + \mathfrak{a}_2) + \mathfrak{b}_2)
  }\prod_{i = 1}^2 E_1 \begin{bmatrix}
    -1 \\ a_iq^{-1/2}
  \end{bmatrix} \ ,\nonumber
\end{align}
where $Q_i = 1 - E_1 [\substack{-1 \\ a_iq^{-1/2}}]^{-1}$. The integration contour is given by boundary toroidal contours $\mathcal{C}_1, \mathcal{C}_2$
\begin{equation}
  \mathcal{C}_1 : \{|a_1| = 1\} \cup \{|a_1| = |q|\}^-, \qquad
  \mathcal{C}_2 : \{|a_2| = 1\} \cup \{|a_2| = |q|\}^- \ .
\end{equation}
The Eisenstein series are the necessary price to pay when transforming the original unit-circle integral into the boundary of the (real) four-dimensional annulus.  This integral is reminiscent of the $\mathcal{C}$-integral appearing in the Bethe ansatz method. We will compute the integration via two methods: sequential integration and the multivariate residue theorem. First we consider $|a_2| = 1$ integral. The $a_1$ integration inside picks up residues from poles
\begin{equation}
  \mathfrak{a}_1 = \tau, \quad \frac{\tau}{2} + \mathfrak{a}_2 \pm \mathfrak{b}_1, 
  \quad \frac{\tau}{2} - \mathfrak{a}_2 \pm \mathfrak{b}_2 \ ,
\end{equation}
giving 
\begin{align}
&
\frac{
  \vartheta_1(\mathfrak{a}_2)^4\,
}{
  \vartheta_4(-\mathfrak{a}_2+\mathfrak{b}_1)\,
  \vartheta_4(\mathfrak{a}_2+\mathfrak{b}_1)\,
  \vartheta_4(-\mathfrak{a}_2+\mathfrak{b}_2)\,
  \vartheta_4(\mathfrak{a}_2+\mathfrak{b}_2)
}E_1\begin{bmatrix}
    -1 \\
    \frac{a_2}{\sqrt{q}}
  \end{bmatrix}
\nonumber\\
&\quad
+
\frac{
  i\,\vartheta_4(\mathfrak{b}_1)^2\,
  \vartheta_4(-2\mathfrak{a}_2+\mathfrak{b}_1)^2\,
}{
  \eta(\tau)^3\,
  \vartheta_1(2\mathfrak{b}_1)\,
  \vartheta_1(2\mathfrak{a}_2-\mathfrak{b}_1+\mathfrak{b}_2)\,
  \vartheta_1(-2\mathfrak{a}_2+\mathfrak{b}_1+\mathfrak{b}_2)
}E_1\begin{bmatrix}
    -1 \\
    \frac{a_2}{\sqrt{q}}
  \end{bmatrix}
  E_1\begin{bmatrix}
    -1 \\
    \frac{a_2}{b_1}
  \end{bmatrix}
\nonumber\\
&\quad
-
\frac{
  i\,\vartheta_4(\mathfrak{b}_1)^2\,
  \vartheta_4(2\mathfrak{a}_2+\mathfrak{b}_1)^2\,
}{
  \eta(\tau)^3\,
  \vartheta_1(2\mathfrak{b}_1)\,
  \vartheta_1(-2\mathfrak{a}_2-\mathfrak{b}_1+\mathfrak{b}_2)\,
  \vartheta_1(2\mathfrak{a}_2+\mathfrak{b}_1+\mathfrak{b}_2)
}E_1\begin{bmatrix}
    -1 \\
    \frac{a_2}{\sqrt{q}}
  \end{bmatrix}
  E_1\begin{bmatrix}
    -1 \\
    a_2 b_1
  \end{bmatrix}
\nonumber\\
&\quad
-
\frac{
  i\,\vartheta_4(\mathfrak{b}_2)^2\,
  \vartheta_4(-2\mathfrak{a}_2+\mathfrak{b}_2)^2\,
}{
  \eta(\tau)^3\,
  \vartheta_1(2\mathfrak{b}_2)\,
  \vartheta_1(-2\mathfrak{a}_2-\mathfrak{b}_1+\mathfrak{b}_2)\,
  \vartheta_1(-2\mathfrak{a}_2+\mathfrak{b}_1+\mathfrak{b}_2)
}E_1\begin{bmatrix}
    -1 \\
    \frac{a_2}{\sqrt{q}}
  \end{bmatrix}
  E_1\begin{bmatrix}
    -1 \\
    \frac{a_2}{b_2}
  \end{bmatrix}
\nonumber\\
&\quad
+
\frac{
  i\,\vartheta_4(\mathfrak{b}_2)^2\,
  \vartheta_4(2\mathfrak{a}_2+\mathfrak{b}_2)^2\,
}{
  \eta(\tau)^3\,
  \vartheta_1(2\mathfrak{b}_2)\,
  \vartheta_1(2\mathfrak{a}_2-\mathfrak{b}_1+\mathfrak{b}_2)\,
  \vartheta_1(2\mathfrak{a}_2+\mathfrak{b}_1+\mathfrak{b}_2)
}E_1\begin{bmatrix}
    -1 \\
    \frac{a_2}{\sqrt{q}}
  \end{bmatrix}
  E_1\begin{bmatrix}
    -1 \\
    a_2 b_2
  \end{bmatrix} \ .
\end{align}
Then we look at the one with $|a_2| = |q|$. The inner $a_1$ integration picks up residues from poles
\begin{equation}
  \mathfrak{a}_1 = \tau, \quad - \frac{\tau}{2} + \mathfrak{a}_2 \pm \mathfrak{b}_1, \quad \frac{3\tau}{2} - \mathfrak{a}_2 \pm \mathfrak{b}_2 \ ,
\end{equation}
which are inside the region $|q| \le |a_1| < 1$, giving
\begin{align}
&\frac{
\vartheta_1(\mathfrak{a}_2)^4\,
}{
\vartheta_4(\mathfrak{a}_2-\mathfrak{b}_1)\,
\vartheta_4(\mathfrak{a}_2+\mathfrak{b}_1)\,
\vartheta_4(\mathfrak{a}_2-\mathfrak{b}_2)\,
\vartheta_4(\mathfrak{a}_2+\mathfrak{b}_2)
}E_1\begin{bmatrix}
-1 \\
a_2/\sqrt{q}
\end{bmatrix}
\\[6pt]
&
+\sum_{i=1}^{2}\sum_{\alpha=\pm 1}
\frac{
i\,\alpha\,
\vartheta_4(\mathfrak{b}_i)^2\,
\vartheta_4(2\mathfrak{a}_2+\alpha\mathfrak{b}_i)^2\,
}{
\eta(\tau)^3\,
\vartheta_1(2\mathfrak{b}_i)\,
\prod_{\beta=\pm 1}
\vartheta_1(2\mathfrak{a}_2+\alpha\mathfrak{b}_i+\beta\mathfrak{b}_{3-i})
}E_1\begin{bmatrix}
-1 \\
a_2/\sqrt{q}
\end{bmatrix}
\left(
1+
E_1\begin{bmatrix}
-1 \\
a_2 b_i^{\alpha}
\end{bmatrix}
\right)\ . \nonumber
\end{align}

With the inner $a_1$-integration done, we will carry out the outer $a_2$ integrals over $|a_2| = 1$ and $|a_2| = |q|$. Note that at this stage, the two $a_2$-integrand in the $|a_2| = 1$ and $|a_2| = |q|$ integrals are different: although the original integrand (\ref{eq:example-original}) along the two-dimensional unit tori is doubly elliptic, the integrand in the rewritten form (\ref{eq:example-rewrite}) is no longer elliptic due to the Eisenstein series. Therefore we cannot combine the two integrals to apply residue theorem. Instead, we need to compute the two circle integration carefully. The exact computation method is discussed in the next section, and here we simply present the results.  The $|a_2| = 1$ integral gives 
\begin{align}
  \frac{\vartheta_4(\mathfrak{b}_1)^2 \vartheta_4(\mathfrak{b}_2)^2}{\eta(\tau)^6\prod_{i = 1,2}\vartheta_1(2 \mathfrak{b}_i)} \sum_{\alpha = \pm} \bigg(
      E_2 \begin{bmatrix}
        \alpha \\ b_2^{1/2} b_1^{1/2 }
      \end{bmatrix}
      - E_2 \begin{bmatrix}
        \alpha \\ b_2^{1/2} b_1^{-1/2}
      \end{bmatrix}
    \bigg) 
\end{align}
The $|a_2| = |q|$ integral gives
\begin{align}
  & \ - \frac{2i}{\eta(\tau)^3 \vartheta_1(\mathfrak{b}_1 \pm \mathfrak{b}_2)}\bigg(
    \frac{\vartheta_4(\mathfrak{b}_1)^4}{\vartheta_1(2\mathfrak{b}_1)}E_1
    \begin{bmatrix}
      -1 \\ b_1
    \end{bmatrix}
    - \frac{\vartheta_4(\mathfrak{b}_2)^4}{\vartheta_1(2\mathfrak{b}_2)}E_1
    \begin{bmatrix}
      -1 \\ b_2
    \end{bmatrix}
    \bigg) \nonumber \\
    & \ - \frac{\vartheta_4(\mathfrak{b}_1)^2 \vartheta_4(\mathfrak{b}_2)^2}{\eta(\tau)^6\prod_{i = 1,2}\vartheta_1(2 \mathfrak{b}_i)} \sum_{\alpha = \pm} \bigg(
      E_2 \begin{bmatrix}
        \alpha \\ b_2^{1/2} b_1^{1/2}
      \end{bmatrix}
      - E_2 \begin{bmatrix}
        \alpha \\ b_2^{1/2} b_1^{-1/2}
      \end{bmatrix}
    \bigg)  \ .
\end{align}
Combining the above, the final result is then given by the difference of the two,
\begin{align}
  \mathcal{I} = & \ \frac{2i}{\eta(\tau)^3 \vartheta_1(\mathfrak{b}_1 \pm \mathfrak{b}_2)}\bigg(
    \frac{\vartheta_4(\mathfrak{b}_1)^4}{\vartheta_1(2\mathfrak{b}_1)}E_1
    \begin{bmatrix}
      -1 \\ b_1
    \end{bmatrix}
    - \frac{\vartheta_4(\mathfrak{b}_2)^4}{\vartheta_1(2\mathfrak{b}_2)}E_1
    \begin{bmatrix}
      -1 \\ b_2
    \end{bmatrix}
    \bigg) \nonumber \\
    & \ + \frac{2\vartheta_4(\mathfrak{b}_1)^2 \vartheta_4(\mathfrak{b}_2)^2}{\eta(\tau)^6\prod_{i = 1,2}\vartheta_1(2 \mathfrak{b}_i)} \sum_{\alpha, \epsilon = \pm} \bigg(
      E_2 \begin{bmatrix}
        \alpha \\ b_2^{1/2} b_1^{1/2}
      \end{bmatrix}
      - E_2 \begin{bmatrix}
        \alpha \\ b_2^{1/2} b_1^{-1/2}
      \end{bmatrix}
    \bigg) \ .
\end{align}
As $q$-series expansion, 
\begin{equation}
  \mathcal{I} = 4 q^{\frac{1}{2}} - 4(b_1 + b_1^{-1} + b_2 + b_2^{-1}) q 
  + 4(8 + b_1^{\pm 2} + b_2^{\pm2} + (b_1b_2)^\pm + (b_1/b_2)^\pm)q^{\frac{3}{2}} + \cdots \ .
\end{equation}
This is precisely the same as the standard constant term computation.

Let us try to collect the Grothendieck residue of the poles within the annulus region. The relevant poles are given by
\begin{align}
  & \ (\tau, \tau), (\tau, \frac{\tau}{2} + {\mathfrak{b}_1}), (\tau, \frac{\tau}{2} - {\mathfrak{b}_1}), (\tau, \frac{\tau}{2} + {\mathfrak{b}_2}), (\tau, \tau/2 - {\mathfrak{b}_2}), \nonumber\\
   & \ (\frac{\tau}{2} - {\mathfrak{b}_1}, \tau), (\frac{\tau}{2} + {\mathfrak{b}_1}, \tau), (\frac{\tau}{2} + {\mathfrak{b}_2}, \tau), (\frac{\tau}{2} - {\mathfrak{b}_2}, \tau), \nonumber\\
   & \ (\tau - {\mathfrak{b}_1}/2 + {\frac{\mathfrak{b}_2}{2}}, \frac{\tau}{2} + {\mathfrak{b}_1}/2 + {\frac{\mathfrak{b}_2}{2}}), (\tau/2 - {\mathfrak{b}_1}/
   2 - {\frac{\mathfrak{b}_2}{2}}, \tau + {\mathfrak{b}_1}/2 - {\mathfrak{b}_2}/
   2), \nonumber\\
   & \ (\tau/2 + {\mathfrak{b}_1}/2 + {\mathfrak{b}_2}/
   2, \tau - {\mathfrak{b}_1}/2 + {\mathfrak{b}_2}/
   2), (\tau + {\mathfrak{b}_1}/2 - {\frac{\mathfrak{b}_2}{2}}, \frac{\tau}{2} - {\mathfrak{b}_1}/2 - {\frac{\mathfrak{b}_2}{2}}) \ .
\end{align}
The multivariate residue theorem gives a sum of Grothendieck residues,
\begin{equation}
  - \frac{i\vartheta_4(\mathfrak{b}_1)^4}{2\pi^2 \eta(\tau)^3\vartheta_1(2 \mathfrak{b}_1)\vartheta_1(\mathfrak{b}_1 \pm \mathfrak{b}_2)} E_1 \begin{bmatrix}
    -1 \\ b_1
  \end{bmatrix}
  + \frac{i\vartheta_4(\mathfrak{b}_2)^4}{2\pi^2 \eta(\tau)^3\vartheta_1(2 \mathfrak{b}_2)\vartheta_1(\mathfrak{b}_1 \pm \mathfrak{b}_2)} E_1 \begin{bmatrix}
    -1 \\ b_2
  \end{bmatrix} \ .
\end{equation}
However, this is different from the expected result above, off by the $E_2$ terms.

To summarize, the idea in \cite{Benini:2018mlo} turning iterative contour integral into a ``residue theorem'' type problem is possible, provided one considers all the poles from the intersections of the singularities the integration contour ``links''. Unfortunately, to completely account for such higher dimensional linking we need to carefully analyze the homology of the integration contour and the singularities, which is a highly nontrivial task. In the next section we propose an intermediate approach: instead of working out the exact higher dimensional homology, we simply identify the homology structure one dimension at a time. In the end, the analysis produces the exact closed-form integral.

\section{Difference equation and integration formula}\label{sec:integration-formula}
\subsection{Single variable integration formula}

As discussed in detail in \cite{Pan:2021mrw}, the multivariate integral can be performed iteratively and the first integral of an elliptic function is straightforward. The only problem comes after the first integral, where the subsequent integrands inevitably contain Eisenstein series, which are not elliptic. 

Let us first reexamine the first integral. Consider $f(a)$ such that $f(aq) = f(a)$. Viewing $a = e^{2\pi i \mathfrak{a}}$ and $f$ as a function of $\mathfrak{a}$, $f$ is elliptic, $f(\mathfrak{a} + 1) = f(\mathfrak{a} + \tau) = f(\mathfrak{a})$. As usual we assume $\operatorname{Im}\tau > 0$ and therefore $|q| < 1$. Due to the ellipticity, poles of $f$ generically form infinite sequence that converges to the origin, leading to a non-isolated singularity. The presence of such a singularity is one of the main difficulties in evaluating the integral by applying the standard residue theorem. However, it is possible to ``shield'' the integral from such bad behavior. 

Consider the contour integral
\begin{equation}
  \mathcal{I} \coloneqq \oint_{|a| = 1} \frac{da}{2\pi i a} f(a) \ .
\end{equation}
Take a meromorphic function $Q(a)$, and consider adding a trivial factor into the integral, completely similar to the \cite{Benini:2018mlo},
\begin{equation}
  \mathcal{I} = \oint_{|a| = 1} \frac{da}{2\pi i a} \frac{1 - Q(a)}{1 - Q(a)} f(a)
  = \oint_{|a| = 1} \frac{da}{2\pi i a} \frac{f(a)}{1 - Q(a)} 
  - \oint_{|a| = 1} \frac{da}{2\pi i a} \frac{Q(a)f(a)}{1 - Q(a)} \ .
\end{equation}
Note that the $Q$'s appearing in \cite{Benini:2018mlo} come from the large gauge transformation of the integrand $\mathcal{Z}$, whereas here $Q$ is just some unknown function introduced by hand.

We would like to absorb the factor $Q(a)$ into a shift of integration contour, such that the following integral equation holds,
\begin{equation}
  \oint_{|a = 1|} \frac{da}{2\pi i a} \frac{Q(a)f(a)}{1 - Q(a)}
  = \oint_{|a| = 1} \frac{da}{2\pi i a} \frac{f(aq)}{1 - Q(aq)}
  = \oint_{|a| = q} \frac{da}{2\pi i} \frac{1}{a} \frac{f(a)}{1 - Q(a)} \ .
\end{equation}
Here, the first equation is the intended relation, while the second equality is simply a change of variable $a \to aq$ and therefore a shift of contour. If this is achieved, the original integral becomes
\begin{equation}
  \mathcal{I} = \left({\oint_{|a| = 1} - \oint_{|a| = |q|}}\right) \frac{da}{2\pi i} \frac{1}{a}\frac{f(a)}{1 - Q(a)}
  = \sum_{J} \operatorname{Res}_{a = a_J} \frac{1}{a}\frac{f(a)}{1 - Q(a)} \ ,
\end{equation}
where we collect poles $a_J$ of the full product $f(a)/(1 - Q(a))$ within the annulus $|q| \le |a| < 1$ (note that poles on $|q| = |a|$ but not $|a| = 1$ are included, a different choice from \cite{Pan:2021mrw}). Recall that in \cite{Benini:2018mlo} the same integral equality also holds, but in a different manner: roughly, the integrand $f$ there is non-elliptic but $Q$ is,
\begin{equation}
  f(aq) = Q(a) f(a) , \qquad
  Q(aq) = Q(a) \ .
\end{equation}
Also note that here we need all poles from both $(1 - Q(a))$ and $f(a)$.

The intended equality is an integral equation, which can be trivially solved if the integrands are equal
\begin{equation}
  \frac{Q(a)f(a)}{1 - Q(a)} = \frac{f(aq)}{1 - Q(aq)} = \frac{f(a)}{1 - Q(aq)} \ .
\end{equation}
We have used the ellipticity of $f$ in the second equality. This leads to a functional equation for $Q(a)$,
\begin{equation}
  \frac{Q(a)}{1 - Q(a)} = \frac{1}{1 - Q(aq)} \ .
\end{equation}
Interestingly, this functional equation has a family of meromorphic solutions
\begin{equation}
  Q(a) = 1 - \frac{1}{E_1 \Big[\substack{-1 \\ a / (bq^{\frac{1}{2}})}\Big]} , \qquad
  \Rightarrow \qquad
  \frac{1}{1 - Q(a)} = E_1 \Big[\substack{-1 \\ a / (bq^{\frac{1}{2}})}\Big] \ ,
\end{equation}
where $b$ is an arbitrary parameter. The above residue formula then leads to the following integration formula
\begin{equation}
  \mathcal{I} = \sum_{J} \operatorname{Res}_{a = a_J} \frac{1}{a} f(a) E_1 \begin{bmatrix}
    -1 \\ a / (bq^{\frac{1}{2}})
  \end{bmatrix} \ .
\end{equation}

Again we stress that the poles $a_J$ come from the product $\frac{1}{a}f(a)E_1\Big[\substack{-1 \\ a / (bq^{\frac{1}{2}})} \Big]$. The factor $\frac{1}{a} E_1\Big[\substack{-1 \\ a / (bq^{\frac{1}{2}})} \Big]$ has a simple pole at $a/ (bq^{\frac{1}{2}}) = q^{+\frac{1}{2}}$ (\emph{i.e., $a = bq$}) with unit residue,
\begin{equation}
  \operatorname{Res}_{a = bq} \frac{1}{a} E_1 \begin{bmatrix}
    -1 \\ a/ (bq^{\frac{1}{2}})
  \end{bmatrix} = 1 \ .
\end{equation}
Therefore, assuming the function $f$ has only simple poles at $a_j$, the integral reads
\begin{equation}
  \mathcal{I} = f(bq) + \sum_{j} R_j E_1 \begin{bmatrix}
    -1 \\ a_j / (bq^{\frac{1}{2}})
  \end{bmatrix}, \qquad
  R_j = \operatorname{Res}_{a = a_j} \frac{1}{a} f(a) \ ,
\end{equation}
where $a_j$ denotes poles of $f$. This is precisely the formula derived in \cite{Pan:2021mrw}, where $b$ is the arbitrary reference value, and we have migrated all real pole contributions to those with $|a| = |q|$, treated as imaginary.

Given the freedom of choosing $b$, we can choose to set $b$ to collide with one of the poles of $f$, say $a_1$. In this case adding the $n(a)$ does not change the pole,
\begin{equation}
  \mathcal{I} = \sum_{j} \mathop{\operatorname{Res}}_{a = a_j} \frac{1}{a} f(a) E_1 \begin{bmatrix}
    -1 \\ a_j / (a_1 )q^{\frac{1}{2}}
  \end{bmatrix}\ .
\end{equation}
However, one must be careful that the pole $a_1$ is a double pole in computing the residue; but this is not difficult to handle.

\subsection{Multivariate integration formula}

Next we extend the above formula to the case of multiple variables. After the first integral, Eisenstein series are produced, and we expect the further integrals will produce polynomials of Eisenstein series. Hence we focus on the following type of integrals,
\begin{equation}
  \mathcal{I} = \oint \frac{da}{2\pi i a}f(a) P(a)  \ ,
\end{equation}
where $P(a)$ denotes some monomial
\begin{equation}
  P(a) = E_{k_1}\begin{bmatrix}
    \pm 1 \\ a^{K_1} b_1
  \end{bmatrix} \cdots
  E_{k_n}\begin{bmatrix}
    \pm 1 \\ a^{K_n} b_n
  \end{bmatrix} \ .
\end{equation}
Here the $\pm 1$ are arbitrary independent signs which do not have to be the same. When some $b_i$ and $K_i$ are identical, we continue to treat the corresponding Eisenstein series as separate factors in the subsequent algorithm.

Again, we consider a similar manipulation,
\begin{equation}
  \mathcal{I} 
  = \oint_{|a| = 1} \frac{da}{2\pi i a} \frac{f(a)P(a)}{1 - Q(a)}
  - \oint_{|a| = 1} \frac{da}{2\pi i a} \frac{Q(a) f(a)P(a)}{1 - Q(a)} \ ,
\end{equation}
and we expect the $Q(a)$ can be absorbed by performing a shift of contour,
\begin{equation}
  \frac{Q(a)f(a) P(a)}{1 - Q(a)}
  = \frac{f(aq) P(a q)}{1 - Q(aq)}
  = \frac{f(a) P(a q)}{1 - Q(aq)} \ ,
\end{equation}
where we have used the ellipticity of $f$ in the second equality. This leads to a functional equation for $Q(a)$,
\begin{equation}
  \frac{Q(a)P(a)}{1 - Q(a)} = \frac{P(aq)}{1 - Q(aq)} \ .
\end{equation}
Since $P(a)$ is known, all we need to do is find the right function $Q(a)$. To simplify the problem, we denote $Q(a) = 1 + P(a)/n(a)$. The functional equation then becomes a simple difference equation,
\begin{equation}
  \frac{Q(a)P(a)}{1 - Q(a)} - \frac{P(aq)}{1 - Q(aq)} = n(aq) - n(a) - P(a) = 0 \ .
  \end{equation}
The main observation is that the solution $n(a)$ and therefore $Q(a)$ can be obtained algorithmically, and the solution is non-unique. Before describing the algorithm, we first finish the residue computation. Once the functional equality is satisfied, using $\frac{P(a)}{1-Q(a)} = -n(a)$, we have
\begin{equation}
  \mathcal{I} = \oint _{|a| = 1} \frac{da}{2\pi i a} \frac{f(a)P(a)}{1 - Q(a)}
  - \oint_{|a| = |q|} \frac{da}{2\pi i a} \frac{f(a)P(a)}{1 - Q(a)}
  = - \sum_{J} \operatorname{Res}_{a = a_J} \left({\frac{1}{a} f(a)n(a)}\right)\ .
\end{equation}
Here $a_J$ denotes poles of the product $a^{-1}f(a)n(a)$ within the annulus $|q| \le |a| < 1$. The final result is independent of the specific choice of the solution $n(a)$.

Next we describe one algorithm to solve the functional equation. First we define ``complexity'' $C(M)$ of a monomial $M = E_{k_1}\Big[\substack{\pm 1 \\ a b_1}\Big] \cdots E_{k_\ell}\Big[\substack{\pm 1 \\ a b_\ell}\Big]$,
\begin{align}
  C(M) \coloneqq & \ (\ell(M), k(M), m(M)) \ , \\
  \ell(M) \coloneqq & \ \ell, \quad
  k(M) \coloneqq \sum_{i=1}^{\ell} k_i, \quad
  m(M) \coloneqq \min \{k_i\}_{i=1}^{\ell} \ , \qquad k_i \in \mathbb{N} \ .
\end{align}
The ordering of complexity is defined to be lexicographic, namely one compares $\ell$ first, then $k$ and finally $m$. Define a projection $\pi$ that maps a polynomial $P = \sum_{i=1} M_i$ of Eisenstein series to the monomial term with the highest complexity,
\begin{equation}
  \pi(P) = M, \qquad C(M) = \max_{i} C(M_i) \ .
\end{equation}
When there are multiple monomial terms sharing the same largest complexity, $\pi$ gives the sum of these terms. Finally, we define a linear operator $\operatorname{L}$ that takes a monomial $M = E_{k_1} \Big[ \substack{\pm 1 \\ a^{K_1} b_1}\Big] \cdots E_{k_\text{max}} \Big[\substack{\pm 1 \\ a^{K_{k_\text{max}}} b_{k_\text{max}}}\Big] \cdots E_{k_n}\Big[\substack{\pm 1 \\ a^{K_n} b_n}\Big]$ and ``lifts'' it to a new monomial
\begin{equation}
  \operatorname{L}(M) = \frac{1}{K_{k_\text{max}}}E_{k_1} \Big[ \substack{\pm 1 \\ a^{K_1} b_1}\Big] \cdots E_{k_\text{max} + 1} \Big[\substack{\pm 1 \\ a^{K_{k_\text{max} + 1}} b_{k_\text{max}}}\Big] \cdots E_{k_\ell}\Big[\substack{\pm 1 \\ a^{K_\ell} b_\ell}\Big] \ ,
\end{equation}
where $k_\text{max} = \max_{i} k_i$. When several identical Eisenstein factors are present with the same $k_\text{max}$ and $b_{k_\text{max}}$, we only lift one of them and keep the rest unchanged.

The algorithm to find solution is an iterative procedure using $\pi, \operatorname{L}$. Since the difference equation is linear, we may only consider a monomial $P(a) = \prod_{i} E_{k_i}\Big[\substack{\pm 1 \\ a^{K_i} b_i}\Big]$. Start with the initial value $n_0(a) \coloneqq \operatorname{L}(P)$. Then we consider the following iteration,
\begin{equation}
  n_{i + 1}(a) = n_i(a) - \operatorname{L} \circ \pi \Big(n_i(aq) - n_i(a) \Big) \ .
\end{equation}
The iteration ends in finite steps and stabilizes\footnote{We are unable to prove the eventual stability in general.}, giving the solution to the difference equation,
\begin{equation}
  n(a) = \lim_{i \to \infty} n_i(a) \ .
\end{equation}
Sometimes the iteration stops at a solution $n(a)$ satisfying $n(aq) - n(a) = P(a) + r$, with a constant $r$. One simply adds $r E_1 [\substack{-1 \\ a c}]$ with an arbitrary parameter $c$ to $n(a)$ to get the solution to the original difference equation. In particular, one may choose $c = q^{-1/2}$, or choose it such that the $n(a)$ share the same pole as the elliptic function.


\subsection{Closed-form solutions to the difference equation}

In the above we have provided one of many simple algorithms to find the solution $n(a)$. However, the solution $n(a)$ is far from unique. We can exploit the $q$-periodic property of the Eisenstein series to acquire a closed-form general solution. Recall that the Eisenstein series satisfies
\begin{equation}
  E_k \begin{bmatrix}
    \pm 1 \\ (a q)^K
  \end{bmatrix}
  = \sum_{\ell = 0}^{k} \frac{K^{\ell}}{\ell!} E_{k - \ell} \begin{bmatrix}
    \pm 1 \\ a^K
  \end{bmatrix}
  = e^{K\partial} E_k \begin{bmatrix}
    \pm 1 \\ a^K
  \end{bmatrix} \ ,
\end{equation}
where $\partial$ denotes the \emph{weight-lowering operator} of $E_k[\substack{\pm 1 \\ a^K}]$,
\begin{equation}
  \partial E_k \begin{bmatrix}
    \pm 1 \\ a^K
  \end{bmatrix} \coloneqq E_{k - 1} \begin{bmatrix}
    \pm 1 \\ a^K
  \end{bmatrix} \ , \qquad
  \partial^{-1} E_k \begin{bmatrix}
    \pm 1 \\ a^K
  \end{bmatrix} \coloneqq E_{k + 1} \begin{bmatrix}
    \pm 1 \\ a^K
  \end{bmatrix} \ .
\end{equation}
We require the operator $\partial$ to be linear and satisfy the Leibniz rule, $\partial (M_1 M_2) = (\partial M_1) M_2 + M_1 (\partial M_2)$, where $M_i$ are monomials of Eisenstein series. Due to the convention $E_0[\substack{\pm 1 \\ a^K}] = -1$, the operator $\partial$ is nilpotent when acting on the space of polynomials of $E_k$.

Now we have
\begin{equation}
  E_k \begin{bmatrix}
    \pm 1 \\ (aq)^K
  \end{bmatrix}
  - E_k \begin{bmatrix}
    \pm 1 \\ a^K
  \end{bmatrix}
  = (e^{K \partial} - 1) E_k \begin{bmatrix}
    \pm 1 \\ a^K
  \end{bmatrix} \ .
\end{equation}
For a monomial $M = \prod_{i} E_{k_i}\Big[\substack{\pm 1 \\ a^{K_i} b_i}\Big]$ with respect to the variable $a$, we also have
\begin{equation}
  M(aq) - M(a) = \left(e^{\sum_i K_i \partial_i} - 1\right) M(a) \ ,
\end{equation}
where $\partial_i$ is the weight-lowering operator only for the $i$-th Eisenstein series. It is then straightforward to extend the behavior to any polynomial $n(a)$ of Eisenstein series,
\begin{equation}
  n(aq) - n(a) = (e^{\partial} - 1) n(a) \ ,
\end{equation}
where $\partial = \sum_i K_i \partial_i$ denotes automatic linear combination of $\partial_i$'s (with coefficients $K_i$ taking care of the individual exponents of $a$) depending on the monomial structure. The difference equation for $n(a)$ then can be solved formally,
\begin{equation}
  (e^{\partial} - 1) n(a) = P(a) \ \qquad \ \Rightarrow \qquad
  n(a) = \frac{1}{e^{\partial} - 1} P(a) = \frac{\partial}{e^{\partial} - 1} \partial^{-1}P(a) \ .
\end{equation}
The operator $\frac{\partial}{e^{\partial} - 1}$ can be expanded in terms of Bernoulli numbers $B_r$,
\begin{equation}
  \frac{\partial}{e^{\partial} - 1} = \sum_{r = 0}^{\infty} \frac{B_r}{r!} \partial^r \ \qquad \Rightarrow \qquad
  n(a) = \sum_{r = 0}^{\infty} \frac{B_r}{r!} \partial^{r - 1} P(a) 
\end{equation}
Explicitly, when $P(a)$ is a single Eisenstein series, the solution reads
\begin{align}
  & P(a) = E_k \begin{bmatrix}
    \pm1 \\ a^K b
  \end{bmatrix} \quad
  \Rightarrow \quad
  n(a)
  =
  \sum_{r=0}^{k}
  \frac{B_r K^{r - 1}}{r!}
  E_{k+1-r}\left[\begin{matrix}\pm\\ a^K b\end{matrix}\right]
\end{align}
When $P(a)$ is a product of two Eisenstein series, the solution reads
\begin{align}
  & P(a) = E_{k_1} \begin{bmatrix}
    \pm1 \\ a^{K_1} b_1
  \end{bmatrix}E_{k_2} \begin{bmatrix}
    \pm1 \\ a^{K_2} b_2
  \end{bmatrix} \\
  \Rightarrow & \ 
  n(a)
  =
  \sum_{r = 0}^{k_1 + k_2} \frac{B_r}{r!}
  \sum_{u = 0}^{r}
  \sum_{s = 0}^{k_2} (-1)^s
  \begin{pmatrix}
    r\\
    u
  \end{pmatrix}
  \frac{K_2^s}{K_1^{s + 1}}
  K_1^u K_2^{r - u}
  E_{k_1 + 1 + s - u} \begin{bmatrix}
    \pm1 \\ a^{K_1} b_1
  \end{bmatrix}
  E_{k_2 - s - r + u} \begin{bmatrix}
    \pm1 \\ a^{K_2} b_2
  \end{bmatrix} \nonumber \ .
\end{align}
Apparently, the sum over $u$ comes from the action of $\partial^r$,
\begin{equation}
  \partial^r = (K_1 \partial_1 + K_2 \partial_2)^r = \sum_{u = 0}^{r} \begin{pmatrix}
    r\\
    u
  \end{pmatrix} K_1^u K_2^{r - u} \partial_1^u \partial_2^{r - u} \ .
\end{equation}
The inner sum over $s$ comes from the action of $\partial^{-1}$,
\begin{equation}
  \partial^{-1} P(a) = \sum_{s = 0}^{k_2}(-1)^s \frac{K_2^s}{K_1^{s + 1}} E_{k_1 + 1 + s } \begin{bmatrix}
    \pm1 \\ a^{K_1}b_1
  \end{bmatrix}
  E_{k_2 - s} \begin{bmatrix}
    \pm1 \\ a^{K_2}b_2
  \end{bmatrix}
\end{equation}
This can be formally understood as a kind of geometric series expansion,
\begin{equation}
  \partial^{-1} = \frac{1}{K_1 \partial_1 + K_2 \partial_2}
  \sim \frac{1}{K_1 \partial_1} \frac{1}{1 + \frac{K_2}{K_1} \frac{\partial_2}{\partial_1}}
  = \frac{1}{K_1} \partial_1^{-1} \sum_{s = 0}^{\infty} (-1)^s \frac{K_2^s}{K_1^{s + 1}} \frac{\partial_2^s}{\partial_1^{s + 1}} \ .
\end{equation}

With these basic examples, we can write down the closed form of solution $n(a)$ for any monomial $P(a) = \prod_{i} E_{k_i}\Big[\substack{\pm 1 \\ a^{K_i} b_i}\Big]$, 
\begin{align}
  n(a) = \sum_{r=0}^{K} & \ B_r \sum_{r_1+\dots+r_m=r} \frac{\prod_{i=1}^m K_i^{r_i}}{r_1! \dots r_m!} \\
  & \ \left[ \sum_{s_2=0}^{k_2} \dots \sum_{s_m=0}^{k_m} (-1)^{|s|} \binom{|s|}{s_2, \dots, s_m} \frac{\prod_{i=2}^m K_i^{s_i}}{K_1^{|s|+1}} E^{(1)}_{k_1 + 1 + |s| - r_1} \prod_{i=2}^m E^{(i)}_{k_i - s_i - r_i} \right]  \ . \nonumber
\end{align}
where we abbreviate 
\begin{equation}
  E^{(i)}_{k} = E_k \begin{bmatrix}
    \pm 1 \\ a^{K_i} b_i
  \end{bmatrix} \ .
\end{equation}

For any solution $n(a)$, $n'(a) = n(a) + \epsilon(a)$ is also a solution assuming $\epsilon(aq) = \epsilon(a)$, or equivalently, $\epsilon$ is elliptic with respect to $\mathfrak{a}$. Looking only at polynomials of Eisenstein series, we can try to characterize $\epsilon(a)$ in general. Again, as an Eisenstein polynomial, 
\begin{equation}
  \epsilon(aq) = e^\partial \epsilon(a) = \epsilon(a) \qquad \Rightarrow \qquad (e^\partial - 1) \epsilon(a) = 0 \ .
\end{equation}
The operator in front can be formally expanded,
\begin{equation}
  e^\partial - 1 = \partial + \frac{\partial^2}{2!} + \frac{\partial^3}{3!} + \cdots = \bigg(
    1 + \frac{\partial}{2!} + \frac{\partial^2}{3!} + \cdots
  \bigg) \partial \ .
\end{equation}
The crucial observation is that the operator in the parentheses is invertible, therefore the solution to the above equation is simply $\partial \epsilon(a) = 0$. For example,
\begin{equation}
  \epsilon(a) = 
    \frac{K_2}{K_1}E_{2} \begin{bmatrix}
      -1 \\ a^{K_1}b_1
    \end{bmatrix}
    + \frac{K_1}{K_2} E_{2} \begin{bmatrix}
      -1 \\ a^{K_2}b_2
    \end{bmatrix}
    +  E_1 \begin{bmatrix}
      -1 \\ a^{K_1}b_1
    \end{bmatrix}
    E_1 \begin{bmatrix}
      -1 \\ a^{K_2}b_2
    \end{bmatrix}
\end{equation}
sits in the kernel of $\partial$, 
\begin{align}
  & \ \partial \left(
    \frac{K_2}{K_1}E_{2} \begin{bmatrix}
      -1 \\ a^{K_1}b_1
    \end{bmatrix}
    + \frac{K_1}{K_2} E_{2} \begin{bmatrix}
      -1 \\ a^{K_2}b_2
    \end{bmatrix}
    +  E_1 \begin{bmatrix}
      -1 \\ a^{K_1}b_1
    \end{bmatrix}
    E_1 \begin{bmatrix}
      -1 \\ a^{K_2}b_2
    \end{bmatrix}
  \right)\\
  = & \ K_2 E_1 \begin{bmatrix}
    -1 \\ a^{K_1}b_1
  \end{bmatrix}
  + K_1 E_1 \begin{bmatrix}
    -1 \\ a^{K_2}b_2
  \end{bmatrix}
  - K_2 E_1 \begin{bmatrix}
    -1 \\ a^{K_1}b_1
  \end{bmatrix}
  - K_1 E_1 \begin{bmatrix}
    -1 \\ a^{K_2}b_2
  \end{bmatrix} = 0 \ .
\end{align}
Hence, we expect $\epsilon(a)$ to be elliptic, and indeed it is, which is easily verified.

\subsection{Bethe ansatz-like formula}

The introduction of $n(a)$ in general modifies the pole structure of the integrand, and the integral in general picks up contributions from these new poles. However, the freedom of choosing $n(a)$ at each step allows us to avoid adding any new pole to the game. Consider a two variable integral,
\begin{equation}
  \oint_{|w| = 1} \frac{dw}{2\pi i w} \oint_{|z| = 1} \frac{dz}{2\pi i z} f(z,w) \ ,
\end{equation}
where $f(z,w)$ is doubly elliptic with respect to $\mathfrak{z}, \mathfrak{w}$, and we assume it to have only simple poles. Suppose $z_j$ are poles of $f$ with respect to $z$. Then we may choose
\begin{equation}
  n(z) = E_1 \begin{bmatrix}
    -1 \\ z / (z_1 )q^{\frac{1}{2}}
  \end{bmatrix} \ ,
\end{equation}
to finish the $z$-integration. We are left with
\begin{equation}
  \oint \frac{dw}{2\pi i w} \sum_{j} \mathop{\operatorname{Res}}_{z = z_j} \frac{1}{z} f(z, w)  E_1 \begin{bmatrix}
    -1 \\ z / (z_1 q^{\frac{1}{2}})
  \end{bmatrix} \ ,
\end{equation}
Note that $z = z_1$ is a double pole, and it contributes
\begin{equation}
  \frac{1}{2\pi \eta(\tau)^3} \frac{d}{d\mathfrak{z}} \Big[\frac{\vartheta_1(\mathfrak{z} - \mathfrak{z_1}(w))}{\sqrt{z/z_1}} f(\mathfrak{z}, w) \Big] \ .
\end{equation}

The contribution from $z = z_2, ...$ are of the form
\begin{equation}
  R_j E_1 \begin{bmatrix}
    -1 \\ z_j / (z_1 q^{\frac{1}{2}})
  \end{bmatrix} \ , \qquad R_j = \operatorname{Res}_{z = z_j} \frac{1}{z} f(z,w) \ .
\end{equation}
Note that $z_j$ are in general functions of $w$, and the Eisenstein series $E_1[\substack{-1 \\ z_j/(z_1 q^{\frac{1}{2}})}]$ has a pole at precisely
\begin{equation}
  \frac{z_j(w)}{z_1(w)} = 1 \ .
\end{equation}
But this is also a pole of $R_j$, since originally
\begin{equation}
  f(z, w) \sim \frac{1}{(\mathfrak{z} - \mathfrak{z_1}(w))(\mathfrak{z} - \mathfrak{z_j}(w)) \cdots} \qquad \Rightarrow \qquad R_j(w) \sim \frac{1}{(\mathfrak{z_j}(w) - \mathfrak{z_1}(w)) \cdots} \ .
\end{equation}
Subsequently, the $w$-integration requires solving the difference equations
\begin{equation}
  n(wq) - n(w) = E_1 \begin{bmatrix}
    -1 \\ z_j/(z_1q^{1/2})
  \end{bmatrix}
\end{equation}
which can also be chosen to share the same pole as $R_j(w)$. Such procedure produces in the end a sum over simultaneous poles of $f(z, w)$,
\begin{equation}
  \mathcal{I} = \sum_{\{z_*, w_*\}} E(z_*, w_*)  \ ,
\end{equation}
In particular, $|q| \le |z_*| < 1$ and $|q| \le |w_*|< 1$, precisely the poles considered in \cite{Benini:2018mlo}. What is different is that the contribution $E(z_*, w_*)$ from each simultaneous pole is not simply the residue of $f$ at that pole, but rather a combination of the residue and the Eisenstein series generated along the way. In a sense, $E$ is a kind of ``generalized residue'' that captures the higher dimensional homology, keeping track of how the contour is deformed to encircle the poles in the $z_1, z_2$ planes sequentially. It would be interesting to characterize the nature of $E$ more geometrically/topologically, which we leave for future work.

In general, for a rank-$r$ Schur index integrand (assuming only simple poles)
\begin{equation}
  \mathcal{Z} = \frac{Z_\text{VM}(\mathfrak{a})}{\prod_i \prod_{w \in \mathcal{R}_f} \prod_{\rho \in \mathcal{R}_i} \vartheta_4(\rho(\mathfrak{a}) + w(\mathfrak{b}))}
\end{equation}
where the denominator accounts for the hypermultiplet contributions, the Schur index can be written as a sum over Bethe ansatz like solutions,
\begin{equation}
  \mathcal{I} = \sum_{\mathfrak{a}_* \in \text{BAE}} \left({\operatorname{Res}_{\mathfrak{a}_*}\mathcal{Z}}\right) P(\mathfrak{a}_*)
\end{equation}
where $\operatorname{Res}$ denotes the Grothendieck residue, 
\begin{equation}
  \text{BAE} = \{\mathfrak{a}_* \mid \mathbf{Q}_A(\mathfrak{a}_*) = 1, \quad A = 1, ..., r\} \ ,
\end{equation} 
where
\begin{equation}
  \{\mathbf{Q}_1, ..., \mathbf{Q}_r\} \subset \cup_{i, w, \rho} \Big\{\frac{\vartheta_1(\rho(\mathfrak{a}) + w(\mathfrak{b}))}{\vartheta_1(\frac{\tau}{2})} \Big\} \ .
\end{equation}

\subsection{Wilson line index}

When computing Wilson line index of Lagrangian theories, we need to evaluate integrals of the form
\begin{equation}
  \oint \frac{da}{2\pi i a} a^m f(a) P(a) \ , \qquad m \in \mathbb{Z}_{\ne 0} \ .
\end{equation}
where $a^m$ comes from the gauge group character insertion into the integral, and $f(a)$ is elliptic, $P(a)$ is a polynomial of Eisenstein series. The above algorithm can be generalized to the cases in the presence of insertion of monomial of integration variables $a$. We would like the integral to equal
\begin{equation}
  \oint \frac{da}{2\pi i a} \frac{1 - Q(a)}{1 - Q(a)} a^m f(a) P(a)
  = \left({\oint_{|a| = 1} - \oint_{|a| = |q|}}\right) \frac{da}{2\pi i a} f(a) P(a) \frac{a^m}{1 - Q(a)} \ ,
\end{equation}
which can be achieved by solving the following functional equation,
\begin{equation}
  \frac{Q(a)P(a)a^m}{1 - Q(a)} = \frac{P(aq)(aq)^m}{1 - Q(aq)} \ .
\end{equation}
Setting again $Q(a) = 1 + \frac{P(a)}{n(a)}$, $n(a)$ is constrained by the difference equation
\begin{equation}
  q^mn(aq) - n(a) = P(a) \ .
\end{equation}
Once the solution $n(a)$ is found, the integral can be evaluated by collecting residues of poles of $a^{m - 1}f(a)n(a)$ within the annulus $|q| \le |a| < 1$,
\begin{equation}
  \oint \frac{da}{2\pi i a} a^m f(a) P(a) = - \sum_{J} \operatorname{Res}_{a = a_J} \left({\frac{1}{a} a^m f(a)n(a)}\right) \ .
\end{equation}
Here $a_J$ denotes poles of the product $a^{m - 1}f(a)n(a)$ within the annulus $|q| \le |a| < 1$. The final result is independent of the specific choice of the solution $n(a)$.

When $P(a) = 1$, then the equation reads
\begin{equation}
  q^m n(aq) - n(a) = 1 \qquad \Rightarrow \qquad
  n(a) = \frac{-1}{1-q^m}\  .
\end{equation}
For more general $P(a) = \prod_{i} E_{k_i}\Big[\substack{\pm 1 \\ a^{K_i} b_i}\Big]$, it is easy to construct one closed-form solution. Using $\partial$, the difference equation can be written as
\begin{equation}
  (q^m e^{\partial} - 1) n(a) = P(a) \quad \Rightarrow \quad
  n(a) = \frac{1}{q^m e^{\partial} - 1} P(a) = \frac{\partial}{q^m e^{\partial} - 1} \partial^{-1}P(a) \ .
\end{equation}
Hence formally
\begin{equation}
  n(a) = \sum_{r = 0}^{+\infty} \frac{C_r(q^{m})}{r!} \partial^r P(a) \ , \qquad
  \frac{1}{\lambda e^z - 1} = \sum_{r = 0}^{+\infty} \frac{C_r(\lambda)}{r!} z^r \ .
\end{equation}
Again, $\partial$ acts on $P$ linearly and distributes automatically to each Eisenstein factor through Leibniz rule. The first few coefficients $C_r(\lambda)$ are given by
\begin{align}
  C_0(\lambda) = & \ \frac{1}{\lambda - 1}, \qquad
  C_1(\lambda) = -\frac{\lambda}{(\lambda - 1)^2}, \qquad
  C_2(\lambda) = \frac{\lambda(\lambda + 1)}{(\lambda - 1)^3}, \\
  C_3(\lambda) = & \  -\frac{\lambda(\lambda^2 + 4\lambda + 1)}{(\lambda - 1)^4} \ .
\end{align}

For example, when $P(a) = E_k \Big[\substack{\pm 1 \\ a^K b}\Big]$, the solution reads
\begin{equation}
  n(a) = \sum_{r = 0}^{k} \frac{C_r(q^m) K^{r}}{r!} E_{k - r} \Big[\substack{\pm 1 \\ a^K b}\Big] \ .
\end{equation}
When $P(a) = E_{k_1}\big[\begin{smallmatrix}\pm_1 1 \\ a^{K_1}b_1\end{smallmatrix}\big]E_{k_2}\big[\begin{smallmatrix}\pm_2 1 \\ a^{K_2}b_2\end{smallmatrix}\big]$, the solution reads
\begin{equation}
  n(a)=
\sum_{r=0}^{k_1+k_2}\frac{C_r(q^m)}{r!}
\sum_{u=0}^{r}\binom{r}{u}K_1^uK_2^{r-u}
E_{k_1-u}\begin{bmatrix}
\pm_1 1 \\ a^{K_1}b_1
\end{bmatrix}
E_{k_2-r+u}\begin{bmatrix}
\pm_2 1 \\ a^{K_2}b_2
\end{bmatrix} \ .
\end{equation}

\subsection{An illustrative example}

Let us return to the example of elliptic function integral and show how the above algorithm works in practice,
\begin{equation*}
  \oint \frac{da_1}{2\pi i a_1} \frac{da_2}{2\pi i a_2}\frac{\vartheta_1(\mathfrak{a}_1 - \mathfrak{a}_2)^2 \vartheta_1(\mathfrak{a}_1 + \mathfrak{a}_2)^2}{\vartheta_4(\mathfrak{a}_1 - \mathfrak{a}_2 + \mathfrak{b}_1)\vartheta_4(- \mathfrak{a}_1 + \mathfrak{a}_2 + \mathfrak{b}_1)
  \vartheta_4(\mathfrak{a}_1 + \mathfrak{a}_2 + \mathfrak{b}_2) \vartheta_4(- \mathfrak{a}_1 - \mathfrak{a}_2 + \mathfrak{b}_2)} \ .
\end{equation*}
The $a_1$ integral begins with solving the functional equation
\begin{equation}
  n(a_1 q) - n(a_1) = 1 \qquad \Rightarrow \qquad
  n(a_1) = E_1 \begin{bmatrix}
  -1 \\ a_1 q^{-\frac{1}{2}}
  \end{bmatrix} \ .
\end{equation}
Here we choose to have $n(a)$ to have a pole at $a_1 = q$, or equivalently, $\mathfrak{a}_1 = \tau$. The integral picks up $\mathfrak{a}_1$ poles at $\mathfrak{a}_1 = \tau, \pm(\mathfrak{a}_2 - \mathfrak{b}_1) + \tau, \pm(- \mathfrak{a}_2 - \mathfrak{b}_2) + \tau$ within the region $|q| \le |a_1| < 1$, leading to the sum of residues,
\begin{align}
= &\frac{
\vartheta_1(\mathfrak{a}_2)^4
}{
\vartheta_4(\mathfrak{a}_2-\mathfrak{b}_1)\,
\vartheta_4(\mathfrak{a}_2+\mathfrak{b}_1)\,
\vartheta_4(\mathfrak{a}_2-\mathfrak{b}_2)\,
\vartheta_4(\mathfrak{a}_2+\mathfrak{b}_2)
}
\nonumber\\[0.8em]
&\quad
-
\frac{
\mathrm{i}\,
\vartheta_4(2\mathfrak{a}_2-\mathfrak{b}_1)^2\,
\vartheta_4(\mathfrak{b}_1)^2
}{
\eta(\tau)^3\,
\vartheta_1(2\mathfrak{b}_1)\,
\vartheta_1(2\mathfrak{a}_2-\mathfrak{b}_1-\mathfrak{b}_2)\,
\vartheta_1(2\mathfrak{a}_2-\mathfrak{b}_1+\mathfrak{b}_2)
}
\,E_1\begin{bmatrix}
-1\\
\frac{a_2}{b_1}
\end{bmatrix}
\nonumber\\[0.8em]
&\quad
+
\frac{
\mathrm{i}\,
\vartheta_4(\mathfrak{b}_1)^2\,
\vartheta_4(2\mathfrak{a}_2+\mathfrak{b}_1)^2
}{
\eta(\tau)^3\,
\vartheta_1(2\mathfrak{b}_1)\,
\vartheta_1(2\mathfrak{a}_2+\mathfrak{b}_1-\mathfrak{b}_2)\,
\vartheta_1(2\mathfrak{a}_2+\mathfrak{b}_1+\mathfrak{b}_2)
}
\,E_1\begin{bmatrix}
-1\\
a_2b_1
\end{bmatrix}
\nonumber\\[0.8em]
&\quad
-
\frac{
\mathrm{i}\,
\vartheta_4(2\mathfrak{a}_2-\mathfrak{b}_2)^2\,
\vartheta_4(\mathfrak{b}_2)^2
}{
\eta(\tau)^3\,
\vartheta_1(2\mathfrak{a}_2-\mathfrak{b}_1-\mathfrak{b}_2)\,
\vartheta_1(2\mathfrak{a}_2+\mathfrak{b}_1-\mathfrak{b}_2)\,
\vartheta_1(2\mathfrak{b}_2)
}
\,E_1\begin{bmatrix}
-1\\
\frac{a_2}{b_2}
\end{bmatrix}
\nonumber\\[0.8em]
&\quad
+
\frac{
\mathrm{i}\,
\vartheta_4(\mathfrak{b}_2)^2\,
\vartheta_4(2\mathfrak{a}_2+\mathfrak{b}_2)^2
}{
\eta(\tau)^3\,
\vartheta_1(2\mathfrak{b}_2)\,
\vartheta_1(2\mathfrak{a}_2-\mathfrak{b}_1+\mathfrak{b}_2)\,
\vartheta_1(2\mathfrak{a}_2+\mathfrak{b}_1+\mathfrak{b}_2)
}
\,E_1\begin{bmatrix}
-1\\
a_2b_2
\end{bmatrix}.
\end{align}
The $a_2$ integration first requires solving the functional equations
\begin{equation}
  n(a q) - n(a) = E_1 \begin{bmatrix}
    1 \\ a b
  \end{bmatrix} \qquad \Rightarrow \qquad
  n(a) = 1 - E_1 \begin{bmatrix}
    -1 \\ aq^{-1/2}
  \end{bmatrix}
  + E_2 \begin{bmatrix}
    -1 \\ abq^{1/2}
  \end{bmatrix}\ .  
\end{equation}
Here we choose the second term to add a pole at $a_2 = q$. Subsequently the $a_2$ integration picks up poles at the following locations,
\begin{align}
& \ \tau , \quad + \mathfrak{b}_1+\tau , \quad - \mathfrak{b}_1+\tau , \quad +\mathfrak{b}_2+\tau , \quad -\mathfrak{b}_2+\tau \\
& \ \tau, \quad + \frac{\mathfrak{b}_1 + \mathfrak{b}_2}{2} + \frac{m}{2} + \frac{n}{2}\tau, \quad
\frac{\mathfrak{b}_1 - \mathfrak{b}_2}{2} + \frac{m}{2} + \frac{n}{2}\tau, \qquad
m, n = 1, 2 \\
& \ \tau, \quad - \frac{\mathfrak{b}_1 + \mathfrak{b}_2}{2} + \frac{m}{2} + \frac{n}{2}\tau, \quad
\frac{\mathfrak{b}_2 - \mathfrak{b}_1}{2} + \frac{m}{2} + \frac{n}{2}\tau, \qquad
m, n = 1, 2 \ .
\end{align}
The final result is given by
\begin{align}
= & \ \frac{\vartheta_4(0)^4}
{\vartheta_1(\mathfrak{b}_1)^2 \vartheta_1(\mathfrak{b}_2)^2}
+ \frac{2i}{\eta(\tau)^3 \vartheta_1(\mathfrak{b}_1 \pm \mathfrak{b}_2)}\left({-\frac{
\vartheta_4(\mathfrak{b}_2)^4
}{
\vartheta_1(2\mathfrak{b}_2)
}E_1 \begin{bmatrix} 1 \\ b_2 \end{bmatrix}
+ \frac{
\vartheta_4(\mathfrak{b}_1)^4
}{
\vartheta_1(2\mathfrak{b}_1)
}E_1 \begin{bmatrix} 1 \\ b_1 \end{bmatrix}}\right)
\nonumber
\\
&\quad
-\frac{
2\vartheta_4(\mathfrak{b}_1)^2
\vartheta_4(\mathfrak{b}_2)^2
}{
\eta(\tau)^6
\vartheta_1(2\mathfrak{b}_2)
\vartheta_1(2\mathfrak{b}_1)
}\sum_{\alpha, \beta = \pm} \bigg(E_2 \begin{bmatrix}
  \alpha \\ \beta \sqrt{\frac{b_1}{b_2}}
\end{bmatrix} 
- E_2 \begin{bmatrix}
  \alpha \\ \beta \sqrt{b_1 b_2}
\end{bmatrix} \bigg) \ .
\end{align}
Terms from the first line come from the added poles when solving the difference equation. The second line comes from the simultaneous poles of the original integrand, namely solutions to the equations
\begin{align}
  \mathfrak{a}_1 - \mathfrak{a}_2  + \mathfrak{b}_1 = \frac{\tau}{2}, \quad
  - \mathfrak{a}_1 + \mathfrak{a}_2  + \mathfrak{b}_1 = \frac{\tau}{2}, \quad\\
  \mathfrak{a}_1 + \mathfrak{a}_2  + \mathfrak{b}_2 = \frac{\tau}{2}, \quad
  - \mathfrak{a}_1 - \mathfrak{a}_2  + \mathfrak{b}_2 = \frac{\tau}{2}, \quad
\end{align}
suitably shifted into the $|q| < |a_1| < 1$ and $|q| < |a_2| < 1$ regions. They are given by
\begin{align}
  (\mathfrak{a}_1, \mathfrak{a}_2) = & \ (\frac{\tau}{2} + \frac{\mathfrak{b}_1}{2} + \frac{\mathfrak{b}_2}{2}, -(\frac{\mathfrak{b}_1}{2}) + \frac{\mathfrak{b}_2}{2}), 
(\frac{\tau}{2} + \frac{\mathfrak{b}_1}{2} - \frac{\mathfrak{b}_2}{2}, -\frac{\mathfrak{b}_1}{2} - \frac{\mathfrak{b}_2}{2}), \nonumber\\
& \ (\frac{\tau}{2} - \frac{\mathfrak{b}_1}{2} + \frac{\mathfrak{b}_2}{2}, \frac{\mathfrak{b}_1}{2} + \frac{\mathfrak{b}_2}{2}),
(\frac{\tau}{2} - \frac{\mathfrak{b}_1}{2} - \frac{\mathfrak{b}_2}{2}, \frac{\mathfrak{b}_1}{2} - \frac{\mathfrak{b}_2}{2})
\end{align}
In particular, the coefficient
\begin{equation}
  \frac{\vartheta_4(\mathfrak{b}_1)^2 \vartheta_4(\mathfrak{b}_2)^2}{\eta(\tau)^6\vartheta_1(2\mathfrak{b}_1) \vartheta_1(2\mathfrak{b}_2)}
\end{equation}
is simply the Grothendieck residue of the original integrand at these poles (up to a sign).

The solutions of $n(a_1)$ or $n(a_2)$ are not unique. When integrating $a_1$, we may choose
\begin{equation}
  n(a) = E_1 \begin{bmatrix}
    -1 \\ \frac{a_1}{a_2 b_1}
  \end{bmatrix} \ ,
\end{equation}
which shares the same pole $a_1 = a_2 b_1q^{\frac{1}{2}}$ as the elliptic integrand. The $a_1$ integration is a bit more complicated, as it involves taking derivative of the entire elliptic integrand. The $a_2$ integration also involves solving some difference equations, which we can also tune it to share pole with the elliptic function in front. In the end, we find the following result,
\begin{align}
\frac{
\vartheta_4(\mathfrak{b}_1)^2
\vartheta_4(\mathfrak{b}_2)^2
}{
\eta(\tau)^6
\vartheta_1(2\mathfrak{b}_1)
\vartheta_1(2\mathfrak{b}_2)
}E_1\begin{bmatrix} -1 \\ b_1 \end{bmatrix}
\Bigg(
- E_1\begin{bmatrix} -1 \\ -b_2 \end{bmatrix}
-5E_1\begin{bmatrix} -1 \\ b_2 \end{bmatrix}&
-E_1\begin{bmatrix} 1 \\ -b_2 \end{bmatrix}\\
& \ -E_1\begin{bmatrix} 1 \\ b_2 \end{bmatrix}
+2E_1\begin{bmatrix} 1 \\ b_2^2 \end{bmatrix}\bigg) \ . \nonumber
\end{align}
This form of result is a sum over simultaneous poles of the original integrand, the overall factor in front is also the same as the standard simultaneous residue of the original integrand, however, accompanied by a non-trivial polynomial of Eisenstein series.

In fact, in the second integration, we may rename $\mathfrak{a}_2 \to \frac{1}{2}\mathfrak{a}_2$, and the integration produces the simplest form
\begin{align}
= -
\frac{
4\vartheta_4(\mathfrak{b}_1)^2
\vartheta_4(\mathfrak{b}_2)^2
}{
\eta(\tau)^6
\vartheta_1(2\mathfrak{b}_1)
\vartheta_1(2\mathfrak{b}_2)
} E_1\begin{bmatrix} -1 \\ b_1 \end{bmatrix}
E_1\begin{bmatrix} -1 \\ b_2 \end{bmatrix} \ .
\end{align}
This shows that there is a lot of freedom in performing these type of integrals, generating different but equivalent expressions, which are related by a vast amount of identities between Eisenstein series.


\section{Gauge theory examples}\label{sec:examples}

In this section we apply the integration formula developed above to compute the flavored Schur indices of various Lagrangian $\mathcal{N} = 2$ SCFTs, ranging from BC type SQCD, BC type $\mathcal{N} = 4$ theories, BC type quiver theories to $SU(k)$ quiver gauge theories. These examples demonstrate the wide validity of our computation technique. However, we should point out that although the technique is (almost) always applicable, the result may not always have elegant compact form. One may need to try different solutions $n(a)$ at each step, and/or make extensive use of identities involving Eisenstein series, Jacobi theta functions and the eta function to bring the result into a more manageable form before further analysis.

\subsection{\texorpdfstring{$SO(N), USp(2N)$ SQCD}{}}

We begin with the simple case of BCD-type SQCD. The $SO(2r)$ SQCD with $2r-2$ hypermultiplets in the vector representation has Schur index given by the following contour integral
\begin{equation}
  \mathcal{I} = \oint \prod_{i=1}^r \frac{da_i}{2\pi i a_i} \Delta(a)
  \operatorname{PE}\bigg[
    - \frac{2q}{1- q} \chi_\text{adj}(a) + \frac{q^{\frac{1}{2}}}{1 - q}\chi_\text{vec}(a) \sum_{i=1}^{2r-2} (b_i + b_i^{-1})
   \bigg] \ ,
\end{equation}
where the Haar measure
\begin{equation}
  \Delta(a) = \frac{1}{2^{r-1} r!} \prod_{1 \le i < j \le r} (1 - a_i a_j)(1 - a_i^{-1} a_j^{-1})(1 - a_i a_j^{-1})(1 - a_i^{-1} a_j) \ ,
\end{equation}
the adjoint character
\begin{equation}
  \chi_\text{adj}(a) = r + \sum_{1 \le i < j \le r} (a_i a_j + a_i^{-1} a_j^{-1} + a_i a_j^{-1} + a_i^{-1} a_j) \ ,
\end{equation}
and the vector character
\begin{equation}
  \chi_\text{vec}(a) = \sum_{i=1}^r (a_i + a_i^{-1}) \ .
\end{equation}
The vector representation is real, therefore the flavor symmetry is given by $USp(2(2r - 2))$, whose fundamental character is $\chi_\text{fund}^{USp(2(2r-2))}(b) = \sum_{i=1}^{2r-2} (b_i + b_i^{-1})$. Written in terms of the Jacobi theta functions, the integrand reads
\begin{equation}
  \mathcal{Z} = \frac{\eta(\tau)^{2r^2}}{2^{r-1}r!} \frac{
    \prod_{1 \le i < j \le r} \vartheta_1(\mathfrak{a}_i - \mathfrak{a}_j)^2 \vartheta_1(\mathfrak{a}_i + \mathfrak{a}_j)^2
  }{
    \prod_{s = 1}^{2r-2} \prod_{i=1}^r \vartheta_4(\pm \mathfrak{a}_i + \mathfrak{b}_s)
  }
\end{equation}

We start with the simplest cases. Consider $r = 2$. The $a_1$ integration gives
\begin{align}
&\frac{
\mathrm{i} \eta(\tau)^5 
\vartheta_4(\mathfrak{a}_2-\mathfrak{b}_1) 
\vartheta_4(\mathfrak{a}_2+\mathfrak{b}_1)
}{
2 
\vartheta_1(2\mathfrak{b}_1) 
\vartheta_1(-\mathfrak{b}_1+\mathfrak{b}_2) 
\vartheta_1(\mathfrak{b}_1+\mathfrak{b}_2) 
\vartheta_4(\mathfrak{a}_2-\mathfrak{b}_2) 
\vartheta_4(\mathfrak{a}_2+\mathfrak{b}_2)
}
 E_1\begin{bmatrix}
-1\\
\frac{a_2}{b_1}
\end{bmatrix}\nonumber\\
& \ -
\frac{
\mathrm{i} \eta(\tau)^5 
\vartheta_4(\mathfrak{a}_2-\mathfrak{b}_1) 
\vartheta_4(\mathfrak{a}_2+\mathfrak{b}_1)
}{
2 
\vartheta_1(2\mathfrak{b}_1) 
\vartheta_1(-\mathfrak{b}_1+\mathfrak{b}_2) 
\vartheta_1(\mathfrak{b}_1+\mathfrak{b}_2) 
\vartheta_4(\mathfrak{a}_2-\mathfrak{b}_2) 
\vartheta_4(\mathfrak{a}_2+\mathfrak{b}_2)
}
 E_1\begin{bmatrix}
-1\\
a_2b_1
\end{bmatrix}
\nonumber\\[0.8em]
&\quad
+
\frac{
\mathrm{i} \eta(\tau)^5 
\vartheta_4(\mathfrak{a}_2-\mathfrak{b}_1) 
\vartheta_4(\mathfrak{a}_2+\mathfrak{b}_1)
}{
4 
\vartheta_1(2\mathfrak{b}_1) 
\vartheta_1(-\mathfrak{b}_1+\mathfrak{b}_2) 
\vartheta_1(\mathfrak{b}_1+\mathfrak{b}_2) 
\vartheta_4(\mathfrak{a}_2-\mathfrak{b}_2) 
\vartheta_4(\mathfrak{a}_2+\mathfrak{b}_2)
}
 E_1\begin{bmatrix}
1\\
\frac{b_1}{b_2}
\end{bmatrix}\nonumber\\
& \ -
\frac{
\mathrm{i} \eta(\tau)^5 
\vartheta_4(\mathfrak{a}_2-\mathfrak{b}_2) 
\vartheta_4(\mathfrak{a}_2+\mathfrak{b}_2)
}{
4 
\vartheta_1(2\mathfrak{b}_2) 
\vartheta_1(-\mathfrak{b}_1+\mathfrak{b}_2) 
\vartheta_1(\mathfrak{b}_1+\mathfrak{b}_2) 
\vartheta_4(\mathfrak{a}_2-\mathfrak{b}_1) 
\vartheta_4(\mathfrak{a}_2+\mathfrak{b}_1)
}
 E_1\begin{bmatrix}
1\\
\frac{b_1}{b_2}
\end{bmatrix}
\nonumber\\[0.8em]
&\quad
+
\frac{
\mathrm{i} \eta(\tau)^5 
\vartheta_4(\mathfrak{a}_2-\mathfrak{b}_1) 
\vartheta_4(\mathfrak{a}_2+\mathfrak{b}_1)
}{
4 
\vartheta_1(2\mathfrak{b}_1) 
\vartheta_1(-\mathfrak{b}_1+\mathfrak{b}_2) 
\vartheta_1(\mathfrak{b}_1+\mathfrak{b}_2) 
\vartheta_4(\mathfrak{a}_2-\mathfrak{b}_2) 
\vartheta_4(\mathfrak{a}_2+\mathfrak{b}_2)
}
 E_1\begin{bmatrix}
1\\
b_1b_2
\end{bmatrix}\nonumber\\
& \ +
\frac{
\mathrm{i} \eta(\tau)^5 
\vartheta_4(\mathfrak{a}_2-\mathfrak{b}_2) 
\vartheta_4(\mathfrak{a}_2+\mathfrak{b}_2)
}{
4 
\vartheta_1(2\mathfrak{b}_2) 
\vartheta_1(-\mathfrak{b}_1+\mathfrak{b}_2) 
\vartheta_1(\mathfrak{b}_1+\mathfrak{b}_2) 
\vartheta_4(\mathfrak{a}_2-\mathfrak{b}_1) 
\vartheta_4(\mathfrak{a}_2+\mathfrak{b}_1)
}
 E_1\begin{bmatrix}
1\\
b_1b_2
\end{bmatrix}.
\end{align}
Here we choose the solution to difference equation
\begin{equation}
  n(a) = E_1 \begin{bmatrix}
    -1 \\ a_1/b_1
  \end{bmatrix} \ .
\end{equation}
so that the pole of $n(a)$ coincides with the pole $\mathfrak{a}_1 = \mathfrak{b}_1 + \frac{\tau}{2}$ of the elliptic integrand. The subsequent $a_2$ integration is straightforward, and we also ensure all the $n(a)$ do not generate new poles. The final result is fairly simple,
\begin{equation}
  \mathcal{I} = \frac{\eta(\tau)^2}{\vartheta_1(2 \mathfrak{b}_1) \vartheta_1(2 \mathfrak{b}_2)} \bigg(
    E_2 \begin{bmatrix}
      1 \\ b_1/b_2
    \end{bmatrix}
    - E_2 \begin{bmatrix}
      1 \\ b_1 b_2
    \end{bmatrix}
  \bigg) \ .
\end{equation}
Note that this is identical to the $A_1$ class-$\mathcal{S}$ index $\mathcal{I}_{1,2}$ corresponding to the genus-one surface with two punctures \cite{Pan:2021mrw}. If we choose $n(a_i)$ to always have pole at $a_i = q$, then the final result is given by a slightly less recognizable form,
\begin{align}
\mathcal{I} = & \ -\frac{
i \eta(\tau)^5 \vartheta_4(\mathfrak{b}_1)^2 
}{ 
\vartheta_1(2\mathfrak{b}_1) 
\vartheta_1(-\mathfrak{b}_1+\mathfrak{b}_2) 
\vartheta_1(\mathfrak{b}_1+\mathfrak{b}_2) 
\vartheta_4(\mathfrak{b}_2)^2
}E_1\begin{bmatrix}
-1 \\
b_1
\end{bmatrix}
\nonumber\\
&\quad
+\frac{
i \eta(\tau)^5 \vartheta_4(\mathfrak{b}_2)^2 
}{ 
\vartheta_1(2\mathfrak{b}_2) 
\vartheta_1(-\mathfrak{b}_1+\mathfrak{b}_2) 
\vartheta_1(\mathfrak{b}_1+\mathfrak{b}_2) 
\vartheta_4(\mathfrak{b}_1)^2
}E_1\begin{bmatrix}
-1 \\
b_2
\end{bmatrix}
\nonumber\\
&\quad
+\frac{
2 \eta(\tau)^2 
}{ 
\vartheta_1(2\mathfrak{b}_1) 
\vartheta_1(2\mathfrak{b}_2)
}E_1\begin{bmatrix}
-1 \\
b_1
\end{bmatrix}
E_1\begin{bmatrix}
-1 \\
b_2
\end{bmatrix} \ .
\end{align}

Moving on to $r = 3$. For uniformity we always choose the $n(a)$ to have pole at $a_i = q$. The result reads
\begin{equation}
  \mathcal I_{SO(6),N_f=4}
=
\sum_{1\leq i<j\leq 4}\mathcal A_{ij}
+
\sum_{1\leq i<j<k\leq 4}\mathcal B_{ijk}.
\end{equation}
where
\begin{equation}
  \mathcal A_{ij}
=
\frac{
\eta(\tau)^{12}
\vartheta_4(\mathfrak b_i)^2
\vartheta_4(\mathfrak b_j)^2
E_iE_j
}{
\vartheta_1(2\mathfrak b_i)
\vartheta_1(2\mathfrak b_j)
\prod_{\substack{m=1\\ m\neq i,j}}^{4}
\vartheta_4(\mathfrak b_m)^2
\vartheta_1(\mathfrak b_i\pm\mathfrak b_m)
\vartheta_1(\mathfrak b_j\pm\mathfrak b_m)
}
\end{equation}

\begin{equation}
  \mathcal B_{ijk}
=
\frac{
2i 
\eta(\tau)^9
E_iE_jE_k
}{
\vartheta_1(2\mathfrak b_i)
\vartheta_1(2\mathfrak b_j)
\vartheta_1(2\mathfrak b_k)
\prod_{\substack{m=1\\ m\neq i,j,k}}^{4}
\vartheta_1(\mathfrak b_i\pm\mathfrak b_m)
\vartheta_1(\mathfrak b_j\pm\mathfrak b_m)
\vartheta_1(\mathfrak b_k\pm\mathfrak b_m)
},
\end{equation}
where $E_i \coloneqq E_1[\substack{-1 \\ b_i}]$.

The $SO(6)$ result can be easily generalized to all $SO(N = 2r)$ theories with $N_f = 2r - 2$ flavors. Since the vector representation is real, the flavor symmetry is given by
$USp(2N_f)$, and we parametrize its maximal torus by
$b_s=e^{2\pi i\mathfrak b_s}$, $s=1,\ldots,2r-2$.  Define
\begin{equation}
  F=\{1,\ldots,2r-2\},
\qquad
E_s
=
E_1\left[\begin{matrix}-1\\ b_s\end{matrix}\right](q).
\end{equation}
Then the Schur index can be written as
\begin{equation}
  \mathcal I_{SO(2r), N_f=2r-2}
=
\sum_{\substack{S\subset F\\ |S|=r-1}}
\mathcal A_S
+
\sum_{\substack{S\subset F\\ |S|=r}}
\mathcal B_S ,
\end{equation}
where
\begin{equation}
 \mathcal A_S
=
\frac{
i^{ r-1}
(-1)^{\frac{(r-1)(r-2)}2}
\eta(\tau)^{2r^2-3r+3}
\prod_{s\in S}
\vartheta_4(\mathfrak b_s)^2
E_s
}{
(
\prod_{s\in S}
\vartheta_1(2\mathfrak b_s)
)
(
\prod_{t\in F\setminus S}
\vartheta_4(\mathfrak b_t)^2
)
(
\prod_{\substack{s\in S\\ t\in F\setminus S}}
\vartheta_1(\mathfrak b_s \pm \mathfrak b_t)
)
},
\end{equation}
and
\begin{equation}
  \mathcal B_S
=
\frac{
2i^{ r}
(-1)^{\frac{r(r-1)}2}
\eta(\tau)^{r(2r-3)}
\prod_{s\in S}
E_s
}{
(
\prod_{s\in S}
\vartheta_1(2\mathfrak b_s)
)
(
\prod_{\substack{s\in S\\ t\in F\setminus S}}
\vartheta_1(\mathfrak b_s \pm\mathfrak b_t)
)
},
\end{equation}

Next we consider the $SO(2r+1)$ SQCD with $2r-1$ flavors. The Schur index is given by the following contour integral
\begin{equation}
  \mathcal{I}_{SO(2r+1)}
=
\oint \left[{\frac{da}{2\pi i a}}\right]\frac{
\eta(\tau)^{2r^2+2r-1}
}{
2^r r!
}
\frac{
\prod_{i=1}^{r}\vartheta_1(\mathfrak a_i)^2
\prod_{1\leq i< j\leq r}
\vartheta_1(\mathfrak a_i-\mathfrak a_j)^2
\vartheta_1(\mathfrak a_i+\mathfrak a_j)^2
}{
\prod_{s=1}^{2r-1}\vartheta_4(\mathfrak b_s)
\prod_{i=1}^{r}\prod_{s=1}^{2r-1}
\vartheta_4(\mathfrak a_i-\mathfrak b_s)
\vartheta_4(\mathfrak a_i+\mathfrak b_s)
} \ . \nonumber
\end{equation}
When $r = 1$, $SO(3) = SU(2)$ and the vector representation is the same as the adjoint representation of $SU(2)$, hence it is simply the $\mathcal{N} = 4$ $SU(2)$ index,
\begin{equation}
  \mathcal{I}_{SO(3)} = \frac{i \vartheta_4(\mathfrak{b})}{\vartheta_1(2 \mathfrak{b}_1)} E_1 \begin{bmatrix}
    -1 \\ b_1
  \end{bmatrix} \ .
\end{equation}
When $r = 2$, choosing the solution to $n(aq) - n(a) = 1$ as $E_1[\substack{-1 \\ aq^{-1/2}}]$ leads to the closed form
\begin{align}
\mathcal I_{SO(5)}
= & \
\eta(\tau)^5
\Bigg(
\frac{
\vartheta_4(\mathfrak b_1)\vartheta_4(\mathfrak b_2) 
E_1 \left[\begin{smallmatrix}-1\\ b_1\end{smallmatrix}\right]
E_1 \left[\begin{smallmatrix}-1\\ b_2\end{smallmatrix}\right]
}{
\vartheta_4(\mathfrak b_3) 
\vartheta_1(2\mathfrak b_1)\vartheta_1(2\mathfrak b_2) 
\vartheta_1(\mathfrak b_1-\mathfrak b_3)
\vartheta_1(\mathfrak b_1+\mathfrak b_3)
\vartheta_1(\mathfrak b_2-\mathfrak b_3)
\vartheta_1(\mathfrak b_2+\mathfrak b_3)
}
\nonumber\\[0.8em]
&+
\frac{
\vartheta_4(\mathfrak b_1)\vartheta_4(\mathfrak b_3) 
E_1 \left[\begin{smallmatrix}-1\\ b_1\end{smallmatrix}\right]
E_1 \left[\begin{smallmatrix}-1\\ b_3\end{smallmatrix}\right]
}{
\vartheta_4(\mathfrak b_2) 
\vartheta_1(2\mathfrak b_1)\vartheta_1(2\mathfrak b_3) 
\vartheta_1(\mathfrak b_1-\mathfrak b_2)
\vartheta_1(\mathfrak b_1+\mathfrak b_2)
\vartheta_1(\mathfrak b_3-\mathfrak b_2)
\vartheta_1(\mathfrak b_3+\mathfrak b_2)
}
\nonumber\\[0.8em]
&+
\frac{
\vartheta_4(\mathfrak b_2)\vartheta_4(\mathfrak b_3) 
E_1 \left[\begin{smallmatrix}-1\\ b_2\end{smallmatrix}\right]
E_1 \left[\begin{smallmatrix}-1\\ b_3\end{smallmatrix}\right]
}{
\vartheta_4(\mathfrak b_1) 
\vartheta_1(2\mathfrak b_2)\vartheta_1(2\mathfrak b_3) 
\vartheta_1(\mathfrak b_2-\mathfrak b_1)
\vartheta_1(\mathfrak b_2+\mathfrak b_1)
\vartheta_1(\mathfrak b_3-\mathfrak b_1)
\vartheta_1(\mathfrak b_3+\mathfrak b_1)
}
\Bigg).
\end{align}

The computation of $SO(5)$ can be easily generalized to all $SO(2r+1)$ theories with $N_f = 2r-1$ flavors. The final result is given by
\begin{align}
  \mathcal I_{SO(2r+1)}
=
\sum_{\substack{S\subset F\\ |S|=r}}
\frac{
i^{r^2}\eta(\tau)^{2r^2-r-1}\prod_{s\in S}
\vartheta_4(\mathfrak b_s) 
E_1 \left[\begin{smallmatrix}-1\\ b_s\end{smallmatrix}\right]
}{
\prod_{t\in F\setminus S}\vartheta_4(\mathfrak b_t)
\prod_{s\in S}\vartheta_1(2\mathfrak b_s)
\prod_{\substack{s\in S\\ t\in F\setminus S}}
\vartheta_1(\mathfrak b_s-\mathfrak b_t)
\vartheta_1(\mathfrak b_s+\mathfrak b_t)
} \ ,
\end{align}
where $F = \{1, 2, \cdots, 2r - 1\}$.

As the last example of BCD-type SQCD theories, we consider the $USp(2r)$ SQCD with $N_f = 2r + 2$ fundamental flavors. The fundamental representation of $USp(2r)$ is pseudoreal, hence the flavor symmetry is given by $SO(4r + 4)$. The Schur index is given by the following contour integral
\begin{align}
  \mathcal{I}
  =
  \oint \left[{\frac{da}{2\pi i a}}\right]\frac{\eta(\tau)^{2r(r+3)}
  }{2^r r!}
  \frac{
  \prod_{A=1}^r\vartheta_1(2\mathfrak a_A)^2
  \prod_{A < B}
  \vartheta_1(\mathfrak a_A-\mathfrak a_B)^2
  \vartheta_1(\mathfrak a_A+\mathfrak a_B)^2
  }{
  \prod_{A=1}^r\prod_{j=1}^{2r+2}
  \vartheta_4(\mathfrak a_A-\mathfrak b_j)
  \vartheta_4(\mathfrak a_A+\mathfrak b_j)
  }.
\end{align}
The Schur index in closed form can be written in a very elegant form,
\begin{align}
\mathcal{I}_{USp(2r)}
=
\sum_{\substack{S\subset\{1,\dots,2r+2\}\\ |S|=r}}
\frac{
\prod_{i\in S}
i^{r^2}\eta(\tau)^{r(2r+3)}\vartheta_1(2\mathfrak b_i) 
}{
\prod_{i\in S}\prod_{j\notin S}
\vartheta_1(\mathfrak b_i-\mathfrak b_j)
\vartheta_1(\mathfrak b_i+\mathfrak b_j)
}E_1\left[\begin{matrix}-1\\ b_i\end{matrix}\right]\ .
\end{align}

This family of theories has appeared in the construction of generalized partition function \cite{Deb:2025ypl}, and later analyzed in more detail in \cite{Deb:2025ddc,Chandra:2025qpv}. In particular, \cite{Chandra:2025qpv} shows that the unflavored Schur index can be reorganized into the special integral
\begin{equation}
  \oint_0^\lambda \cdots \int_0^\lambda \prod_{j = 1}^r dt_j \prod_{i = 1}^r (t_i (1 - t_i)(\lambda - t_i))^{\alpha - \frac{1}{2}} \prod_{1 \le i < j \le r} (t_i - t_j)^{2\alpha}\ ,
\end{equation}
where $t_i = \lambda \operatorname{sn}(v_i, k)^2$, $v_i = \pi \vartheta_3(0)^2 \mathfrak{a}_i$, $k^2 = \lambda = \vartheta_2(\tau)^4/\vartheta_3(\tau)^4$, implying that the index satisfies an $(r + 1)$th-order MLDE with zero Wronskian index. It would be very useful to generalize their method to include flavor fugacities, and to the $SO(N)$ SQCDs as well. Further understanding these equations directly from the VOA representation theory point of view would also be very interesting. When $r = 2$, \cite{Chandra:2025qpv} conjectures the associated VOA to be $SO(12)_{-4}$. It would be interesting to confirm this conjecture and generalize to the entire series of $USp(2r)$ SQCD, which we leave to future work.

\subsection{\texorpdfstring{$\mathcal{N} = 4$ $SO(N)$ Theory}{}}

Next we focus on the $\mathcal{N} = 4$ $SO(N)$ theory. When $N = 2r$, the Schur index is given by the following contour integral
\begin{equation}
  \mathcal{I}_{SO(2r)} = \oint \prod_{i = 1}^r \frac{da_i}{2\pi i a_i}\frac{1}{|W|} \eta(\tau)^{2r} \left({\frac{\eta(\tau)}{\vartheta_4(\mathfrak{b})}}\right)^r \prod_{\alpha, \beta = \pm} \prod_{\substack{i, j = 1\\ i<j}}^r
  \frac{\vartheta_1(\alpha \mathfrak{a}_i + \beta \mathfrak{a}_j)}{\vartheta_4(\alpha \mathfrak{a}_i + \beta \mathfrak{a}_j + \mathfrak{b})}
\end{equation}
while when $N = 2r + 1$, the integral takes the form
\begin{align}
  \mathcal{I}_{SO(2r + 1)}
=
\oint
\prod_{i=1}^{r}\frac{da_i}{2\pi i a_i} 
\frac{1}{|W|} 
\eta(\tau)^{2r} 
\left(\frac{\eta(\tau)}{\vartheta_4(\mathfrak b)}\right)^r
\prod_{i=1}^{r}
\frac{
\vartheta_1(\mathfrak a_i)^2
}{
\vartheta_4(\mathfrak a_i+\mathfrak b)
\vartheta_4(-\mathfrak a_i+\mathfrak b)
}\nonumber\\
\times \prod_{1\le i < j \le r}
\prod_{\alpha,\beta=\pm}
\frac{
\vartheta_1(\alpha\mathfrak a_i+\beta\mathfrak a_j)
}{
\vartheta_4(\alpha\mathfrak a_i+\beta\mathfrak a_j+\mathfrak b)
}
\ .
\end{align}
Here $|W|$ is the order of the Weyl group of $SO(N)$. The analytic computation of the integral is straightforward, but we will see that the intermediate results grows rapidly in complexity as the rank $r$ increases. We will only present $SO(6), SO(7), SO(8)$ examples in the following discussions.

\subsubsection{\texorpdfstring{$SO(6)$}{}}

We start with the $SO(6)$ case. We will evaluate the integral by integrating $\mathfrak{a}_1, \mathfrak{a}_2, \mathfrak{a}_3$ in order. The first integration can be viewed as an integral with a trivial polynomial $1$ of Eisenstein series, and the solution to the corresponding difference equation is
\begin{align}
  n(qa_1) - n(a_1) = 1 \Rightarrow n(a_1) = - E_1 \begin{bmatrix}
    -1 \\ a_1 q^{- 1/2}
  \end{bmatrix} \ .
\end{align}
The residue theorem leads to the poles
\begin{align}
  \alpha \mathfrak{b} + \beta \mathfrak{a}_2 + \frac{\tau}{2}, \qquad
  \alpha \mathfrak{b} + \beta \mathfrak{a}_3 + \frac{\tau}{2}, \qquad
  \tau \ .
\end{align}
The last pole comes from $E_1 \big[ \substack{-1 \\ a_1 q^{-1/2}} \big]$. Collecting the residues, we have
\begin{align}
= & \ 
\frac{
\eta(\tau)^9 
\prod_{i=2}^{3}
\vartheta_1(\mathfrak{a}_i)^4 
\prod_{\alpha=\pm 1}
\vartheta_1(\alpha\mathfrak{a}_2+\mathfrak{a}_3)^2
}{
4 
\vartheta_4(\mathfrak{b})^3 
\prod_{\alpha=\pm 1}
\prod_{i=2}^{3}
\vartheta_4(\mathfrak{b}+\alpha\mathfrak{a}_i)^2 
\prod_{\alpha,\beta=\pm 1}
\vartheta_4(\mathfrak{b}+\alpha\mathfrak{a}_2+\beta\mathfrak{a}_3)
} \\
& \ 
- \sum_{\alpha=\pm 1}
\frac{
 i \eta(\tau)^6 
 \vartheta_4(\mathfrak{b}+2\alpha \mathfrak{a}_2)^2
 \prod_{\beta=\pm 1}
 \vartheta_1(\mathfrak{a}_3+\alpha\beta \mathfrak{a}_2) 
 \vartheta_4(\mathfrak{b}+\beta \mathfrak{a}_3+\alpha \mathfrak{a}_2)
 E_1
\big[
\substack{-1 \\
a_2 b^\alpha}
\big]
}{
 2 
 \vartheta_1(2\mathfrak{b}) 
 \vartheta_4(\mathfrak{b}) 
 \vartheta_1(2\mathfrak{b}+2\alpha \mathfrak{a}_2) 
 \vartheta_1(2\mathfrak{a}_2)
 \prod_{\pm}
 \vartheta_1(2\mathfrak{b}\pm \mathfrak{a}_3+\alpha \mathfrak{a}_2) 
 \vartheta_4(\mathfrak{b}\pm \mathfrak{a}_3-\alpha \mathfrak{a}_2)
}
 \nonumber\\
& \ - \sum_{\alpha=\pm 1}
\frac{
 i \eta(\tau)^6 
 \vartheta_4(\mathfrak{b}+2\alpha \mathfrak{a}_3)^2
 \prod_{\beta=\pm 1}
 \vartheta_1(\mathfrak{a}_2+\alpha\beta \mathfrak{a}_3) 
 \vartheta_4(\mathfrak{b}+\beta \mathfrak{a}_2+\alpha \mathfrak{a}_3)
 E_1
\big[
\substack{-1 \\
a_3 b^\alpha}
\big]
}{
 2 
 \vartheta_1(2\mathfrak{b}) 
 \vartheta_4(\mathfrak{b}) 
 \vartheta_1(2\mathfrak{b}+2\alpha \mathfrak{a}_3) 
 \vartheta_1(2\mathfrak{a}_3)
 \prod_{\pm}
 \vartheta_1(2\mathfrak{b}\pm \mathfrak{a}_2+\alpha \mathfrak{a}_3) 
 \vartheta_4(\mathfrak{b}\pm \mathfrak{a}_2-\alpha \mathfrak{a}_3)
} \ . \nonumber
\end{align}
The resulting Eisenstein series leads to the following difference equations for the subsequent integrations,
\begin{align}
n(a q) - n(a) = & \ E_1 \begin{bmatrix}
  -1 \\ a b^\alpha
\end{bmatrix}
\quad\Rightarrow
\quad
n(a) = 1 - E_1 \begin{bmatrix}
  -1 \\ a
\end{bmatrix} + E_2 \begin{bmatrix}
  1 \\ b^\alpha a q^{\frac{1}{2}}
\end{bmatrix} \ .
\end{align}
The subsequent integrations generate lengthy and uninspiring results, which we do not demonstrate here. The final result of the index receives contributions from five terms,
\begin{align}
  \mathcal{I}_{SO(6)}
  = - \frac{11}{6} \frac{\eta(\tau)^3\vartheta_4(\mathfrak{b})}{\vartheta_1(2\mathfrak{b})^2} \frac{\vartheta_4(\mathfrak{b})}{\vartheta_4(3 \mathfrak{b})}+ \frac{\eta(\tau)^3\vartheta_4(\mathfrak{b})}{\vartheta_1(2\mathfrak{b})^2} \sum_{i = 2}^{4} \frac{\vartheta_i(0)^2}{\vartheta_i(2 \mathfrak{b})} \mathcal{E}_i
  + \frac{\vartheta_4(\mathfrak{b})}{\vartheta_1(4\mathfrak{b})} \mathcal{E}_4 \ .
\end{align}
Here $\mathcal{E}_i$ are polynomials of Eisenstein series. Explicitly, the index $\mathcal{I}$ is given by
\begin{align}
&\frac{1}{48}
\frac{\vartheta_4(\mathfrak{b})\eta(\tau)^3}
{\vartheta_1(2\mathfrak{b})^2}
\Biggl( -\frac{88\vartheta_4(\mathfrak{b})}{\vartheta_4(3\mathfrak{b})} \nonumber\\
& \ 
+\frac{
9\vartheta_2(0)^2
}{\vartheta_2(2\mathfrak{b})^2}
\bigg(-1-8E_2\begin{bmatrix}1\\-1\end{bmatrix}
+8E_2\begin{bmatrix}-1\\-b\end{bmatrix}\bigg)
+\frac{
9\vartheta_3(0)^2
}{\vartheta_3(2\mathfrak{b})^2}\bigg(
1-8E_2\begin{bmatrix}-1\\-1\end{bmatrix}
+8E_2\begin{bmatrix}1\\-b\end{bmatrix}
\bigg) \nonumber\\
& \ \qquad +\frac{
9\vartheta_4(0)^2
}{\vartheta_4(2\mathfrak{b})^2}\bigg(
1-8E_2\begin{bmatrix}-1\\1\end{bmatrix}
+8E_2\begin{bmatrix}1\\b\end{bmatrix}
\bigg)
\Biggr)
\nonumber\\
&\quad
+\frac{i}{12}
\frac{\vartheta_4(\mathfrak{b})}{\vartheta_1(4\mathfrak{b})}
\Biggl[
\sum_{\alpha, \beta = \pm}
\Biggl(
\left(
22-\frac{19(1-\alpha)(1+\beta)}{4}
\right)
E_1\begin{bmatrix}\alpha\\ \beta b\end{bmatrix}
-\frac{19+9\alpha}{2}
E_1\begin{bmatrix}\alpha\\ \beta b^2\end{bmatrix}
\nonumber\\
&\qquad\qquad
+\left(
48+9(1-\alpha)(1+\beta)
\right)
E_3\begin{bmatrix}\alpha\\ \beta b\end{bmatrix}
-\left(
24+9(1+\alpha)(1+\beta)
\right)
E_3\begin{bmatrix}\alpha\\ \beta b^2\end{bmatrix}
\nonumber\\
&\qquad\qquad
+24E_1\begin{bmatrix}\alpha\\ \beta b^2\end{bmatrix}
\Biggl(
\frac{1-\alpha}{2}
\left(
E_2\begin{bmatrix}1\\ \beta b\end{bmatrix}
-
E_2\begin{bmatrix}-1\\ \beta\end{bmatrix}
\right)
\nonumber\\
&\qquad\qquad\qquad
+\frac{(1+\alpha)(1-\beta)}{4}
\left(
E_2\begin{bmatrix}-1\\ \beta b\end{bmatrix}
-
E_2\begin{bmatrix}1\\ \beta\end{bmatrix}
+ 2E_2\begin{bmatrix}-1\\ \beta b\end{bmatrix}
-\frac{1}{2}
E_2\begin{bmatrix}1\\ \beta b^2\end{bmatrix}
\right)\Biggr)
\Biggr)
\nonumber\\
&\qquad
+\sum_{\alpha=-1,1}
\Biggl(
12E_1\begin{bmatrix}-1\\ b\end{bmatrix}
\Biggl(
\frac{5\alpha-1}{2}
E_2\begin{bmatrix}
\alpha\\
b^{\frac{3+\alpha}{2}}
\end{bmatrix}
+\frac{1-3\alpha}{2}
E_2\begin{bmatrix}
\alpha\\
b^{\frac{7+\alpha}{2}}
\end{bmatrix}
\Biggr)
\nonumber\\
&\qquad\qquad
+\alpha
E_1\begin{bmatrix}
\alpha\\
b^{\frac{7+\alpha}{2}}
\end{bmatrix}
\Biggl(
19+24
\left(
E_2\begin{bmatrix}-1\\ b\end{bmatrix}
-
E_2\begin{bmatrix}1\\ b^2\end{bmatrix}
\right)
\Biggr)
\Biggr)
\nonumber\\
&\qquad
+12E_1\begin{bmatrix}-1\\ b\end{bmatrix}
E_1\begin{bmatrix}1\\ b^2\end{bmatrix}
\left(
2E_1\begin{bmatrix}1\\ b^2\end{bmatrix}
-3E_1\begin{bmatrix}-1\\ b\end{bmatrix}
\right)
\Biggr].
\end{align}
The unflavored index can be written as
\begin{align}
  \mathcal{I} = \frac{\vartheta_4(0)}{4\eta(\tau)^3}\bigg(
    &E_2(\tau)
+3E_2(\tau)^2
-12E_4(\tau)
+\frac{1}{2}E_2\begin{bmatrix}-1\\-1\end{bmatrix}
+3E_2(\tau)E_2\begin{bmatrix}-1\\-1\end{bmatrix}
+\frac{5}{2}E_2\begin{bmatrix}-1\\-1\end{bmatrix}^2
\nonumber\\
&\quad
+\frac{5}{2}E_2\begin{bmatrix}-1\\1\end{bmatrix}
+15E_2(\tau)E_2\begin{bmatrix}-1\\1\end{bmatrix}
-6E_2\begin{bmatrix}-1\\1\end{bmatrix}^2
+\frac{1}{2}E_2\begin{bmatrix}1\\-1\end{bmatrix}
\nonumber\\
&\quad
+3E_2(\tau)E_2\begin{bmatrix}1\\-1\end{bmatrix}
-8E_2\begin{bmatrix}-1\\-1\end{bmatrix}
E_2\begin{bmatrix}1\\-1\end{bmatrix}
+\frac{5}{2}E_2\begin{bmatrix}1\\-1\end{bmatrix}^2
\nonumber\\
&\quad
-9E_4\begin{bmatrix}-1\\-1\end{bmatrix}
-9E_4\begin{bmatrix}1\\-1\end{bmatrix}
  \bigg) \ .
\end{align}

The Schur index of the $SO(6)$ theory is related to that of the non-Lagrangian $\widehat{E}_6(SO(6))$ by a simple replacement \cite{Kang:2021lic}
\begin{equation}
  \mathcal{I}_{\widehat{E}_6(SO(6))}(q) = \mathcal{I}_{SO(6)}(b \to q^{3}, q\to q^{\frac{3}{2} - 1}) \ .
\end{equation}
Using shift properties like
\begin{align}
  \vartheta_1(\tau|3\tau) = & \ i q^{-1/8}\vartheta_4(\frac{\tau}{2}|3\tau)\ ,
  & \frac{1}{12\pi^2} \frac{\vartheta_1'''(0|3\tau)}{\vartheta_1'(0|3\tau)}  = & \ E_2(3\tau) , \\
  -i q^{\frac{1}{6}}\vartheta_1(\tau|3\tau) = & \ q^{\frac{1}{24}}\vartheta_4(\frac{\tau}{2}|3\tau)  = \eta(\tau) \ , 
  \qquad
  & -2\pi i q^{1/8}\prod_{i = 1}^3\vartheta_i(\frac{\tau}{2}|3\tau)
  = & \ \vartheta_1'(0|3\tau) \ ,
\end{align}
the latter index can be written in closed-form. Define
\begin{equation}
\mathcal C_1=
\left[\begin{matrix}1\\ \sqrt q\end{matrix}\right],
\qquad
\mathcal C_2=
\left[\begin{matrix}1\\ -\sqrt q\end{matrix}\right],
\qquad
\mathcal C_3=
\left[\begin{matrix}-1\\ -\sqrt q\end{matrix}\right],
\qquad
\mathcal C_4=
\left[\begin{matrix}-1\\ \sqrt q\end{matrix}\right],
\end{equation}
and
\begin{equation}
  \mathcal D_2=
\left[\begin{matrix}1\\ -1\end{matrix}\right],
\qquad
\mathcal D_3=
\left[\begin{matrix}-1\\ -1\end{matrix}\right],
\qquad
\mathcal D_4=
\left[\begin{matrix}-1\\ 1\end{matrix}\right] \ ,
\end{equation}
and set the coefficients
\begin{equation}
  (\rho_{i= 1,2,3})=(1,-1,-1),
\quad
(\sigma_{i = 1,2,3})=(1,-1,1),
\quad
(\lambda_{i=1,2,3})=(-129,-129,33).
\end{equation}
Then the Schur index of $\widehat{E}_6(SO(6))$ is $\mathcal{I}_{\widehat{E}_6(SO(6))}=q^{5/8}A+B,$ where
\begin{align}
& A
=
\frac{1}{12}E_2(3\tau)
+
\frac{1}{1296}
\Bigg[
32\sum_{i=1}^{3}E_1[\mathcal C_i](3\tau)^3
\nonumber\\
&
+
24\sum_{i=1}^{3}\sum_{j=1}^{3}
(1-\delta_{ij})E_1[\mathcal C_i](3\tau)^2E_1[\mathcal C_j](3\tau)
-
456\prod_{i=1}^{3}E_1[\mathcal C_i](3\tau)
\nonumber\\
&
-
84\sum_{i=1}^{3}E_1[\mathcal C_i](3\tau)^2
-
240\sum_{1\leq i < j\leq 3}
E_1[\mathcal C_i](3\tau)E_1[\mathcal C_j](3\tau)
\nonumber\\
&
+
\sum_{i=1}^{3}
E_1[\mathcal C_i](3\tau)
(
432E_2[\mathcal D_{5-i}](3\tau)
-
360E_2[\mathcal C_i](3\tau)
+
504\sum_{j=1}^{3}(1-\delta_{ij})E_2[\mathcal C_j](3\tau)
+
\lambda_i
)
\nonumber\\
&
+
72\sum_{i=1}^{3}E_2[\mathcal C_i](3\tau)
+
864\sum_{i=1}^{3}E_3[\mathcal C_i](3\tau)
-
49
\Bigg],
\end{align}
and
\begin{align}
B
={}&
\frac{\sqrt q}{96}
\Bigg[
q^{1/8}
\left(
8\sum_{i=2}^{4}
\left(
E_2(3\tau)+2E_2[\mathcal D_i](3\tau)
\right)
-
16\sum_{i=1}^{3}
\left(
E_2(3\tau)+2E_2[\mathcal C_i](3\tau)
\right)
\right)
\nonumber\\
&\qquad
+
\sum_{i=1}^{3}
\frac{
3\eta(3\tau)^3\vartheta_{5-i}(0|3\tau)^2
}{
\vartheta_i(\tau/2|3\tau)^2\vartheta_4(\tau/2|3\tau)
}
\left(
8\rho_i
\left(
E_2[\mathcal C_i](3\tau)
-
E_2[\mathcal D_{5-i}](3\tau)
\right)
+
\sigma_i
\right)
\Bigg].
\end{align}

\subsubsection{\texorpdfstring{$SO(7)$}{}}

The $\mathcal{N} = 4$ $SO(7)$ Schur index can be written as the following contour integral
\begin{align}
  \mathcal{I}_{SO(7)}
=
\oint
\prod_{i=1}^{3}\frac{da_i}{2\pi i a_i} 
\frac{1}{48} 
\eta(\tau)^6
\left(\frac{\eta(\tau)}{\vartheta_4(\mathfrak b)}\right)^3
\prod_{i=1}^{3}
\frac{
\vartheta_1(\mathfrak a_i)^2
}{
\vartheta_4(\mathfrak a_i+\mathfrak b)
\vartheta_4(-\mathfrak a_i+\mathfrak b)
}\nonumber\\
\times \prod_{1\le i < j \le 3}
\prod_{\alpha,\beta=\pm}
\frac{
\vartheta_1(\alpha\mathfrak a_i+\beta\mathfrak a_j)
}{
\vartheta_4(\alpha\mathfrak a_i+\beta\mathfrak a_j+\mathfrak b)
}
\ .
\end{align}
The flavored Schur index of the theory was first computed in \cite{Guo:2023mkn}. Here we apply the new technique to compute the index in simpler form, receiving contributions from nine terms, each term containing a theta-ratio prefactor and a polynomial of Eisenstein series,
\begin{align}
  & \ \mathcal{I}_{SO(7)} \nonumber \\
= & \ \frac{\eta(\tau)^3 \vartheta_4(\mathfrak{b})^4}{6  \vartheta_1(2\mathfrak{b})^4 \vartheta_4(3\mathfrak{b})} \left(11 + 6 \left(E_1 \begin{bsmallmatrix} -1 \\ b \end{bsmallmatrix}\right)^{ 2}\right) \\
& \ + \frac{3 \mathrm{i}  \vartheta_2(0)  \vartheta_3(\mathfrak{b})  \vartheta_4(\mathfrak{b})^3}{8  \vartheta_1(2\mathfrak{b})^3 \vartheta_2(2\mathfrak{b})  \vartheta_3(3\mathfrak{b})}   E_1 \begin{bsmallmatrix} -1 \\ b \end{bsmallmatrix} \left(-1 + 8 E_2 \begin{bsmallmatrix} -1 \\ -b \end{bsmallmatrix} - 8 E_2 \begin{bsmallmatrix} 1 \\ -1 \end{bsmallmatrix}\right) \nonumber\\
& \ + \frac{3 \mathrm{i}  \vartheta_2(\mathfrak{b})  \vartheta_3(0)  \vartheta_4(\mathfrak{b})^3}{8  \vartheta_1(2\mathfrak{b})^3 \vartheta_2(3\mathfrak{b})  \vartheta_3(2\mathfrak{b})}   E_1 \begin{bsmallmatrix} -1 \\ b \end{bsmallmatrix} \left(1 - 8 E_2 \begin{bsmallmatrix} -1 \\ -1 \end{bsmallmatrix} + 8 E_2 \begin{bsmallmatrix} 1 \\ -b \end{bsmallmatrix}\right) \nonumber\\
& \ + \frac{3 \mathrm{i}  \vartheta_1(\mathfrak{b})  \vartheta_4(0)  \vartheta_4(\mathfrak{b})^3}{8  \vartheta_1(2\mathfrak{b})^3 \vartheta_1(3\mathfrak{b})  \vartheta_4(2\mathfrak{b})}   E_1 \begin{bsmallmatrix} -1 \\ b \end{bsmallmatrix} \left(1 - 8 E_2 \begin{bsmallmatrix} -1 \\ 1 \end{bsmallmatrix} + 8 E_2 \begin{bsmallmatrix} 1 \\ b \end{bsmallmatrix}\right) \nonumber\\
& \ + \frac{\eta(\tau)^3 \vartheta_1(2\mathfrak{b})}{6  \vartheta_1(4\mathfrak{b})  \vartheta_4(3\mathfrak{b})} \left(-7 + 12 E_1 \begin{bsmallmatrix} -1 \\ b \end{bsmallmatrix} E_1 \begin{bsmallmatrix} 1 \\ b^2 \end{bsmallmatrix} + 12 E_2 \begin{bsmallmatrix} -1 \\ b \end{bsmallmatrix} - 12 E_2 \begin{bsmallmatrix} 1 \\ b^2 \end{bsmallmatrix}\right) \nonumber\\
& \ + \frac{\mathrm{i}  \vartheta_3(0)  \vartheta_3(2\mathfrak{b})  \vartheta_4(\mathfrak{b})}{12  \vartheta_1(4\mathfrak{b})  \vartheta_2(\mathfrak{b})  \vartheta_2(3\mathfrak{b})} \bigg(22 E_1 \begin{bsmallmatrix} 1 \nonumber\\ -b \end{bsmallmatrix} + E_1 \begin{bsmallmatrix} -1 \\ -b^2 \end{bsmallmatrix} \left(-5 - 24 E_2 \begin{bsmallmatrix} -1 \\ -1 \end{bsmallmatrix} + 24 E_2 \begin{bsmallmatrix} 1 \\ -b \end{bsmallmatrix}\right) \nonumber\\
& \ \qquad \qquad - 24 E_3 \begin{bsmallmatrix} -1 \\ -b^2 \end{bsmallmatrix} + 48 E_3 \begin{bsmallmatrix} 1 \\ -b \end{bsmallmatrix}\bigg) \nonumber\\
& \ - \frac{\mathrm{i}  \vartheta_4(0)  \vartheta_4(\mathfrak{b})  \vartheta_4(2\mathfrak{b})}{12  \vartheta_1(\mathfrak{b})  \vartheta_1(3\mathfrak{b})  \vartheta_1(4\mathfrak{b})} \bigg(22 E_1 \begin{bsmallmatrix} 1 \\ + b \end{bsmallmatrix} + E_1 \begin{bsmallmatrix} -1 \\ + b^2 \end{bsmallmatrix} \left(- 5 -  24 E_2 \begin{bsmallmatrix} -1 \\ 1 \end{bsmallmatrix} + 24 E_2 \begin{bsmallmatrix} 1 \\ b \end{bsmallmatrix}\right) \nonumber\\
& \ \qquad \qquad - 24 E_3 \begin{bsmallmatrix} -1 \\ b^2 \end{bsmallmatrix} + 48 E_3 \begin{bsmallmatrix} 1 \\ b \end{bsmallmatrix} \bigg) \nonumber\\
& \ + \frac{\mathrm{i}  \vartheta_2(0)  \vartheta_2(2\mathfrak{b})  \vartheta_4(\mathfrak{b})}{12  \vartheta_1(4\mathfrak{b})  \vartheta_3(\mathfrak{b})  \vartheta_3(3\mathfrak{b})} \bigg(22 E_1 \begin{bsmallmatrix} -1 \\ -b \end{bsmallmatrix} + E_1 \begin{bsmallmatrix} 1 \\ -b^2 \end{bsmallmatrix} \left(-14 + 24 E_2 \begin{bsmallmatrix} -1 \\ -b \end{bsmallmatrix} - 24 E_2 \begin{bsmallmatrix} 1 \\ -1 \end{bsmallmatrix}\right) \nonumber \nonumber\\
& \ \qquad \qquad+ 24 E_3 \begin{bsmallmatrix} -1 \\ -b \end{bsmallmatrix} - 12 E_3 \begin{bsmallmatrix} 1 \\ -b^2 \end{bsmallmatrix}\bigg) \nonumber\\
& \ - \frac{\mathrm{i}  \vartheta_4(\mathfrak{b})  \vartheta_4(3\mathfrak{b})  \vartheta_4(5\mathfrak{b})}{12  \vartheta_1(2\mathfrak{b})  \vartheta_1(4\mathfrak{b})  \vartheta_1(6\mathfrak{b})} \Big(19 E_1 \begin{bsmallmatrix} -1 \\ b \end{bsmallmatrix} + 10 E_1 \begin{bsmallmatrix} -1 \\ b^3 \end{bsmallmatrix} - 44 E_1 \begin{bsmallmatrix} 1 \\ b^2 \end{bsmallmatrix} \nonumber\\
&\quad + 24 E_1 \begin{bsmallmatrix} -1 \\ b \end{bsmallmatrix} E_1 \begin{bsmallmatrix} -1 \\ b^3 \end{bsmallmatrix} E_1 \begin{bsmallmatrix} 1 \\ b^2 \end{bsmallmatrix} + 48 E_1 \begin{bsmallmatrix} -1 \\ b^3 \end{bsmallmatrix} E_2 \begin{bsmallmatrix} -1 \\ b \end{bsmallmatrix} - 24 E_1 \begin{bsmallmatrix} -1 \\ b \end{bsmallmatrix} E_2 \begin{bsmallmatrix} -1 \\ b^3 \end{bsmallmatrix} \nonumber\\
&\quad + 24 E_1 \begin{bsmallmatrix} -1 \\ b \end{bsmallmatrix} E_2 \begin{bsmallmatrix} 1 \\ b^2 \end{bsmallmatrix} - 48 E_1 \begin{bsmallmatrix} -1 \\ b^3 \end{bsmallmatrix} E_2 \begin{bsmallmatrix} 1 \\ b^2 \end{bsmallmatrix} + 48 E_3 \begin{bsmallmatrix} -1 \\ b \end{bsmallmatrix} + 48 E_3 \begin{bsmallmatrix} -1 \\ b^3 \end{bsmallmatrix} - 96 E_3 \begin{bsmallmatrix} 1 \\ b^2 \end{bsmallmatrix}\Big) \ . \nonumber 
\end{align}

Finally, it is straightforward to take the unflavored limit $\mathfrak{b} \to 0$ in the above expression. Making use of a few identities of Eisenstein series and theta functions collected in appendix \ref{app:useful-identities}, we arrive at a relatively simple closed-form expression for the unflavored Schur index,
\begin{align}\label{eq:N4SO7unflavored}
\mathcal{I}_{SO(7)}^\text{unflavored} = &\frac{2 E_2(\tau)^3}{3 \vartheta_2(0)^3 \vartheta_3(0)^3}
+ \frac{5 \vartheta_2(0)^5}{3456 \vartheta_3(0)^3}
+ \frac{\vartheta_2(0)^9}{2592 \vartheta_3(0)^3}
\\
&- \frac{11 \vartheta_2(0) \vartheta_3(0)}{3456}
- \frac{5 \vartheta_2(0)^5 \vartheta_3(0)}{3456}
+ \frac{\vartheta_3(0)^5}{108 \vartheta_2(0)^3}
- \frac{5 \vartheta_2(0) \vartheta_3(0)^5}{3456}
+ \frac{\vartheta_3(0)^9}{2592 \vartheta_2(0)^3}
\nonumber\\
&+ E_2(\tau)^2 \left(
\frac{\vartheta_2(0)}{6 \vartheta_3(0)^3}
+ \frac{\vartheta_3(0)}{6 \vartheta_2(0)^3}
\right)\nonumber
\\
&+ E_2(\tau) \left(
\frac{5 \vartheta_2(0)}{288 \vartheta_3(0)^3}
+ \frac{\vartheta_2(0)^5}{72 \vartheta_3(0)^3}
+ \frac{\vartheta_3(0)}{9 \vartheta_2(0)^3}
- \frac{1}{288} \vartheta_2(0) \vartheta_3(0)
+ \frac{\vartheta_3(0)^5}{72 \vartheta_2(0)^3}
\right) \ . \nonumber
\end{align}
We will see that this is identical to the $USp(6)$ unflavored index.

\subsubsection{\texorpdfstring{$SO(8)$}{}}

The $\mathcal{N} = 4$ $SO(8)$ Schur index is given by the following contour integral
\begin{equation}
  \mathcal{I}_{SO(8)} = \oint \prod_{i = 1}^4 \frac{da_i}{2\pi i a_i} \frac{1}{|W|} \eta(\tau)^8 \left({\frac{\eta}{\vartheta_4(\mathfrak{b})}}\right)^4 \prod_{\alpha, \beta = \pm} \prod_{\substack{i, j = 1\\ i<j}}^4
  \frac{\vartheta_1(\alpha \mathfrak{a}_i + \beta \mathfrak{a}_j)}{\vartheta_4(\alpha \mathfrak{a}_i + \beta \mathfrak{a}_j + \mathfrak{b})} \ . \nonumber
\end{equation}
The integral can be evaluated by integrating $\mathfrak{a}_1, \mathfrak{a}_2, \mathfrak{a}_3, \mathfrak{a}_4$ in order. The final result of the index can be written in Eisenstein series, which is quite lengthy and we will not present it here. Although the closed-form is somewhat complicated, the overall structure of $\mathcal{I}_{SO(8)}$ is quite simple, receiving just eight contributions,
\begin{align}
  & \ \mathcal{I}_{SO(8)} = - \frac{22 \eta(\tau)^6 \vartheta_1(\mathfrak{b})^4 \vartheta_4(\mathfrak{b})}{3\vartheta_1(2\mathfrak{b})^2 \vartheta_4(2 \mathfrak{b})^4 \vartheta_4(3 \mathfrak{b})}\nonumber\\
  & \ + \frac{i \eta(\tau)^3 \vartheta_3(0)^2 \vartheta_3(2 \mathfrak{b})^2}{
    3 \vartheta_2(\mathfrak{b})^2
    \vartheta_2(3 \mathfrak{b})^2
    \vartheta_1(4 \mathfrak{b}) 
  } \mathcal{E}_1
  + \frac{i\eta(\tau)^3 \vartheta_4(0)^2 \vartheta_4(2 \mathfrak{b})^2}{
    3 \vartheta_1(\mathfrak{b})^2
    \vartheta_1(3 \mathfrak{b})^2
    \vartheta_1(4 \mathfrak{b}) 
  } \mathcal{E}_2
  -\frac{i \eta (\tau)^3 \vartheta_2(0)^2 \vartheta_2(2 \mathfrak{b})^2}{3 \vartheta_1(4 \mathfrak{b}) \vartheta_3(\mathfrak{b})^2 \vartheta_3(3 \mathfrak{b})^2} \mathcal{E}_3\nonumber\\
  & \ + \frac{
    \vartheta_3(0)^2 
    \vartheta_2(\mathfrak{b})
    \vartheta_4(\mathfrak{b})^4}{4
  \vartheta_1(2 \mathfrak{b})^4
  \vartheta_3(2\mathfrak{b})^2
  \vartheta_2(3\mathfrak{b})} \mathcal{E}_4
  + \frac{
    \vartheta_2(0)^2 \vartheta_3(\mathfrak{b})\vartheta_4(\mathfrak{b})^4
  }{
    \vartheta_1(2 \mathfrak{b})^4
    \vartheta_2(2\mathfrak{b})^2
    \vartheta_3(3 \mathfrak{b})
  }\mathcal{E}_6\nonumber\\
  & \ -\frac{\vartheta_4(\mathfrak{b})^3 \vartheta_4(3 \mathfrak{b})}{12 \vartheta_1(2 \mathfrak{b})^2 \vartheta_1(4 \mathfrak{b})^2}\mathcal{E}_5
  + \frac{\vartheta_4(\mathfrak{b}) \vartheta_4(3 \mathfrak{b})^2 \vartheta_4(5 \mathfrak{b})}{12 \vartheta_1(2 \mathfrak{b}) \vartheta_1(4 \mathfrak{b})^2 \vartheta_1(6 \mathfrak{b})}\mathcal{E}_7 \ ,
\end{align}
where $\mathcal{E}_i$ are polynomials of Eisenstein series. The explicit expressions of $\mathcal{E}_i$ are computed, but are quite lengthy and will not be presented here. Making use of a few identities of Eisenstein series and Jacobi theta functions, we can write down the unflavored Schur index in the following form
\begin{align}
\mathcal{I}_{SO(8)}^\text{unflavored} = & \ \frac{1}{144}E_2(\tau)
+\frac{1}{48}E_2(\tau)^2
+\frac{E_2(\tau)^2}{12\vartheta_2(0)^4}
+\frac{E_2(\tau)^3}{12\vartheta_2(0)^4}
+\frac{1}{576}E_2(\tau)\vartheta_2(0)^4
+\frac{\vartheta_2(0)^8}{20736}
\nonumber\\
&\quad
+\frac{E_2(\tau)^2}{12\vartheta_3(0)^4}
+\frac{E_2(\tau)^3}{12\vartheta_3(0)^4}
+\frac{E_2(\tau)^4}{4\vartheta_2(0)^4\vartheta_3(0)^4}
+\frac{\vartheta_2(0)^4}{512\vartheta_3(0)^4}
\nonumber\\
&\quad
+\frac{E_2(\tau)\vartheta_2(0)^4}{72\vartheta_3(0)^4}
+\frac{E_2(\tau)^2\vartheta_2(0)^4}{96\vartheta_3(0)^4}
+\frac{\vartheta_2(0)^8}{1728\vartheta_3(0)^4}
+\frac{E_2(\tau)\vartheta_2(0)^8}{1728\vartheta_3(0)^4}
\\
&\quad
+\frac{\vartheta_2(0)^{12}}{82944\vartheta_3(0)^4}
+\frac{1}{576}E_2(\tau)\vartheta_3(0)^4
+\frac{E_2(\tau)\vartheta_3(0)^4}{72\vartheta_2(0)^4}
+\frac{E_2(\tau)^2\vartheta_3(0)^4}{96\vartheta_2(0)^4}
\nonumber\\
&\quad
-\frac{\vartheta_2(0)^4\vartheta_3(0)^4}{1728}
+\frac{\vartheta_3(0)^8}{20736}
+\frac{\vartheta_3(0)^8}{1728\vartheta_2(0)^4}
+\frac{E_2(\tau)\vartheta_3(0)^8}{1728\vartheta_2(0)^4}
+\frac{\vartheta_3(0)^{12}}{82944\vartheta_2(0)^4} \ . \nonumber
\end{align}
It would be interesting to match analytically the above $SO(N)$ unflavored Schur index with the results in \cite{Du:2023kfu}.

\subsection{\texorpdfstring{$\mathcal{N} = 4$ $USp(6)$ Theory}{}}

The Schur index of $\mathcal{N} = 4$ $USp(6)$ theory can be written as the following contour integral
\begin{equation}
  \mathcal{I}_{USp(6)} = \oint \prod_{i = 1}^3 \frac{da_i}{2\pi i a_i} \frac{1}{8} \eta(\tau)^6 \left({\frac{\eta}{\vartheta_4(\mathfrak{b})}}\right)^3 \prod_{\alpha, \beta = \pm} \prod_{\substack{i, j = 1\\ i<j}}^3
  \frac{\vartheta_1(\alpha \mathfrak{a}_i + \beta \mathfrak{a}_j)}{\vartheta_4(\alpha \mathfrak{a}_i + \beta \mathfrak{a}_j + \mathfrak{b})}
  \prod_{i = 1}^3 \frac{\vartheta_1(2\mathfrak{a}_i)}{\vartheta_4(2\mathfrak{a}_i + \mathfrak{b})} \ . \nonumber
\end{equation}
By S-duality, the Schur index is expected to coincide with that of the $USp(6)^\vee = SO(7)$ $\mathcal{N} = 4$ theory. To simplify computation, we define a new set of variables $a'$
\begin{equation}
  a'_1 = a_1 a_2, \qquad
  a'_2 = a_2 a_3, \qquad
  a'_3 = a_3 a_1 \ .
\end{equation}
The integral can be evaluated by integrating $\mathfrak{a}'_1, \mathfrak{a}'_2, \mathfrak{a}'_3$ in order. Unfortunately, the intermediate results are fairly tedious and do not fit in the page, hence we will only present the final result of the index in Eisenstein series. The overall structure of $\mathcal{I}_{USp(6)}$ is quite simple, receiving nine contributions,
\begin{align}
  \mathcal{I}_{USp(6)}
  = -\frac{
  i \vartheta_2(0)\vartheta_3(0)\vartheta_4(0)
  }{
  32 \vartheta_2(2\mathfrak{b})\vartheta_3(2\mathfrak{b})\vartheta_4(2\mathfrak{b})
  }
  \frac{\vartheta_4(\mathfrak{b})^3}{\vartheta_1(2\mathfrak{b})^3}\mathcal{E}_1
  + \sum_{i = 2}^{4} \frac{\eta(\tau)^3\vartheta_4(\mathfrak{b})^3}{64\vartheta_1(2 \mathfrak{b})^4}
  \frac{\vartheta_i(0)^2}{\vartheta_i(2\mathfrak{b})^2}\mathcal{E}_{i} \nonumber \\
  + \sum_{i = 1}^{4} \frac{i \vartheta_4(\mathfrak{b})^2 \vartheta_4(3 \mathfrak{b})}{192 \vartheta_1(2 \mathfrak{b})^2 \vartheta_1(4 \mathfrak{b})} \frac{\vartheta_i(\mathfrak{b})}{\vartheta_i(3 \mathfrak{b})} \mathcal{E}'_i
  + \frac{
  i \vartheta_4(\mathfrak{b})\vartheta_4(3\mathfrak{b})\vartheta_4(5\mathfrak{b})
  }{
  96 \vartheta_1(2\mathfrak{b})\vartheta_1(4\mathfrak{b})\vartheta_1(6\mathfrak{b})
  } \mathcal{E}_5 \ .
\end{align}
The $\mathcal{E}_i$ are polynomials of Eisenstein series. 
\begin{align}
\mathcal{E}_1 = & \ 4E_1\begin{bmatrix}1\\ b\end{bmatrix}
\left(
E_2\begin{bmatrix}-1\\ -1\end{bmatrix}
-
E_2\begin{bmatrix}-1\\ -b\end{bmatrix}
+
E_2\begin{bmatrix}1\\ -1\end{bmatrix}
-
E_2\begin{bmatrix}1\\ -b\end{bmatrix}
\right)
\nonumber\\
&\quad
+4E_1\begin{bmatrix}1\\ -b\end{bmatrix}
\left(
E_2\begin{bmatrix}-1\\ 1\end{bmatrix}
-
E_2\begin{bmatrix}-1\\ -b\end{bmatrix}
+
E_2\begin{bmatrix}1\\ -1\end{bmatrix}
-
E_2\begin{bmatrix}1\\ b\end{bmatrix}
\right)
\nonumber\\
&\quad
+4E_1\begin{bmatrix}-1\\ -b\end{bmatrix}
\left(
E_2\begin{bmatrix}-1\\ -1\end{bmatrix}
-
E_2\begin{bmatrix}1\\ -b\end{bmatrix}
+
E_2\begin{bmatrix}-1\\ 1\end{bmatrix}
-
E_2\begin{bmatrix}1\\ b\end{bmatrix}
\right)
-3E_1\begin{bmatrix}-1\\ -b\end{bmatrix} ,
\end{align}
\begin{align}
  \mathcal{E}_2 = 1 - 8 E_2 \begin{bmatrix}
    -1 \\ - b
  \end{bmatrix} + 8 E_2 \begin{bmatrix}
    1 \\ - 1
  \end{bmatrix}, \quad
  \mathcal{E}_3 = -1 - 8 E_2 \begin{bmatrix}
    -1 \\ - b
  \end{bmatrix} + 8 E_2 \begin{bmatrix}
    -1 \\ -1
  \end{bmatrix}, \\
  \mathcal{E}_4 = - 1 - 8 E_2 \begin{bmatrix}
    1 \\ b
  \end{bmatrix} + 8 E_2 \begin{bmatrix}
    -1 \\ 1
  \end{bmatrix} \ .
\end{align} 
The remaining $\mathcal{E}_5$ and $\mathcal{E}'_i$ are slightly more complicated,
\begin{align}
\mathcal{E}'_1 = & \ 6E_1\begin{bmatrix}-1\\ -b\end{bmatrix}
+24E_1\begin{bmatrix}-1\\ b\end{bmatrix}
-97E_1\begin{bmatrix}-1\\ b^2\end{bmatrix}
-6E_1\begin{bmatrix}1\\ -b\end{bmatrix}
+260E_1\begin{bmatrix}1\\ b\end{bmatrix}
+18E_1\begin{bmatrix}1\\ b^2\end{bmatrix}
\nonumber\\
&\quad
-96E_1\begin{bmatrix}-1\\ b\end{bmatrix}
E_2\begin{bmatrix}-1\\ 1\end{bmatrix}
-48E_1\begin{bmatrix}-1\\ b^2\end{bmatrix}
E_2\begin{bmatrix}-1\\ 1\end{bmatrix}
-48E_1\begin{bmatrix}1\\ b\end{bmatrix}
E_2\begin{bmatrix}-1\\ 1\end{bmatrix}
\nonumber\\
&\quad
-48E_1\begin{bmatrix}1\\ b^2\end{bmatrix}
E_2\begin{bmatrix}-1\\ 1\end{bmatrix}
-48E_1\begin{bmatrix}-1\\ b^2\end{bmatrix}
E_2\begin{bmatrix}-1\\ b\end{bmatrix}
+48E_1\begin{bmatrix}1\\ b\end{bmatrix}
E_2\begin{bmatrix}-1\\ b\end{bmatrix}
\nonumber\\
&\quad
+48E_1\begin{bmatrix}-1\\ b\end{bmatrix}
E_2\begin{bmatrix}-1\\ b^2\end{bmatrix}
+48E_1\begin{bmatrix}-1\\ b\end{bmatrix}
E_2\begin{bmatrix}1\\ b\end{bmatrix}
+48E_1\begin{bmatrix}-1\\ b^2\end{bmatrix}
E_2\begin{bmatrix}1\\ b\end{bmatrix}
\nonumber\\
&\quad
+48E_1\begin{bmatrix}1\\ b\end{bmatrix}
E_2\begin{bmatrix}1\\ b\end{bmatrix}
+48E_1\begin{bmatrix}1\\ b^2\end{bmatrix}
E_2\begin{bmatrix}1\\ b\end{bmatrix}
+24E_1\begin{bmatrix}-1\\ b^2\end{bmatrix}
E_2\begin{bmatrix}1\\ b^2\end{bmatrix}
\nonumber\\
&\quad
-192E_3\begin{bmatrix}-1\\ b^2\end{bmatrix}
+384E_3\begin{bmatrix}1\\ b\end{bmatrix} \ ,
\end{align}

\begin{align}
\mathcal{E}'_2 = & \ 6E_1\begin{bmatrix}-1\\ -b\end{bmatrix}
+24E_1\begin{bmatrix}-1\\ b\end{bmatrix}
-97E_1\begin{bmatrix}-1\\ -b^2\end{bmatrix}
+260E_1\begin{bmatrix}1\\ -b\end{bmatrix}
-6E_1\begin{bmatrix}1\\ b\end{bmatrix}
+18E_1\begin{bmatrix}1\\ b^2\end{bmatrix}
\nonumber\\
&\quad
-96E_1\begin{bmatrix}-1\\ b\end{bmatrix}
E_2\begin{bmatrix}-1\\ -1\end{bmatrix}
-48E_1\begin{bmatrix}-1\\ -b^2\end{bmatrix}
E_2\begin{bmatrix}-1\\ -1\end{bmatrix}
-48E_1\begin{bmatrix}1\\ -b\end{bmatrix}
E_2\begin{bmatrix}-1\\ -1\end{bmatrix}
\nonumber\\
&\quad
-48E_1\begin{bmatrix}1\\ b^2\end{bmatrix}
E_2\begin{bmatrix}-1\\ -1\end{bmatrix}
-48E_1\begin{bmatrix}-1\\ -b^2\end{bmatrix}
E_2\begin{bmatrix}-1\\ b\end{bmatrix}
+48E_1\begin{bmatrix}1\\ -b\end{bmatrix}
E_2\begin{bmatrix}-1\\ b\end{bmatrix}
\nonumber\\
&\quad
+48E_1\begin{bmatrix}-1\\ b\end{bmatrix}
E_2\begin{bmatrix}-1\\ -b^2\end{bmatrix}
+48E_1\begin{bmatrix}-1\\ b\end{bmatrix}
E_2\begin{bmatrix}1\\ -b\end{bmatrix}
+48E_1\begin{bmatrix}-1\\ -b^2\end{bmatrix}
E_2\begin{bmatrix}1\\ -b\end{bmatrix}
\nonumber\\
&\quad
+48E_1\begin{bmatrix}1\\ -b\end{bmatrix}
E_2\begin{bmatrix}1\\ -b\end{bmatrix}
+48E_1\begin{bmatrix}1\\ b^2\end{bmatrix}
E_2\begin{bmatrix}1\\ -b\end{bmatrix}
+24E_1\begin{bmatrix}-1\\ -b^2\end{bmatrix}
E_2\begin{bmatrix}1\\ b^2\end{bmatrix}
\nonumber\\
&\quad
-192E_3\begin{bmatrix}-1\\ -b^2\end{bmatrix}
+384E_3\begin{bmatrix}1\\ -b\end{bmatrix} \ ,
\end{align}

\begin{align}
\mathcal{E}'_3 = & \ 224E_1\begin{bmatrix}-1\\ -b\end{bmatrix}
-24E_1\begin{bmatrix}-1\\ b\end{bmatrix}
+6E_1\begin{bmatrix}1\\ -b\end{bmatrix}
+6E_1\begin{bmatrix}1\\ b\end{bmatrix}
-133E_1\begin{bmatrix}1\\ -b^2\end{bmatrix}
-18E_1\begin{bmatrix}1\\ b^2\end{bmatrix}
\nonumber\\
&\quad
+48E_1\begin{bmatrix}-1\\ -b\end{bmatrix}
E_2\begin{bmatrix}-1\\ -b\end{bmatrix}
+48E_1\begin{bmatrix}-1\\ b\end{bmatrix}
E_2\begin{bmatrix}-1\\ -b\end{bmatrix}
+48E_1\begin{bmatrix}1\\ -b^2\end{bmatrix}
E_2\begin{bmatrix}-1\\ -b\end{bmatrix}
\nonumber\\
&\quad
+48E_1\begin{bmatrix}1\\ b^2\end{bmatrix}
E_2\begin{bmatrix}-1\\ -b\end{bmatrix}
+48E_1\begin{bmatrix}-1\\ -b\end{bmatrix}
E_2\begin{bmatrix}-1\\ b\end{bmatrix}
-48E_1\begin{bmatrix}1\\ -b^2\end{bmatrix}
E_2\begin{bmatrix}-1\\ b\end{bmatrix}
\nonumber\\
&\quad
-48E_1\begin{bmatrix}-1\\ -b\end{bmatrix}
E_2\begin{bmatrix}1\\ -1\end{bmatrix}
-96E_1\begin{bmatrix}-1\\ b\end{bmatrix}
E_2\begin{bmatrix}1\\ -1\end{bmatrix}
-48E_1\begin{bmatrix}1\\ -b^2\end{bmatrix}
E_2\begin{bmatrix}1\\ -1\end{bmatrix}
\nonumber\\
&\quad
-48E_1\begin{bmatrix}1\\ b^2\end{bmatrix}
E_2\begin{bmatrix}1\\ -1\end{bmatrix}
+48E_1\begin{bmatrix}-1\\ b\end{bmatrix}
E_2\begin{bmatrix}1\\ -b^2\end{bmatrix}
+24E_1\begin{bmatrix}1\\ -b^2\end{bmatrix}
E_2\begin{bmatrix}1\\ b^2\end{bmatrix}
\nonumber\\
&\quad
+384E_3\begin{bmatrix}-1\\ -b\end{bmatrix}
-192E_3\begin{bmatrix}1\\ -b^2\end{bmatrix}.
\end{align}
\begin{align}
\mathcal{E}'_4 = & \ 6E_1\begin{bmatrix}-1\\ -b\end{bmatrix}
+569E_1\begin{bmatrix}-1\\ b\end{bmatrix}
-96E_1\begin{bmatrix}-1\\ b\end{bmatrix}^3
-E_1\begin{bmatrix}-1\\ b^3\end{bmatrix}
-6E_1\begin{bmatrix}1\\ -b\end{bmatrix}
-6E_1\begin{bmatrix}1\\ b\end{bmatrix}
\nonumber\\
&\quad
+17E_1\begin{bmatrix}1\\ b^2\end{bmatrix}
-48E_1\begin{bmatrix}-1\\ b\end{bmatrix}^2
E_1\begin{bmatrix}1\\ b^2\end{bmatrix}
-48E_1\begin{bmatrix}-1\\ b\end{bmatrix}
E_1\begin{bmatrix}-1\\ b^3\end{bmatrix}
E_1\begin{bmatrix}1\\ b^2\end{bmatrix}
\nonumber\\
&\quad
-48E_1\begin{bmatrix}-1\\ b\end{bmatrix}
E_1\begin{bmatrix}1\\ b^2\end{bmatrix}^2
-132E_1\begin{bmatrix}1\\ b^4\end{bmatrix}
+48E_1\begin{bmatrix}-1\\ b\end{bmatrix}
E_1\begin{bmatrix}1\\ b^2\end{bmatrix}
E_1\begin{bmatrix}1\\ b^4\end{bmatrix}
\nonumber\\
&\quad
-96E_1\begin{bmatrix}-1\\ b\end{bmatrix}
E_2\begin{bmatrix}-1\\ b\end{bmatrix}
-48E_1\begin{bmatrix}-1\\ b^3\end{bmatrix}
E_2\begin{bmatrix}-1\\ b\end{bmatrix}
-96E_1\begin{bmatrix}1\\ b^2\end{bmatrix}
E_2\begin{bmatrix}-1\\ b\end{bmatrix}
\nonumber\\
&\quad
+48E_1\begin{bmatrix}1\\ b^4\end{bmatrix}
E_2\begin{bmatrix}-1\\ b\end{bmatrix}
+168E_1\begin{bmatrix}-1\\ b\end{bmatrix}
E_2\begin{bmatrix}1\\ b^2\end{bmatrix}
+72E_1\begin{bmatrix}-1\\ b^3\end{bmatrix}
E_2\begin{bmatrix}1\\ b^2\end{bmatrix}
\nonumber\\
&\quad
+120E_1\begin{bmatrix}1\\ b^2\end{bmatrix}
E_2\begin{bmatrix}1\\ b^2\end{bmatrix}
-96E_1\begin{bmatrix}1\\ b^4\end{bmatrix}
E_2\begin{bmatrix}1\\ b^2\end{bmatrix}
-192E_3\begin{bmatrix}-1\\ b\end{bmatrix}
+96E_3\begin{bmatrix}1\\ b^2\end{bmatrix} \ ,
\end{align}
and finally
\begin{align}
\mathcal{E}_5 = & \ 115E_1\begin{bmatrix}-1\\ b\end{bmatrix}
-6E_1\begin{bmatrix}-1\\ -b^2\end{bmatrix}
-6E_1\begin{bmatrix}-1\\ b^2\end{bmatrix}
\nonumber\\
&\quad
-25E_1\begin{bmatrix}-1\\ b^3\end{bmatrix}
+48E_1\begin{bmatrix}-1\\ b\end{bmatrix}^2
E_1\begin{bmatrix}-1\\ b^3\end{bmatrix}
+6E_1\begin{bmatrix}1\\ -b^2\end{bmatrix}
\nonumber\\
&\quad
-193E_1\begin{bmatrix}1\\ b^2\end{bmatrix}
+24E_1\begin{bmatrix}-1\\ b\end{bmatrix}
E_1\begin{bmatrix}1\\ b^2\end{bmatrix}^2
+49E_1\begin{bmatrix}1\\ b^4\end{bmatrix}
\nonumber\\
&\quad
+24E_1\begin{bmatrix}-1\\ b\end{bmatrix}
E_1\begin{bmatrix}1\\ b^2\end{bmatrix}
E_1\begin{bmatrix}1\\ b^4\end{bmatrix}
+96E_1\begin{bmatrix}-1\\ b\end{bmatrix}
E_2\begin{bmatrix}-1\\ b\end{bmatrix}
+48E_1\begin{bmatrix}-1\\ b^3\end{bmatrix}
E_2\begin{bmatrix}-1\\ b\end{bmatrix}
\nonumber\\
&\quad
+24E_1\begin{bmatrix}1\\ b^2\end{bmatrix}
E_2\begin{bmatrix}-1\\ b\end{bmatrix}
+72E_1\begin{bmatrix}1\\ b^4\end{bmatrix}
E_2\begin{bmatrix}-1\\ b\end{bmatrix}
-48E_1\begin{bmatrix}-1\\ b\end{bmatrix}
E_2\begin{bmatrix}-1\\ b^3\end{bmatrix}
\nonumber\\
&\quad
-24E_1\begin{bmatrix}-1\\ b^3\end{bmatrix}
E_2\begin{bmatrix}1\\ b^2\end{bmatrix}
-72E_1\begin{bmatrix}1\\ b^2\end{bmatrix}
E_2\begin{bmatrix}1\\ b^2\end{bmatrix}
-72E_1\begin{bmatrix}1\\ b^4\end{bmatrix}
E_2\begin{bmatrix}1\\ b^2\end{bmatrix}
\nonumber\\
&\quad
-24E_1\begin{bmatrix}-1\\ b\end{bmatrix}
E_2\begin{bmatrix}1\\ b^4\end{bmatrix}
+192E_3\begin{bmatrix}-1\\ b\end{bmatrix}
+96E_3\begin{bmatrix}-1\\ b^3\end{bmatrix}
\nonumber\\
&\quad
-336E_3\begin{bmatrix}1\\ b^2\end{bmatrix}
+48E_3\begin{bmatrix}1\\ b^4\end{bmatrix}.
\end{align}
Note also that the nine contributions can be further merged into seven, since there are identities among the $\vartheta$-ratio factors,
\begin{equation}
  \prod_{i = 2}^4\vartheta_i(0) = \frac{2 \prod_{i = 1}^4 \vartheta_i(2 \mathfrak{b})}{\vartheta_1(4 \mathfrak{b})} \ ,
\end{equation}
and
\begin{equation}
0 = \sum_{i=1}^{4}
\varepsilon_i 
\frac{
\vartheta_i(\mathfrak{b})\vartheta_4(\mathfrak{b})^2
\vartheta_4(3\mathfrak{b})
}{
\vartheta_1(2\mathfrak{b})^2
\vartheta_i(3\mathfrak{b})
\vartheta_1(4\mathfrak{b})
}
+
\frac{
2\vartheta_4(\mathfrak{b})\vartheta_4(3\mathfrak{b})
\vartheta_4(5\mathfrak{b})
}{
\vartheta_1(2\mathfrak{b})
\vartheta_1(4\mathfrak{b})
\vartheta_1(6\mathfrak{b})
},
\quad
\varepsilon_i=
\begin{cases}
1, & i=1,4,\\
-1, & i=2,3.
\end{cases}
\end{equation}

The flavored $USp(6)$ index should be equal to that of the $SO(7)$ theory by S-duality. Although we can expand both expressions to high order as a $q$-series and verify the zero difference, unfortunately we are unable to prove the equality analytically. Nonetheless, we observe some relations among the theta-ratios appearing in both flavored index,
\begin{align}
0 = & \ - \frac{\vartheta_3(0) \vartheta_3(2\mathfrak{b})}{\vartheta_1(4\mathfrak{b}) \vartheta_2(\mathfrak{b}) \vartheta_2(3\mathfrak{b})}+ \frac{\vartheta_2(\mathfrak{b}) \vartheta_3(0) \vartheta_4(\mathfrak{b})^2}{\vartheta_1(2\mathfrak{b})^3 \vartheta_2(3\mathfrak{b}) \vartheta_3(2\mathfrak{b})} - \frac{\vartheta_1(\mathfrak{b}) \vartheta_4(0) \vartheta_4(\mathfrak{b})^2}{\vartheta_1(2\mathfrak{b})^3 \vartheta_1(3\mathfrak{b}) \vartheta_4(2\mathfrak{b})} \nonumber\\
& - \frac{\vartheta_4(0) \vartheta_4(2\mathfrak{b})}{\vartheta_1(\mathfrak{b}) \vartheta_1(3\mathfrak{b}) \vartheta_1(4\mathfrak{b})}  \ ,\\
0 = &+ \frac{\vartheta_1(\mathfrak{b}) \vartheta_4(\mathfrak{b})^2}{\vartheta_1(2\mathfrak{b})^3 \vartheta_1(3\mathfrak{b}) \vartheta_4(2\mathfrak{b})}
- \frac{\vartheta_2(0) \vartheta_3(0) \vartheta_4(\mathfrak{b})^2}{\vartheta_1(2\mathfrak{b})^3 \vartheta_2(2\mathfrak{b}) \vartheta_3(2\mathfrak{b}) \vartheta_4(2\mathfrak{b})} + \frac{\vartheta_4(2\mathfrak{b})}{\vartheta_1(\mathfrak{b}) \vartheta_1(3\mathfrak{b}) \vartheta_1(4\mathfrak{b})} \ ,\\
0 = & \ \frac{\vartheta_4(\mathfrak{b})^2}{\vartheta_1(2\mathfrak{b})^2 \vartheta_1(4\mathfrak{b})} - \frac{\vartheta_1(\mathfrak{b}) \vartheta_4(0) \vartheta_4(\mathfrak{b})^2}{2 \vartheta_1(2\mathfrak{b})^3 \vartheta_1(3\mathfrak{b}) \vartheta_4(2\mathfrak{b})} 
- \frac{\vartheta_4(0) \vartheta_4(2\mathfrak{b})}{2 \vartheta_1(\mathfrak{b}) \vartheta_1(3\mathfrak{b}) \vartheta_1(4\mathfrak{b})} \\ 
0 = & \ \frac{\vartheta_1(\mathfrak{b}) \vartheta_4(0) \vartheta_4(\mathfrak{b})^2}{2 \vartheta_1(2\mathfrak{b})^3 \vartheta_4(2\mathfrak{b})} 
- \frac{\vartheta_4(0) \vartheta_4(2\mathfrak{b})}{2 \vartheta_1(\mathfrak{b}) \vartheta_1(4\mathfrak{b})} 
+ \frac{\vartheta_1(\mathfrak{b}) \vartheta_4(\mathfrak{b}) \vartheta_4(3\mathfrak{b})}{\vartheta_1(2\mathfrak{b})^2 \vartheta_1(4\mathfrak{b})} \ ,\\
0 = & \ - \frac{\vartheta_3(0) \vartheta_3(2\mathfrak{b})}{\vartheta_1(4\mathfrak{b}) \vartheta_2(\mathfrak{b}) \vartheta_2(3\mathfrak{b})} 
- \frac{\vartheta_1(\mathfrak{b}) \vartheta_4(0) \vartheta_4(\mathfrak{b})^2}{2 \vartheta_1(2\mathfrak{b})^3 \vartheta_1(3\mathfrak{b}) \vartheta_4(2\mathfrak{b})} 
- \frac{\vartheta_4(0) \vartheta_4(2\mathfrak{b})}{2 \vartheta_1(\mathfrak{b}) \vartheta_1(3\mathfrak{b}) \vartheta_1(4\mathfrak{b})} \nonumber\\
&\ + \frac{\vartheta_2(\mathfrak{b}) \vartheta_4(\mathfrak{b}) \vartheta_4(3\mathfrak{b})}{\vartheta_1(2\mathfrak{b})^2 \vartheta_1(4\mathfrak{b}) \vartheta_2(3\mathfrak{b})} \ ,\\
0 = &\ \frac{\vartheta_3(0) \vartheta_3(2\mathfrak{b})}{\vartheta_1(4\mathfrak{b}) \vartheta_2(\mathfrak{b}) \vartheta_2(3\mathfrak{b})} 
+ \frac{\vartheta_2(0) \vartheta_2(2\mathfrak{b})}{\vartheta_1(4\mathfrak{b}) \vartheta_3(\mathfrak{b}) \vartheta_3(3\mathfrak{b})} 
+ \frac{\vartheta_1(\mathfrak{b}) \vartheta_4(0) \vartheta_4(\mathfrak{b})^2}{\vartheta_1(2\mathfrak{b})^3 \vartheta_1(3\mathfrak{b}) \vartheta_4(2\mathfrak{b})}\nonumber \\
& \ - \frac{2 \vartheta_4(3\mathfrak{b}) \vartheta_4(5\mathfrak{b})}{\vartheta_1(2\mathfrak{b}) \vartheta_1(4\mathfrak{b}) \vartheta_1(6\mathfrak{b})} \ ,\\
0 = &\frac{\vartheta_3(0) \vartheta_3(2\mathfrak{b})}{\vartheta_1(4\mathfrak{b}) \vartheta_2(\mathfrak{b}) \vartheta_2(3\mathfrak{b})} 
+ \frac{\vartheta_2(0) \vartheta_3(\mathfrak{b}) \vartheta_4(\mathfrak{b})^2}{\vartheta_1(2\mathfrak{b})^3 \vartheta_2(2\mathfrak{b}) \vartheta_3(3\mathfrak{b})} 
- \frac{\vartheta_4(0) \vartheta_4(2\mathfrak{b})}{\vartheta_1(\mathfrak{b}) \vartheta_1(3\mathfrak{b}) \vartheta_1(4\mathfrak{b})} \nonumber\\
& \ - \frac{2 \vartheta_4(3\mathfrak{b}) \vartheta_4(5\mathfrak{b})}{\vartheta_1(2\mathfrak{b}) \vartheta_1(4\mathfrak{b}) \vartheta_1(6\mathfrak{b})} \ ,\\
0 = & \ \frac{\vartheta_3(0) \vartheta_3(2\mathfrak{b})}{\vartheta_1(4\mathfrak{b}) \vartheta_2(\mathfrak{b}) \vartheta_2(3\mathfrak{b})} 
+ \frac{\vartheta_1(\mathfrak{b}) \vartheta_4(0) \vartheta_4(\mathfrak{b})^2}{2 \vartheta_1(2\mathfrak{b})^3 \vartheta_1(3\mathfrak{b}) \vartheta_4(2\mathfrak{b})} 
- \frac{\vartheta_4(0) \vartheta_4(2\mathfrak{b})}{2 \vartheta_1(\mathfrak{b}) \vartheta_1(3\mathfrak{b}) \vartheta_1(4\mathfrak{b})} \nonumber\\
&\ + \frac{\vartheta_3(\mathfrak{b}) \vartheta_4(\mathfrak{b}) \vartheta_4(3\mathfrak{b})}{\vartheta_1(2\mathfrak{b})^2 \vartheta_1(4\mathfrak{b}) \vartheta_3(3\mathfrak{b})}
- \frac{2 \vartheta_4(3\mathfrak{b}) \vartheta_4(5\mathfrak{b})}{\vartheta_1(2\mathfrak{b}) \vartheta_1(4\mathfrak{b}) \vartheta_1(6\mathfrak{b})} \ .
\end{align}

Although the flavored index is somewhat tedious to write down, the unflavored $USp(6)$ index actually takes a much simpler form,
\begin{align}\label{eq:N4USp6unflavored}
  \mathcal{I}_{USp(6)}^\text{unflavored} = &\ \frac{1}{\vartheta_2(0)^3\vartheta_3(0)^3} \bigg[ \frac{2}{3}E_2(\tau)^3
  +\frac{1}{6}E_2(\tau)^2
  \left(
  \vartheta_2(0)^4+\vartheta_3(0)^4
  \right) \nonumber\\
  &\quad
  +\frac{1}{288}E_2(\tau)
  \left(
  5\vartheta_2(0)^4
  +4\vartheta_2(0)^8
  +32\vartheta_3(0)^4
  -\vartheta_2(0)^4\vartheta_3(0)^4
  +4\vartheta_3(0)^8
  \right) \nonumber\\
  &\quad
  +\frac{1}{10368}
  \left(
  15\vartheta_2(0)^8
  +4\vartheta_2(0)^{12}
  -33\vartheta_2(0)^4\vartheta_3(0)^4
  -15\vartheta_2(0)^8\vartheta_3(0)^4
  \right. \nonumber\\
  &\qquad\qquad\left.
  +96\vartheta_3(0)^8
  -15\vartheta_2(0)^4\vartheta_3(0)^8
  +4\vartheta_3(0)^{12}
  \right)
  \bigg] \ .
\end{align}
Fortunately, this expression is identical to that of the $SO(7)$ theory given in (\ref{eq:N4SO7unflavored}), as predicted from S-duality.

\subsection{\texorpdfstring{$\mathcal{N} = 2$ $[1] - SU(2) \times SU(3)- [4]$ quiver theory}{}}

Consider the $\mathcal{N} = 2$ superconformal quiver $[1]-SU(2)-SU(3)-[4]$, where $[1]$ and $[4]$ denote one and four fundamental hypermultiplets under the $SU(2)$ and $SU(3)$ gauge groups. The Schur index is given by
\begin{align}
  \mathcal{I} = & \ \oint \frac{da_4}{2\pi i a_4} \prod_{A = 1}^2 \frac{da_A}{2\pi i a_A} \mathcal{Z}_\text{VM} \mathcal{Z}_\text{HM} \ ,
\end{align}
where the vector multiplet contribution is
$$
\mathcal{Z}_\text{VM} = \eta(\tau)^{-2} \frac{1}{2}\vartheta_1(\pm 2\mathfrak{a}_4) \frac{1}{3!} \prod_{\substack{A, B = 1 \\ A \ne B}}^3  \vartheta_1(\mathfrak{a}_A - \mathfrak{a}_B)\bigg|_{\mathfrak{a}_3 = -\mathfrak{a}_1 - \mathfrak{a}_2} \ ,
$$
and the hypermultiplet contribution is
\begin{align}
  \mathcal{Z}_\text{HM} = \frac{\eta(\tau)^2}{\vartheta_4(\pm \mathfrak{a}_4 + \mathfrak{b})} \prod_{A = 1}^3 \prod_\pm \frac{\eta(\tau)}{\vartheta_4(\pm \mathfrak{a}_4 - \mathfrak{a}_A + \mathfrak{c})} \prod_{A = 1}^3\prod_{i = 1}^4 \frac{\eta(\tau)}{\vartheta_4(\pm \mathfrak{a}_A + \mathfrak{d}_i)} \ .
\end{align}
The $\mathfrak{a}_4$ denotes the gauge fugacity of $SU(2)$, while $\mathfrak{a}_1, \mathfrak{a}_2$ are the gauge fugacities of $SU(3)$. For simplicity we turn off the fugacities $\mathfrak{b}, \mathfrak{c}$, keeping only the $SU(4)$ flavor fugacities $\mathfrak{d}_i$.

The index can be evaluated by integrating subsequently in the order $\mathfrak{a}_4, \mathfrak{a}_1, \mathfrak{a}_2$. The first integration is straightforward, giving the integrand
\begin{align}
  = & \ \frac{i\eta(\tau)^{17}\vartheta_1(2 \mathfrak{a}_1)\vartheta_1(\mathfrak{a}_1 - \mathfrak{a}_2)\vartheta_1(\mathfrak{a}_1 - \mathfrak{a}_3)\vartheta_1(\mathfrak{a}_2 - \mathfrak{a}_3)^2}{2\vartheta_1(\mathfrak{a}_1)\prod_{j = 1}^4 \vartheta_4(\mathfrak{a}_A + \mathfrak{d}_j)\prod_{A = 1}^{3}\vartheta_1(\mathfrak{a}_A)}
  E_1 \begin{bmatrix}
    -1 \\ a_1
  \end{bmatrix} \nonumber\\
  & \ - \frac{i \eta(\tau)^{17} \vartheta_1(2 \mathfrak{a}_2) \vartheta_1( \mathfrak{a}_1 - \mathfrak{a}_2) \vartheta_1(\mathfrak{a}_1 - \mathfrak{a}_3)^2 \vartheta_1(\mathfrak{a}_2 - \mathfrak{a}_3)}{2 \vartheta_1(\mathfrak{a}_2) \prod_{j = 1}^4 \vartheta_4(\mathfrak{a}_A + \mathfrak{d}_j) \prod_{A = 1}^3 \vartheta_1(\mathfrak{a}_A)} E_1 \begin{bmatrix}
    -1 \\ a_2
  \end{bmatrix} \nonumber\\
  & \ - \frac{i \eta(\tau)^{17} \vartheta_1(2 \mathfrak{a}_3) \vartheta_1(\mathfrak{a}_1 - \mathfrak{a}_2)^2 \vartheta_1(\mathfrak{a}_1 - \mathfrak{a}_3) \vartheta_1(\mathfrak{a}_2 - \mathfrak{a}_3)}{2 \vartheta_1(\mathfrak{a}_3) \prod_{j = 1}^4 \vartheta_4(\mathfrak{a}_A + \mathfrak{d}_j) \prod_{A = 1}^3 \vartheta_1(\mathfrak{a}_A)} E_1 \begin{bmatrix}
    -1 \\ a_1 a_2
  \end{bmatrix} \ ,
\end{align}
where $a_3 = 1/(a_1 a_2)$, $\mathfrak{a}_3 = - \mathfrak{a}_1 - \mathfrak{a}_2$. The two Eisenstein series that depend on $a_1$ gives rise to three solutions to the $n(a_2q) - n(a_2) = E_1 \big[\substack{-1 \\ a_1}\big]$ and $E_1 \big[\substack{-1 \\ a_1 a_2}\big]$,
\begin{align}
  1 - E_1 \begin{bmatrix}
    -1 \\ a_1
  \end{bmatrix} + E_2 \begin{bmatrix}
    1 \\ a_1 q^{\frac{1}{2}}
  \end{bmatrix}, \qquad
  1 - E_1 \begin{bmatrix}
    -1 \\ a_1a_2
  \end{bmatrix} + E_2 \begin{bmatrix}
    1 \\  a_2 q^{\frac{1}{2}}
  \end{bmatrix} \ .
\end{align}

The subsequent $a_1$ and $a_2$ integrations are straightforward but with much longer result. The final result can be written as
\begin{align}
  \mathcal{I} = \sum_{i < j}A_{ij} + \sum_{i = 1}^{4} B_i + \sum_{a\ne b} C_{ab} + \sum_{i = 1}^{4}D_i \ .  
\end{align}
Here
\begin{align}
  A_{ij} = & \ \frac{\vartheta_1(2 \mathfrak{d}_i + 2 \mathfrak{d}_j)}{
    \vartheta_1(\mathfrak{d}_i + \mathfrak{d}_j)^2
    \vartheta_4(\mathfrak{d}_i) \vartheta_4(\mathfrak{d}_j)
    \prod_{k\ne i,j}
      \vartheta_1(\mathfrak{d}_i - \mathfrak{d}_k)
      \vartheta_1(\mathfrak{d}_j - \mathfrak{d}_k)
    \prod_{k\ne i,j}
      \vartheta_1(\mathfrak{d}_i + \mathfrak{d}_j + \mathfrak{d}_k)
  }\nonumber\\
  & \ \times \bigg(
    \frac{5i}{12}
  -\frac{5i}{4}E_1 \left[\begin{matrix}-1\\ d_i\end{matrix}\right]
  -\frac{5i}{4}E_1 \left[\begin{matrix}-1\\ d_j\end{matrix}\right]
  -\frac{13i}{24}E_1 \left[\begin{matrix}-1\\ d_i d_j\end{matrix}\right]
  \nonumber\\
    &\quad
  +\frac{i}{2}E_1 \left[\begin{matrix}-1\\ d_i\end{matrix}\right]
    E_1 \left[\begin{matrix}-1\\ d_i d_j\end{matrix}\right]
  +\frac{i}{2}E_1 \left[\begin{matrix}-1\\ d_j\end{matrix}\right]
    E_1 \left[\begin{matrix}-1\\ d_i d_j\end{matrix}\right]
  \nonumber\\
    &\quad
  -\frac{i}{24}E_1 \left[\begin{matrix}+1\\ d_i\end{matrix}\right]
  -\frac{i}{2}E_1 \left[\begin{matrix}-1\\ d_j\end{matrix}\right]
    E_1 \left[\begin{matrix}+1\\ d_i\end{matrix}\right]
  -\frac{i}{24}E_1 \left[\begin{matrix}+1\\ d_j\end{matrix}\right]
  -\frac{i}{2}E_1 \left[\begin{matrix}-1\\ d_i\end{matrix}\right]
    E_1 \left[\begin{matrix}+1\\ d_j\end{matrix}\right]
  \nonumber\\
    &\quad
  +\frac{i}{2}E_2 \left[\begin{matrix}-1\\ d_i\end{matrix}\right]
  +\frac{i}{2}E_1 \left[\begin{matrix}-1\\ d_i d_j\end{matrix}\right]
    E_2 \left[\begin{matrix}-1\\ d_i\end{matrix}\right]
  +\frac{i}{2}E_2 \left[\begin{matrix}-1\\ d_j\end{matrix}\right]
  +\frac{i}{2}E_1 \left[\begin{matrix}-1\\ d_i d_j\end{matrix}\right]
    E_2 \left[\begin{matrix}-1\\ d_j\end{matrix}\right]
  \nonumber\\
    &\quad
  -\frac{i}{2}E_1 \left[\begin{matrix}-1\\ d_i\end{matrix}\right]
    E_2 \left[\begin{matrix}-1\\ d_i d_j\end{matrix}\right]
  -\frac{i}{2}E_1 \left[\begin{matrix}-1\\ d_j\end{matrix}\right]
    E_2 \left[\begin{matrix}-1\\ d_i d_j\end{matrix}\right]
  \nonumber\\
    &\quad
  +\frac{i}{2}E_3 \left[\begin{matrix}-1\\ d_i\end{matrix}\right]
  +\frac{i}{2}E_3 \left[\begin{matrix}-1\\ d_j\end{matrix}\right]
  +2i E_3 \left[\begin{matrix}-1\\ d_i d_j\end{matrix}\right]
  \bigg) \ ,
\end{align}
\begin{align}
  B_i
  = & \ 
  -
  \frac{
  \vartheta_1(2\mathfrak{d}_i) 
  }{
  \prod_{j\neq i}
  \vartheta_1(\mathfrak{d}_i\pm\mathfrak{d}_j)
  \cdot
  \prod_{k=1}^{4}\vartheta_4(\mathfrak{d}_k) }\bigg(
  \frac{5i}{8}
+\frac{i}{4}\sum_{j=1}^{4}E_1 \left[\begin{matrix}-1\\ d_j\end{matrix}\right]
+i E_1 \left[\begin{matrix}-1\\ d_i\end{matrix}\right]
 \nonumber\\
  &\quad
+\frac{i}{2}\sum_{j=1}^{4}
  E_1 \left[\begin{matrix}-1\\ d_j\end{matrix}\right]
  E_1 \left[\begin{matrix}+1\\ d_i\end{matrix}\right]
+\frac{5i}{2}
  E_1 \left[\begin{matrix}-1\\ d_i\end{matrix}\right]
  E_1 \left[\begin{matrix}+1\\ d_i\end{matrix}\right]
  -\frac{9i}{4}E_1 \left[\begin{matrix}+1\\ d_i^2\end{matrix}\right]
 \nonumber\\
  &\quad
-i E_1 \left[\begin{matrix}+1\\ d_i\end{matrix}\right]
  E_1 \left[\begin{matrix}+1\\ d_i^2\end{matrix}\right]
  -\frac{i}{4}\sum_{j\neq i}E_1 \left[\begin{matrix}+1\\ d_i d_j\end{matrix}\right]
+\frac{i}{2}\sum_{j\neq i}
  E_1 \left[\begin{matrix}+1\\ d_i\end{matrix}\right]
  E_1 \left[\begin{matrix}+1\\ d_i d_j\end{matrix}\right]
  -\frac{i}{2}E_2 \left[\begin{matrix}-1\\ 1\end{matrix}\right]
 \nonumber\\
  &\quad
-5i E_1 \left[\begin{matrix}-1\\ d_i\end{matrix}\right]
  E_2 \left[\begin{matrix}-1\\ 1\end{matrix}\right]
+i E_2 \left[\begin{matrix}-1\\ d_i\end{matrix}\right]
 \\
  &\quad
+\frac{i}{2}E_2 \left[\begin{matrix}+1\\ d_i\end{matrix}\right]
+\frac{13i}{2}
  E_1 \left[\begin{matrix}-1\\ d_i\end{matrix}\right]
  E_2 \left[\begin{matrix}+1\\ d_i\end{matrix}\right]
+i\sum_{j\neq i}
  E_1 \left[\begin{matrix}-1\\ d_j\end{matrix}\right]
  E_2 \left[\begin{matrix}+1\\ d_i\end{matrix}\right]
 \nonumber\\
  &\quad
+i E_1 \left[\begin{matrix}+1\\ d_i^2\end{matrix}\right]
  E_2 \left[\begin{matrix}+1\\ d_i\end{matrix}\right]
-i\sum_{j\neq i}
  E_1 \left[\begin{matrix}+1\\ d_i d_j\end{matrix}\right]
  E_2 \left[\begin{matrix}+1\\ d_i\end{matrix}\right]
 \nonumber\\
  &\quad
+\frac{i}{2}E_2 \left[\begin{matrix}+1\\ d_i^2\end{matrix}\right]
-\frac{i}{2}
  E_1 \left[\begin{matrix}+1\\ d_i\end{matrix}\right]
  E_2 \left[\begin{matrix}+1\\ d_i^2\end{matrix}\right]
  -\frac{i}{2}\sum_{j\neq i}E_2 \left[\begin{matrix}+1\\ d_i/d_j\end{matrix}\right]
 \nonumber\\
  &\quad
-\frac{i}{2}\sum_{j\neq i}
  E_1 \left[\begin{matrix}+1\\ d_i\end{matrix}\right]
  E_2 \left[\begin{matrix}+1\\ d_i/d_j\end{matrix}\right]
  +\frac{i}{2}\sum_{j\neq i}E_2 \left[\begin{matrix}+1\\ d_i d_j\end{matrix}\right]
-\frac{i}{2}\sum_{j\neq i}
  E_1 \left[\begin{matrix}+1\\ d_i\end{matrix}\right]
  E_2 \left[\begin{matrix}+1\\ d_i d_j\end{matrix}\right]
 \nonumber\\
  &\quad
-i E_3 \left[\begin{matrix}-1\\ d_i\end{matrix}\right]
+8i E_3 \left[\begin{matrix}+1\\ d_i\end{matrix}\right]
-\frac{i}{2}E_3 \left[\begin{matrix}+1\\ d_i^2\end{matrix}\right]
-\frac{i}{2}\sum_{j\neq i}E_3 \left[\begin{matrix}+1\\ d_i/d_j\end{matrix}\right]
-\frac{i}{2}\sum_{j\neq i}E_3 \left[\begin{matrix}+1\\ d_i d_j\end{matrix}\right]
  \bigg) \ , \nonumber
\end{align}
\begin{align}
  C_{ab}
  = & \ 
  \frac{
  \vartheta_1(2\mathfrak{d}_a)\,
  \vartheta_4(\mathfrak{d}_a+2\mathfrak{d}_b)\,
  }{
  \vartheta_4(\mathfrak{d}_a)^2
  \vartheta_4(\mathfrak{d}_b)
  \vartheta_1(\mathfrak{d}_a\pm\mathfrak{d}_b)
  \prod_{k\notin\{a,b\}}
  \vartheta_1(\mathfrak{d}_a-\mathfrak{d}_k)
  \vartheta_1(\mathfrak{d}_b-\mathfrak{d}_k)
  \vartheta_4(\mathfrak{d}_a+\mathfrak{d}_b+\mathfrak{d}_k)
  } \nonumber \\
  & \ \times \bigg(
  \frac{5i}{24}
  -\frac{3i}{2}E_1\!\left[\begin{matrix}-1\\ d_b\end{matrix}\right]
  +\frac{i}{24}E_1\!\left[\begin{matrix}-1\\ d_a d_b\end{matrix}\right]
  +\frac{i}{12}E_1\!\left[\begin{matrix}+1\\ d_a\end{matrix}\right]
  -\frac{i}{2}E_1\!\left[\begin{matrix}-1\\ d_b\end{matrix}\right]
  \nonumber\\
  &\quad
    E_1\!\left[\begin{matrix}+1\\ d_a\end{matrix}\right]
  -\frac{i}{4}E_1\!\left[\begin{matrix}+1\\ d_a d_b\end{matrix}\right]
  +\frac{i}{2}E_1\!\left[\begin{matrix}+1\\ d_b\end{matrix}\right]
    E_1\!\left[\begin{matrix}+1\\ d_a d_b\end{matrix}\right]
    -\frac{i}{2}E_2\!\left[\begin{matrix}-1\\ d_a d_b\end{matrix}\right]
  \\
  &\quad
  -\frac{3i}{2}E_1\!\left[\begin{matrix}-1\\ d_b\end{matrix}\right]
    E_2\!\left[\begin{matrix}+1\\ d_a\end{matrix}\right]
  -i\,E_1\!\left[\begin{matrix}+1\\ d_a d_b\end{matrix}\right]
    E_2\!\left[\begin{matrix}+1\\ d_a\end{matrix}\right]
    +\frac{i}{2}E_2\!\left[\begin{matrix}+1\\ d_b\end{matrix}\right]
  \nonumber\\
  &\quad
  +\frac{i}{2}E_2\!\left[\begin{matrix}+1\\ d_a d_b\end{matrix}\right]
  -\frac{i}{2}E_1\!\left[\begin{matrix}+1\\ d_a\end{matrix}\right]
    E_2\!\left[\begin{matrix}+1\\ d_a d_b\end{matrix}\right]
    -i\,E_3\!\left[\begin{matrix}+1\\ d_a\end{matrix}\right]
  -\frac{i}{2}E_3\!\left[\begin{matrix}+1\\ d_a d_b\end{matrix}\right]
  \bigg) \ , \nonumber
\end{align}
and finally
\begin{align}
  D_i
  =
  \frac{
  \vartheta_1(2\mathfrak{d}_i)\,
  \eta(\tau)^3\,
  }{
  \vartheta_4(0)
  \prod_{k=1}^{4}\vartheta_1(\mathfrak{d}_k)
  \prod_{j\neq i}\vartheta_4(\mathfrak{d}_i\pm\mathfrak{d}_j)
  }
  \bigg(
  \frac{1}{24}
  +\frac12 E_1\!\left[\begin{matrix}-1\\ d_i\end{matrix}\right]
  -\frac12 E_1\!\left[\begin{matrix}+1\\ d_i\end{matrix}\right] \bigg) \ .
\end{align}

Although the flavored Schur index is somewhat complicated, it is straightforward to compute the unflavored version of the index, which is much simpler,
\begin{align}
  \mathcal{I} = \frac{2}{135\eta(\tau)^{15}\vartheta_4(0)^4} \bigg(& \ 
    -405 E_2(\tau) E_4(\tau)+1350 E_2(\tau)^2 E_4(\tau) -20250 E_4(\tau)^2\nonumber\\
& \ -4165 E_6(\tau)-1890 E_2(\tau) E_6(\tau)-4785 E_4(\tau) E_2\begin{bmatrix}
  -1 \\ 1
\end{bmatrix} \nonumber\\
& \ +270 E_2(\tau)^3 E_2\begin{bmatrix}
  -1 \\ 1
\end{bmatrix}-8370 E_2(\tau) E_4(\tau) E_2\begin{bmatrix}
  -1 \\ 1
\end{bmatrix} \nonumber\\
& \ -1080 E_2(\tau)^2 E_2\begin{bmatrix}
  -1 \\ 1
\end{bmatrix}^2+13500 E_4(\tau) E_2\begin{bmatrix}
  -1 \\ 1
\end{bmatrix}^2 \\
& \ +1276 E_2\begin{bmatrix}
  -1 \\ 1
\end{bmatrix}^3+2592 E_2(\tau) E_2\begin{bmatrix}
  -1 \\ 1
\end{bmatrix}^3-2160 E_2\begin{bmatrix}
  -1 \\ 1
\end{bmatrix}^4\bigg) \nonumber \ .
\end{align}

\subsection{\texorpdfstring{$[SO(4)]-USp(2)-SO(4)-[USp(2)]$}{[SO(4)]-USp(2)-SO(4)-[USp(2)] quiver theory}}

Consider the quiver gauge theory $[SO(4)]-USp(2)-SO(4)-[USp(2)]$ \cite{Tachikawa:2009rb,Lemos:2012ph} where the $USp(2)\times SO(4)$ denotes the gauge group and the $SO(4)$ and $USp(2)$ in the square brackets denote the flavor symmetry. The Schur index can be written as the following contour integral,
\begin{align}
\mathcal{Z}=
&\frac{\eta(\tau)^{12}}{8}
\frac{
\vartheta_1(2\mathfrak{a}_1)^2
\vartheta_1(\mathfrak{a}_2-\mathfrak{a}_3)^2
\vartheta_1(\mathfrak{a}_2+\mathfrak{a}_3)^2
}{
\prod_{i=1}^{2}
\vartheta_4(-\mathfrak{a}_1+\mathfrak{b}_i)
\vartheta_4(\mathfrak{a}_1+\mathfrak{b}_i)
}\nonumber
\\
&\times
\frac{1}{
\prod_{r=2}^{3}
\vartheta_4(-\mathfrak{a}_1+\mathfrak{a}_r)
\vartheta_4(\mathfrak{a}_1+\mathfrak{a}_r)
\vartheta_4(-\mathfrak{a}_r+\mathfrak{c})
\vartheta_4(\mathfrak{a}_r+\mathfrak{c})
} \ .
\end{align}
We denote the $USp(2) = SU(2)$ fundamental character as $\chi^{SU(2)}_{\text{fund}}(a_1) = a_1 + a_1^{-1}$, and the $SO(4)$ vector and adjoint characters as
\begin{align}
  \operatorname{ch}_\text{vec}^{SO(4)}(a_2, a_3) = & \ a_2 + a_2^{-1} + a_3 + a_3^{-1} \\
  \operatorname{ch}_\text{adj}^{SO(4)}(a_2, a_3) = & \ 2 + a_2 a_3 + (a_2 a_3)^{-1} + \frac{a_2}{a_3} + \frac{a_3}{a_2} \ .
\end{align}

Defining
\begin{equation}
  D_1=
\vartheta_1(\mathfrak{b}_1\pm\mathfrak{b}_2)
\prod_{r=2}^{3}\prod_{i=1}^{2}
\vartheta_1(\pm\mathfrak{a}_r+\mathfrak{b}_i)
\prod_{r=2}^{3}
\vartheta_4(\pm\mathfrak{a}_r+\mathfrak{c}) \ .
\end{equation}
The result of integrating $a_1$ is
\begin{align}
=
& \ \frac{i\eta(\tau)^9}{4}
\frac{
\vartheta_1(\mathfrak{a}_2-\mathfrak{a}_3)
\vartheta_1(\mathfrak{a}_2+\mathfrak{a}_3)
}{D_1}
\bigg[ \nonumber\\
& \ \qquad \vartheta_1(2\mathfrak{a}_2)
\prod_{i=1}^{2}
\vartheta_1(-\mathfrak{a}_3+\mathfrak{b}_i)
\vartheta_1(\mathfrak{a}_3+\mathfrak{b}_i)
\vartheta_1(\mathfrak{b}_1-\mathfrak{b}_2)
\vartheta_1(\mathfrak{b}_1+\mathfrak{b}_2)
E_1\begin{bmatrix}-1\\ a_2\end{bmatrix}
\nonumber\\
& \qquad -
\vartheta_1(2\mathfrak{a}_3)
\prod_{i=1}^{2}
\vartheta_1(-\mathfrak{a}_2+\mathfrak{b}_i)
\vartheta_1(\mathfrak{a}_2+\mathfrak{b}_i)
\vartheta_1(\mathfrak{b}_1-\mathfrak{b}_2)
\vartheta_1(\mathfrak{b}_1+\mathfrak{b}_2)
E_1\begin{bmatrix}-1\\ a_3\end{bmatrix}
\nonumber\\
&\qquad +
\vartheta_1(\mathfrak{a}_2-\mathfrak{a}_3)
\vartheta_1(\mathfrak{a}_2+\mathfrak{a}_3)
\bigg(
\vartheta_1(2\mathfrak{b}_1)
\prod_{r=2}^{3}
\vartheta_1(-\mathfrak{a}_r+\mathfrak{b}_2)
\vartheta_1(\mathfrak{a}_r+\mathfrak{b}_2)
E_1\begin{bmatrix}-1\\ b_1\end{bmatrix}
\nonumber\\
&\hspace{7em}-
\vartheta_1(2\mathfrak{b}_2)
\prod_{r=2}^{3}
\vartheta_1(-\mathfrak{a}_r+\mathfrak{b}_1)
\vartheta_1(\mathfrak{a}_r+\mathfrak{b}_1)
E_1\begin{bmatrix}-1\\ b_2\end{bmatrix}
\bigg)
\bigg].
\end{align}

\begin{align}
  \mathcal{I}
  =& \ 
  \eta(\tau)^6
  \Bigg[
    -\frac{1}{\prod_{i=1}^{2}\vartheta_4(-\mathfrak{b}_i+\mathfrak{c})
  \vartheta_4(\mathfrak{b}_i+\mathfrak{c})} \bigg(\frac{1}{2}+E_2\begin{bmatrix}1\\ c\end{bmatrix}\bigg)
    \nonumber\\
    &\qquad
    + \sum_{i=1}^{2}
    \bigg(
      \frac{(-1)^{i+1}
        \vartheta_1(\mathfrak{b}_i)^2
      }{
        \vartheta_1(\mathfrak{b}_1\pm\mathfrak{b}_2)
        \vartheta_4(\mathfrak{c})^2
        \vartheta_4(\pm\mathfrak{b}_i+\mathfrak{c})
      }\left(\frac{3}{8}+E_1\begin{bmatrix}-1\\ b_i\end{bmatrix}
  E_1\begin{bmatrix}1\\ b_i\end{bmatrix}
  + E_2\begin{bmatrix}-1\\ b_i\end{bmatrix}\right) \nonumber\\
      &\qquad\qquad\qquad
      +\frac{(-1)^{i+1}
        \vartheta_1(2\mathfrak{b}_i)
        \vartheta_4(\mathfrak{c})^2
      }{
        \vartheta_1(\mathfrak{b}_1\pm\mathfrak{b}_2)
        \vartheta_1(\mathfrak{b}_i)^2
        \vartheta_1(2\mathfrak{c})\ 
        \vartheta_4(\pm\mathfrak{b}_i+\mathfrak{c})
      }E_1\begin{bmatrix}-1\\ b_i\end{bmatrix}
        E_1\begin{bmatrix}-1\\ c\end{bmatrix}
  \Bigg]
  \nonumber\\
  &
  -2i \eta(\tau)^3
  \sum_{i=1}^{2}
  \frac{
    (-1)^{i+1}
  }{
    \vartheta_1(\mathfrak{b}_1\pm\mathfrak{b}_2)\ 
    \vartheta_1(2\mathfrak{c})
  }E_1\begin{bmatrix}-1\\ c\end{bmatrix} \Big(E_1\begin{bmatrix}-1\\ b_i\end{bmatrix}
  E_1\begin{bmatrix}1\\ b_i\end{bmatrix}
  + E_2\begin{bmatrix}-1\\ b_i\end{bmatrix} \Big)\ .
\end{align}

Locally, $SO(4) = SU(2) \times SU(2)$, and $USp(2) = SU(2)$. Therefore the $SO$-$USp$ quiver is actually an alternative description of an $A_1$ class-$\mathcal{S}$ theory \cite{Lemos:2012ph}. In this case, the theory corresponds to the $\mathcal{T}_{1,3}$ theory, where the closed form Schur index of the latter is known from \cite{Pan:2021mrw},
\begin{align}
  \mathcal{I}_{1,3} = \frac{i \eta(\tau)^3}{\prod_{i = 1}^3 \vartheta_1(2 \mathfrak{b}_i)}
  \sum_{\alpha_i = \pm} (\alpha_1 \alpha_2 \alpha_3)\bigg(
  \frac{1}{16}E_1 \begin{bmatrix}
    -1 \\ \prod_{i = 1}^3 b_i^{\alpha_i}
  \end{bmatrix}
  - \frac{1}{2}E_3 \begin{bmatrix}
    -1 \\ \prod_{i = 1}^3 b_i^{\alpha_i}
  \end{bmatrix}
  \bigg) \ .
\end{align}
The two expressions are equivalent up to the identification of the flavor fugacities,
\begin{equation}
  b_1 \to b_1 b_2, \qquad
  b_2 \to b_1/b_2, \qquad
  c \to b_3\ .
\end{equation}
We have checked the equality by comparing the $q$-expansions to high order. We may also verify analytically that the two unflavoring limits are identical. Upon replacing $\vartheta_i^{(p)}(0)$ with Eisenstein series, the unflavoring limit of D-type quiver gives
\begin{align}
  -\frac{\eta(\tau)^6}{\vartheta_4(0)^4}\left(\frac18 + 2E_2 + \widetilde E_2\right) + \frac{\left(\frac{E_2^3}{2} + 3E_2^2\widetilde E_2 + 12E_2\widetilde E_2^2 + 16\widetilde E_2^3 - \frac{15}{2}E_2E_4 - 15\widetilde E_2E_4\right)}{\eta(\tau)^6} \ , \nonumber
\end{align}
where $\widetilde{E}_2 = E_2 \Big[\substack{-1\\ 1}\Big]$. The unflavored Schur index of the $\mathcal{T}_{1,3}$ theory reads
\begin{equation}
  \frac{1}{\eta^6}\left(\frac{E_2^3}{2}+3E_2^2\widetilde E_2-12E_2\widetilde E_2^2+4\widetilde E_2^3-\frac32\widetilde E_2^2+\frac{15}{8}E_4+\frac{45}{2}E_2E_4\right) \ .
\end{equation}
The difference is
\begin{equation}
  =-\frac{(1+16E_2+8\widetilde E_2)(\eta(\tau)^{12}+15E_4\vartheta_4(0)^4-12\vartheta_4(0)^4\widetilde E_2^2)}{8\,\eta^6\,\vartheta_4(0)^4} \ .
\end{equation}
Finally, recall that
\begin{align}
  \widetilde{E}_2 = & \ \frac{\vartheta_2(0)^4 + \vartheta_3(0)^4}{24}, &
  E_4 = & \ \frac{\vartheta_2(0)^4 + \vartheta_3(0)^4 + \vartheta_4(0)^4 }{1440},\\
  \eta(\tau)^{12} = & \ \frac{\vartheta_2(0)^4 \vartheta_3(0)^4 \vartheta_4(0)^4}{16} \ , & \vartheta_4(0)^4 = & \ \vartheta_2(0)^4 - \vartheta_3(0)^4 \ ,
\end{align}
the difference automatically vanishes.

\subsection{\texorpdfstring{$[SO(8)]-USp(4)-SO(4)$ quiver theory}{[SO(8)]-USp(4)-SO(4) quiver theory}}

As our final example, consider the $SO(4) \times USp(4)$ gauge theory with bifundamental hypermultiplets and four fundamental hypermultiplets for $USp(4)$. The Schur index is given by the contour integral
\begin{equation}
  \mathcal{I} = \oint \prod_{i = 1}^4 \frac{da_i}{2\pi i a_i} \frac{\eta(\tau)^{20}\prod_{i = 1}^2 \vartheta_1(2 \mathfrak{a}_i)^2 \prod_{\alpha = \pm}\prod_{i= 0,1}\vartheta_1(\mathfrak{a}_{2i+1} + \alpha \mathfrak{a}_{2i+2})^2}{
    32 q^{7/3}\prod_{s = 1}^4\prod_{i = 1,2}\prod_{\alpha = \pm} \vartheta_4(\mathfrak{a}_i + \alpha \mathfrak{b}_s)
    \prod_{\substack{i = 1,2 \\ j = 3,4}}\prod_{\alpha = \pm}\vartheta_4(\mathfrak{a}_i + \alpha \mathfrak{a}_j)
  } \nonumber
\end{equation}
The $a_1$ integration is simple and straightforward, creating the following structure,
\begin{align}
  = & \ \frac{
  i \eta(\tau)^{17}\vartheta_1(2\mathfrak{a}_2)^2
  \vartheta_1(\pm\mathfrak{a}_3+\mathfrak{a}_4)
  }{
  16q^{7/3}
  \prod_{k=1}^{4}
  \vartheta_4(\mathfrak{a}_2 \pm \mathfrak{b}_k)
  }
  +
  \frac{
  i \eta(\tau)^{17}\vartheta_1(2\mathfrak{a}_2)^2
  \vartheta_1(\pm\mathfrak{a}_3+\mathfrak{a}_4)^2
  }{
  32q^{7/3}
  \vartheta_4(\pm\mathfrak{a}_2+\mathfrak{a}_3)
  \vartheta_4(\pm\mathfrak{a}_2+\mathfrak{a}_4)
  }
  \nonumber\\
  &\quad{}\times
  \sum_{j=3}^{4}
  \frac{
  (-1)^j
  \vartheta_1(2\mathfrak{a}_j)
  \vartheta_4(-\mathfrak{a}_2+\mathfrak{a}_j)
  \vartheta_4(\mathfrak{a}_2+\mathfrak{a}_j)
  }{
  \vartheta_4(-\mathfrak{a}_2+\mathfrak{a}_{7-j})
  \vartheta_4(\mathfrak{a}_2+\mathfrak{a}_{7-j})
  \prod_{k=1}^{4}
  \vartheta_1(\mathfrak{a}_j-\mathfrak{b}_k)
  \vartheta_1(\mathfrak{a}_j+\mathfrak{b}_k)
  }E_1\begin{bmatrix}-1\\a_j\end{bmatrix}
  \\
  &\quad{}\times
  \sum_{j=1}^{4}\sum_{s = \pm}
  \frac{
  s \vartheta_1(2s\mathfrak{b}_j)
  \vartheta_4(\mathfrak{a}_2\pm\mathfrak{b}_j)
  E_1\begin{bmatrix}-1\\b_j\end{bmatrix}
  }{
  \prod_{r=\pm}
  \vartheta_1(\mathfrak{a}_3-r\mathfrak{b}_j)
  \vartheta_1(\mathfrak{a}_4-r\mathfrak{b}_j)
  \prod_{\substack{k=1\\k\ne j}}^{4}
  \vartheta_1(s\mathfrak{b}_j\pm\mathfrak{b}_k)
  \prod_{\substack{k=1\\k\ne j}}^{4}
  \vartheta_4(\mathfrak{a}_2\pm\mathfrak{b}_k)
  }. \nonumber
\end{align}
The subsequent integrations yield more tedious expressions, which we do not display here. Fortunately the final result can be reorganized into a more manageable form.
\begin{align}
\mathcal I
=
&\frac{\eta(\tau)^8}{q^{7/3}}
\Bigg\{
\bigg[
\frac{9}{16}
+
\frac{3}{4}
\sum_{r=1}^{4}
\left(
E_1 \Big[\begin{smallmatrix}-1\\ b_r\end{smallmatrix}\Big]
E_1 \Big[\begin{smallmatrix}1\\ b_r\end{smallmatrix}\Big]
+
E_2 \Big[\begin{smallmatrix}-1\\ b_r\end{smallmatrix}\Big]
\right)
\bigg]
\nonumber\\[4pt]
&\qquad\qquad\times
\frac{1}{2}
\sum_{1\leq i < j\leq 4}
\frac{
(-1)^{i+j+1}
}{
\prod_{\substack{1\leq a < b\leq 4\\
\left|\{a,b\}\cap\{i,j\}\right|=1}}
\vartheta_1(-\mathfrak b_a+\mathfrak b_b;q) 
\vartheta_1(\mathfrak b_a+\mathfrak b_b;q)
}
\nonumber\\[6pt]
&
\qquad\qquad+2 \sum_{1\leq i < j\leq 4}
\frac{(-1)^{i+j+1}
}{
\prod_{\substack{1\leq a < b\leq 4\\
\left|\{a,b\}\cap\{i,j\}\right|=1}}
\vartheta_1(-\mathfrak b_a+\mathfrak b_b;q) 
\vartheta_1(\mathfrak b_a+\mathfrak b_b;q)
}
\nonumber\\[-2pt]
&\qquad\qquad\times
\left(
E_1 \Big[\begin{smallmatrix}-1\\ b_i\end{smallmatrix}\Big]
E_1 \Big[\begin{smallmatrix}1\\ b_i\end{smallmatrix}\Big]
+
E_2 \Big[\begin{smallmatrix}-1\\ b_i\end{smallmatrix}\Big]
\right)\left(
E_1 \Big[\begin{smallmatrix}-1\\ b_j\end{smallmatrix}\Big]
E_1 \Big[\begin{smallmatrix}1\\ b_j\end{smallmatrix}\Big]
+
E_2 \Big[\begin{smallmatrix}-1\\ b_j\end{smallmatrix}\Big]
\right) \Bigg\}
\nonumber\\[2pt]
&
+
\frac{i \eta(\tau)^{11}}{q^{7/3}}
\sum_{\substack{1\leq i,j\leq 4\\ i\neq j}}
\frac{
\operatorname{sgn}(j-i)(-1)^{i+j+1} 
\vartheta_1(2\mathfrak b_i;q) 
\vartheta_1(\mathfrak b_j;q)^2
}{
\vartheta_1(\mathfrak b_i;q)^2
\prod_{\substack{1\leq a < b\leq 4\\
\{a,b\}\cap\{i,j\}\neq\varnothing}}
\vartheta_1(-\mathfrak b_a+\mathfrak b_b;q) 
\vartheta_1(\mathfrak b_a+\mathfrak b_b;q)
}
\nonumber\\[-2pt]
&\qquad\times
E_1 \Big[\begin{smallmatrix}-1\\ b_i\end{smallmatrix}\Big]
\left(
\frac{3}{8}
+
E_1 \Big[\begin{smallmatrix}-1\\ b_j\end{smallmatrix}\Big]
E_1 \Big[\begin{smallmatrix}1\\ b_j\end{smallmatrix}\Big]
+
E_2 \Big[\begin{smallmatrix}-1\\ b_j\end{smallmatrix}\Big]
\right) \ .
\end{align}


\section{Discussions}

In this paper we take crucial inspiration from the work \cite{Benini:2018mlo}, identify its core problems, and reorganize the key tricks therein into a new exact and analytic computational technique for Schur index of 4d $\mathcal{N} = 2$ superconformal field theories. It is surprising to see that there exists such a general and yet relatively elementary (independent from intricate special function identities like the Cauchy determinant formula) technique. However, the technique has its drawbacks. The intermediate integrations often receive contribution from a large number of poles, which grows rapidly as the rank of the gauge groups increases. The final result requires simplification using various identities involving the Eisenstein series and the Jacobi theta functions.

One of the problems in the original \cite{Benini:2018mlo} computation is the simultaneous shift of contour and the omitted poles. In our work, we circumvent the problem by performing the computation one integration at a time. The end result is a sum of Grothendieck residues accompanied by polynomials of Eisenstein series. Such a combination can be understood loosely as some ``generalized residue'', if one were to only focus on the poles within the annulus. In some sense, these Eisenstein series can be viewed as ``geometric correction'' to the original Bethe ansatz formula, which simultaneously captures the loss of periodicity at each step of integration, and the topology of the poles in the integrand. It would be very interesting to further clarify the full homological information encoded in the Eisenstein series, and the meaning of generalized residue. The presence of Eisenstein series renders the index some combination of quasi-Jacobi forms. The fact that ordered A-cycle integral leads to a mixed weight quasi-modular form has been discussed in \cite{Li:2020ljm}.

The method proposed in the paper can be applied to a lot more theories. For example, the class-$\mathcal{S}$ Lagrangian theories of type $A/D$, star-shape $SU(N)$ quiver theory, and many more. The only problem is to find suitable simplification of the intermediate and final results. The proposed technique is not limited to Lagrangian theories. For example, using the inversion formula \cite{2004math.....11044S,Gadde:2010te,Razamat:2012uv}, the $E_6$ Schur index can be computed in closed form \cite{Pan:2021mrw} in terms of Eisenstein series and Jacobi theta functions. It is therefore possible to obtain non-Lagrangian $A_2$ class-$\mathcal{S}$ Schur index by further gauging. As shown in \cite{Beem:2023ofp}, A-type Argyres-Douglas theories with exactly marginal coupling can be realized as gauge theories with basic Argyres-Douglas matters (with no exactly marginal coupling). The recent work \cite{Pan:2025vyu} consider $SU(2)$ gauge theories with Argyres-Douglas matters, and computed the corresponding Schur index in compact closed form. Infinite series therein exhibit rather simple large $N$ behavior. It would be interesting to extend the computation to more general Argyres-Douglas theories with higher rank gauge groups, which may help study non-admissible W-algebras \cite{Creutzig:2017qyf,Xie:2026xxg}. Similar logic goes for the generalized partition function computation \cite{Deb:2025ypl,Chandra:2025qpv,Deb:2025ddc,Pan:2025vyu}.

With more closed form Schur index falling within reach using the new technique, it has become possible to carry out more systematic and thorough analysis of the modular properties of the Schur index and defect Schur index. Following the overall picture between associated VOA representation theory and 4d defects \cite{Cordova:2016uwk,Cordova:2017mhb,Pan:2017zie,Dedushenko:2019yiw,Nishinaka:2018zwq}, the full character space of the  $\mathcal{N} = 4$ $SU(N)$ associated VOA has been worked out (conjecturally) in \cite{Li:2025nhc}. The same can be done for other higher rank Lagrangian theories and non-Lagrangian theories with our technique. It may shed more light on the 4d mirror symmetry program \cite{2017arXiv170906142F,Fredrickson:2017yka,Dedushenko:2018bpp,Shan:2023xtw,Pan:2024epf,Pan:2024hcz,Xie:2026xxg}.

Finally, it might be useful to reexamine the original Bethe ansatz formula in \cite{Benini:2018mlo} and its application to $\mathcal{N} = 4$ superconformal index computation, bearing in mind the subtleties that we have discussed in this paper, including the loss of periodicity and the non-trivial topology of the divisors that contribute to the integral.

\section*{Acknowledgments}

We thank Si Li, Yinan Wang, Wenbin Yan for helpful discussions. The work of Y.P. is supported by the National Natural Science Foundation of China (NSFC) under Grant No. 11905301.

\appendix

\section{Grothendieck residue}

In this appendix we collect some elementary facts about the Grothendieck residue. They are
used in section~\ref{sec:BAE}. The discussion is mostly local. In particular, the calculation of local residue and the
problem of which poles are collected by a global contour are two distinct problems. This
distinction is somewhat unimportant for single-variate cases, but is important for multivariate integrals.

\subsection{Local residue}

Let $U$ be a sufficiently small ball around the origin in $\mathbb C^n$ and $\mathcal{O}(U)$ the ring of holomorphic functions on $U$. Consider $n$ holomorphic functions
$f_i\in\mathcal O(U)$, $i=1,\ldots,n$ having an isolated common zero at the origin,
\begin{equation}\label{GR:zero}
  f^{-1}(0)=\{0\} \in \mathbb{C}^n,\qquad f \coloneqq (f_1,\ldots,f_n).
\end{equation}
For any holomorphic function $h(z) \in \mathcal{O}(U)$, consider the meromorphic $n$-form
\begin{equation}\label{GR:form}
                 \omega=\frac{h(z)\,dz_1\wedge\cdots\wedge dz_n}
                 {f_1(z)\cdots f_n(z)}.
\end{equation}
The Grothendieck residue at the origin (an isolated common pole) is defined by
\begin{equation}\label{GR:def}
 \underset{z = 0;f}{\operatorname{Res}}\,\omega
 =\frac{1}{(2\pi i)^n}\int_{\Gamma_f}
   \frac{h(z)\,dz_1\wedge\cdots\wedge dz_n}{f_1(z)\cdots f_n(z)},
\end{equation}
where
\begin{equation}\label{GR:cycle}
        \Gamma_f=\{z\in U:|f_i(z)|=\epsilon_i,\ i=1,\ldots,n\},
\end{equation}
and the orientation is chosen such that
$d\arg f_1\wedge\cdots\wedge d\arg f_n>0$. For sufficiently small positive
$\epsilon_i$, the result does not depend on the particular choice of $\epsilon_i$, or in other words, the integral is invariant under a continuous deformation of the cycle $\Gamma_f$ as long as it does not cross any polar divisor.

Two remarks follow. First, $\Gamma_f$ is not the boundary of a small $2n$-dimensional ball. Instead, it is a real $n$-dimensional torus. Second, its orientation and homology
class depend on the ordered functions $(f_1,\ldots,f_n)$, hence the $f_1, ..., f_n$ underneath the residue symbol: swapping two $f_i$ changes the sign of the residue.

When the common zero of $f$ is non-degenerate, \emph{i.e.} when the Jacobian
\begin{equation}\label{GR:jacobian}
              J_f(0):=\det\left(\frac{\partial f_i}{\partial z_j}(0)\right)
              \neq0,
\end{equation}
the functions $f_i$ can be used as local coordinates in place of $z_i$. The usual Cauchy formula then gives
\begin{equation}\label{GR:jacobian-formula}
 \underset{z = 0;f}{\operatorname{Res}}
 \frac{h(z)\,d^nz}{f_1\cdots f_n}=\frac{h(0)}{J_f(0)},
 \qquad d^nz:=dz_1\wedge\cdots\wedge dz_n \ .
\end{equation}
This is the Jacobian formula used in \cite{Benini:2018mlo}. Taking
$f_i=1-Q_i$, it produces the determinant appearing in the formula of section~\ref{sec:BAE}.

Let
\begin{equation*}
                         I_f=(f_1,\ldots,f_n)\subset\mathcal O_0,
\end{equation*}
where $\mathcal O_0$ is the ring of holomorphic functions at the origin. An
elementary property of the residue is
\begin{equation}\label{GR:ideal}
 h\in I_f\quad\Longrightarrow\quad
 \underset{z = 0;f}{\operatorname{Res}}
 \frac{h(z)\,d^nz}{f_1\cdots f_n}=0.
\end{equation}
Indeed, writing $h=\sum_i a_i f_i$ gives
\begin{equation*}
 \frac{h(z)\,d^nz}{f_1\cdots f_n}
 =\sum_{i=1}^n
 \frac{a_i(z)\,d^nz}{f_1\cdots\widehat{f_i}\cdots f_n}.
\end{equation*}
Every term has at least one polar factor removed, and therefore does not have $n$ independent effective polar directions; we may call the ideal from the reduced set $f_1, \cdots, \hat f_i, ..., f_n$ the effective polar ideal.

It follows that the dependence of residue on $h$ is only through its equivalence class in the local algebra
\begin{equation}\label{GR:algebra}
                       \mathcal A_f=\mathcal O_0/I_f.
\end{equation}
When the common zero is isolated, $\mathcal A_f$ is finite dimensional. This follows
directly from the local analytic Nullstellensatz. If
$\mathfrak m_0=(z_1,\ldots,z_n)$ is the maximal ideal of $\mathcal O_0$, the isolated-zero
condition implies
\begin{equation*}
                         \sqrt{I_f}=\mathfrak m_0,
 \qquad \mathfrak m_0^N\subset I_f
\end{equation*}
for some sufficiently large $N$. Thus all monomials of sufficiently high degree vanish in
$\mathcal A_f$, leaving only a finite number of local directions. The converse is also true:
a finite-dimensional $\mathcal A_f$ implies that the common zero is isolated.

\subsection{Non-isolated zeros}

It is possible to have $J_f = 0$. The first situation is degenerate isolated zeros. For
example, $f_1=x$ and $f_2=y^2$ have an isolated common zero with vanishing Jacobian. The second situation is non-isolated zeros, where the common zero of $f$ has positive dimension. In the following we briefly discuss the latter case. In this case, formula \eqref{GR:jacobian-formula} is no longer applicable, although the
residue remains well-defined by \eqref{GR:def} as long as the small encircling contour does not cross any polar divisor.

The \emph{height}
$\operatorname{ht}J$ of an effective polar ideal $J$ counts, locally, the number of
independent complex constraints supplied by its denominator functions. If
\begin{equation*}
                         \operatorname{ht}J=r<n,
\end{equation*}
the denominators fail to provide all the $n$ transverse polar directions, and their common
zero has positive dimension $d=n-r$, which we refer to as non-isolated common zeros. For instance, in the case of isolated common zero \eqref{GR:ideal}, the original ideal $I_f$ may have
height $n$. However, after canceling $f_i$, the $i$-th term has the effective ideal
\begin{equation*}
                         J_i=(f_1,\ldots,\widehat{f_i},\ldots,f_n),
\end{equation*}
whose height is at most $n-1$ by the Krull height theorem. For a non-isolated common
zero, the original ideal $I_f$ already has height ${\operatorname{ht}I_f} < n$. In both cases the common zero has positive codimension.

This vanishing of residue at non-isolated common poles can be seen directly from a local tube. Let $P$ be a smooth component of
the effective common pole, with complex dimension $d>0$. After a locally invertible
holomorphic change of variables, the holomorphic implicit function theorem allows us to
choose coordinates
\begin{equation*}
             (t_1,\ldots,t_r,s_1,\ldots,s_d),\qquad r+d=n,
\end{equation*}
such that in a sufficiently small (containing no other polar divisors besides $P$) product polydisc
\begin{equation*}
U \coloneqq \Delta_t^r\times\Delta_s^d,
 \qquad     P\cap U=\{t_1=\cdots=t_r=0\}.
\end{equation*}
In other words, the $t_i$ are transverse to $P$, while the $s_a$ parametrize it.
All transverse integrations can be carried out, leaving a holomorphic $d$-form on $P$. Concretely, take the local tubular cycle
\begin{equation*}
 \Gamma_{\epsilon,\delta}=T_t^r(\epsilon)\times T_s^d(\delta),
 \qquad
 T_t^r=\{|t_i|=\epsilon_i\},\quad
 T_s^d=\{|s_a|=\delta_a\},
\end{equation*}
The tube integral then reduces to
\begin{equation*}
 \int_{\Gamma_{\epsilon,\delta}}\omega
 \ \propto\ 
 \int_{T_s^d(\delta)}\rho(s)\,ds_1\wedge\cdots\wedge ds_d=0.
\end{equation*}
The transverse $t_i$-circles surround $t_i=0$ where the common pole lies, so shrinking the $t_i$-circles would cross the divisor and produces the residue (likely of high order pole) captured in $\rho$. The $s_a$-circles, on the
other hand, encounter no divisor inside the polydisc $\Delta^d_s$ and can be
shrunk, hence producing zero.

For example, consider three denominator functions in three variables $x,y, z$,
\begin{equation*}
                         f_1=x,\qquad f_2=y,\qquad f_3=x+y
\end{equation*}
give only two independent constraints ($f_1, f_2 = 0\Rightarrow f_3 = 0$). Their common zero $\{x=y=0\}$ extends along the
$z$ direction, forming a non-isolated common zero. If there is no other $z$-dependent polar divisor in the local neighborhood,
the transverse $x,y$ integrations leave a holomorphic $z$ direction, and the corresponding $x, y, z$ integral gives zero.

\subsection{Residue theorem}

We finally clarify the point relevant for section~\ref{sec:BAE}. Let $D=\bigcup_\alpha\{F_\alpha(z)=0\}$ be the full polar divisor of a multivariate integrand. The integral is invariant under a
deformation of its contour $\mathcal C$ only when the deformation stays inside
$\mathbb C^n-D$. The poles collected by integration along $\mathcal C$ are determined by its
homology class
\begin{equation}\label{GR:homology}
                         [\mathcal C]\in H_n(\mathbb C^n \setminus D,\mathbb Z),
\end{equation}
or the appropriate relative version, rather than by the naive ``annulus region''.

To apply \eqref{GR:def}, the polar components also need to be partitioned into $n$ groups,
\begin{equation}\label{GR:partition}
                         f_i=\prod_{\alpha\in I_i}F_\alpha,
             \qquad \{I_1,\ldots,I_n\}\text{ a partition of all $\alpha$} \ .
\end{equation}
Using the partition one can define local tori $\Gamma_{f, p} \coloneqq \{|f_i - p|=\epsilon_i\}$ where $f(p) = 0$. If one can establish a homology relation in the complement $\mathbb{C}^n \setminus D$
\begin{equation}\label{GR:decomposition}
[\mathcal C]=\sum_{p} [\Gamma_{f,p}] \ ,
\end{equation}
then we have the higher dimensional \emph{Cauchy's residue theorem}
\begin{equation}
  \mathcal{I} = \oint_{\mathcal{C}} \omega = \sum_{p} \oint_{\Gamma_{f,p}} \omega = \sum_{p} \underset{z = p;f}{\operatorname{Res}} \omega \ .
\end{equation}

For the rational example discussed in section~\ref{sec:BAE}, the unit torus $T^2(1,1)$ links all three singular planes $z_1 = a, z_2 = b, z_1 = cz_2$.
Its deformation gives the two local torus cycles around $(a,b)$ and $(bc,b)$, whose residues cancel as shown there.
The fact that $(bc,b)$ lies outside the coordinate-wise annulus does not mean we can discard it.

There is one more caveat. The function $h$ in \eqref{GR:form} is assumed to be holomorphic
near the common zero. If the numerator has poles, these poles are additional
components of $D$ and have to be included before deforming the contour. They are not
discarded by the ideal relation \eqref{GR:ideal}. Applied to the Bethe ansatz derivation of \cite{Benini:2018mlo}, the BAE
determine the candidate local poles and the Jacobian determines their local contribution,
but neither of them alone determine the decomposition \eqref{GR:decomposition}. The absence, cancellation,
or extra contribution from the other singularities of $Z(a)$ has to be checked together
with the contour. This is the qualification to the residue prescription used there.

For the rational example this homology problem involves only finitely many divisors. In the
elliptic example the situation is a little worse. A zero of a theta function is repeated by
all integral $q$-shifts, and hence the polar divisor consists of infinitely many components.
Their intersections give infinitely many isolated common poles, some of which
accumulate at the coordinate boundary. We describe this explicitly below.

\subsection{The elliptic example}

One of the key ideas of \cite{Benini:2018mlo} is that by clever manipulation the unit circle integration is turned into an alternating sum of contour integrals, and therefore only a finite number of poles contribute. Below, we use the example \eqref{eq:example-original} to illustrate that this is unfortunately incorrect: careful tracking of the contour and the polar divisors shows that infinitely many poles outside of the annulus region contribute, including the non-isolated singularity at the origin.

Write
\begin{equation}\label{GR:four-contours}
 \begin{aligned}
 T(R_1,R_2)&=\{|a_1|=R_1,\ |a_2|=R_2\},\\
 \mathcal C&=T(1,1)-T(1,|q|)-T(|q|,1)+T(|q|,|q|) \ .
 \end{aligned}
\end{equation}
The Jacobi theta functions $\vartheta_4$, together with the two $E_1$ factors in \eqref{eq:example-rewrite},
gives six families of polar divisors. For later reference we number the equations, rather
than introducing a new notation for every divisor:
\begin{equation}\label{GR:polar-families}
\begin{alignedat}{3}
 \text{(i)}\quad &a_1=a_2b_1q^{m+1/2},
 &\qquad& m\in\mathbb Z,\\
 \text{(ii)}\quad &a_2=a_1b_1q^{n+1/2},
 && n\in\mathbb Z,\\
 \text{(iii)}\quad &a_1a_2=b_2q^{p+1/2},
 && p\in\mathbb Z,\\
 \text{(iv)}\quad &a_1a_2b_2=q^{r+1/2},
 && r\in\mathbb Z,\\
 \text{(v)}\quad &a_1=q^s,
 && s\in\mathbb Z,\\
 \text{(vi)}\quad &a_2=q^t,
 && t\in\mathbb Z.
\end{alignedat}
\end{equation}
The first four equations come from the theta denominator of \eqref{eq:example-original}, while (v) and (vi)
are additional Eisenstein series of the rewritten integrand \eqref{eq:example-rewrite}. The integers imply the full divisor is an infinite union of divisors differing by $q$-shift.

We now list all the poles after the cancellation among the four contour integrals \eqref{GR:four-contours}. They are produced by first performing the $a_2$ integration then $a_1$, which is different from the order in the main text. A $+$ or $-$ denotes the orientation of the local torus, and not the sign of its
numerical residue. $\varepsilon = \pm 1$ in the coordinates of the poles.

\begin{equation}\label{GR:pole-i-iii}
\begin{aligned}
 \text{(i),(iii)}:\quad
 (a_1,a_2)
 &=\left(\varepsilon\sqrt{b_1b_2}\,q^{(m+p+1)/2},
 \varepsilon\sqrt{\frac{b_2}{b_1}}\,q^{(p-m)/2}\right),\\
 +:&\quad \{m=-1,p\geq1\}\cup\{m\geq1,p=1\},\\
 -:&\quad \{m\geq0,p=0\}\cup\{m=0,p\geq2\}.
\end{aligned}
\end{equation}

\begin{equation}\label{GR:pole-i-iv}
\begin{aligned}
 \text{(i),(iv)}:\quad
 (a_1,a_2)
 &=\left(\varepsilon\sqrt{\frac{b_1}{b_2}}q^{(m+r+1)/2},
 \varepsilon(b_1b_2)^{-1/2}q^{(r-m)/2}\right),\\
 +:&\quad \{m=-1,r\geq1\}\cup\{m\geq1,r=1\},\\
 -:&\quad \{m\geq0,r=0\}\cup\{m=0,r\geq2\}.
\end{aligned}
\end{equation}

\begin{equation}\label{GR:pole-i-v}
\begin{aligned}
 \text{(i),(v)}:\quad
 (a_1,a_2)
 &=(q^s,b_1^{-1}q^{s-m-1/2}),\\
 +:&\quad m=-1,\ s\geq1,\qquad
 -:\quad m=0,\ s\geq2.
\end{aligned}
\end{equation}

\begin{equation}\label{GR:pole-i-vi}
\begin{aligned}
 \text{(i),(vi)}:\quad
 (a_1,a_2)
 &=(b_1q^{m+t+1/2},q^t),\\
 +:&\quad m=-1,\ t\geq2,\qquad
 -:\quad m=0,\ t\geq1.
\end{aligned}
\end{equation}

\begin{equation}\label{GR:pole-ii-iii}
\begin{aligned}
 \text{(ii),(iii)}:\quad
 (a_1,a_2)
 &=\left(\varepsilon\sqrt{\frac{b_2}{b_1}}q^{(p-n)/2},
 \varepsilon\sqrt{b_1b_2}\,q^{(n+p+1)/2}\right),\\
 +:&\quad \{n\leq-2,p=1\}\cup\{n=0,p\geq1\},\\
 -:&\quad \{n\leq-1,p=0\}\cup\{n=-1,p\geq2\}.
\end{aligned}
\end{equation}

\begin{equation}\label{GR:pole-ii-iv}
\begin{aligned}
 \text{(ii),(iv)}:\quad
 (a_1,a_2)
 &=\left(\varepsilon(b_1b_2)^{-1/2}q^{(r-n)/2},
 \varepsilon\sqrt{\frac{b_1}{b_2}}q^{(n+r+1)/2}\right),\\
 +:&\quad \{n\leq-2,r=1\}\cup\{n=0,r\geq1\},\\
 -:&\quad \{n\leq-1,r=0\}\cup\{n=-1,r\geq2\}.
\end{aligned}
\end{equation}

\begin{equation}\label{GR:pole-ii-v}
\begin{aligned}
 \text{(ii),(v)}:\quad
 (a_1,a_2)
 &=(q^s,b_1q^{s+n+1/2}),\\
 +:&\quad n=0,\ s\geq1,\qquad
 -:\quad n=-1,\ s\geq2.
\end{aligned}
\end{equation}

\begin{equation}\label{GR:pole-ii-vi}
\begin{aligned}
 \text{(ii),(vi)}:\quad
 (a_1,a_2)
 &=(b_1^{-1}q^{t-n-1/2},q^t),\\
 +:&\quad n=0,\ t\geq2,\qquad
 -:\quad n=-1,\ t\geq1.
\end{aligned}
\end{equation}

\begin{equation}\label{GR:pole-iii-v}
\begin{aligned}
 \text{(iii),(v)}:\quad
 (a_1,a_2)
 &=(q^s,b_2q^{p+1/2-s}),\\
 +:&\quad p=0,\ s\geq1,\qquad
 -:\quad p=1,\ s\geq2.
\end{aligned}
\end{equation}

\begin{equation}\label{GR:pole-iii-vi}
\begin{aligned}
 \text{(iii),(vi)}:\quad
 (a_1,a_2)
 &=(b_2q^{p+1/2-t},q^t),\\
 +:&\quad p=0,\ t\leq0,\qquad
 -:\quad p=1,\ t\leq1.
\end{aligned}
\end{equation}

\begin{equation}\label{GR:pole-iv-v}
\begin{aligned}
 \text{(iv),(v)}:\quad
 (a_1,a_2)
 &=(q^s,b_2^{-1}q^{r+1/2-s}),\\
 +:&\quad r=0,\ s\geq1,\qquad
 -:\quad r=1,\ s\geq2.
\end{aligned}
\end{equation}

\begin{equation}\label{GR:pole-iv-vi}
\begin{aligned}
 \text{(iv),(vi)}:\quad
 (a_1,a_2)
 &=(b_2^{-1}q^{r+1/2-t},q^t),\\
 +:&\quad r=0,\ t\leq0,\qquad
 -:\quad r=1,\ t\leq1.
\end{aligned}
\end{equation}

\begin{equation}\label{GR:pole-v-vi}
\begin{aligned}
 \text{(v),(vi)}:\quad
 (a_1,a_2)&=(q^s,q^t),\\
 +:&\quad \varnothing,\qquad
 -:\quad s=t=1.
\end{aligned}
\end{equation}

The crucial point we want to make is that, the contour $\mathcal{C}$ ``links'' the infinite number of poles listed in \eqref{GR:pole-i-iii}--\eqref{GR:pole-v-vi}, instead of just the finitely many common poles inside the annulus region. In particular, these poles accumulate at the origin $(0,0)$ where the ordinary Grothendieck residue is undefined. Therefore the higher dimensional residue theorem simply does not apply here, nor can one evaluate the integral by collecting only the poles inside the annulus region.


\section{Special functions}

In this appendix we review definitions and basic properties of the special functions used in the main text. We employ the variable naming convention using the fraktur font, e.g.,
\begin{equation}
	a = e^{2\pi i \mathfrak{a}}, \qquad b = e^{2\pi i \mathfrak{b}}, \qquad y = e^{2\pi i \mathfrak{y}} \ , \qquad z = e^{2\pi i \mathfrak{z}} \ ,
\end{equation}
except for the standard notation $q = e^{2\pi i \tau}$ with $\tau$ in the upper half complex plane. 

The $q$-Pochhammer symbol is defined as $(q;q) \coloneqq \prod_{k = 1}^{+\infty}(1 - q^k)$. Adding a factor of $q^{\frac{1}{24}}$, we obtain the Dedekind $\eta$-function $\eta(\tau) = q^{\frac{1}{24}}(q;q)$.

The Jacobi theta functions are defined as infinite products using the $q$-Pochhammer symbol,
\begin{align}
	\vartheta_1(\mathfrak{z}|\tau) \coloneqq & \ - i z^{\frac{1}{2}}q^{\frac{1}{8}}(q;q)(zq;q)(z^{-1};q) \ ,
	& \vartheta_2(\mathfrak{z}|\tau) \coloneqq & \ z^{1/2} q^{1/8}(q;q)(- zq;q)( - z^{-1};q) \ ,  \nonumber\\
	\vartheta_4(\mathfrak{z}|\tau) \coloneqq & \ (q;q)(zq^{1/2};q)(zq^{-1/2};q) \ ,
	& \vartheta_3(\mathfrak{z}|\tau) \coloneqq & \ (q;q)(-zq^{1/2};q)( - z^{-1}q^{1/2};q) \ . \nonumber
\end{align}
Equivalently, they can be defined as the following $q$-series,
\begin{align}
	\vartheta_1(\mathfrak{z}|\tau) \coloneqq & \ -i \sum_{r \in \mathbb{Z} + \frac{1}{2}} (-1)^{r-\frac{1}{2}} e^{2\pi i r \mathfrak{z}} q^{\frac{r^2}{2}} ,
	& \vartheta_2(\mathfrak{z}|\tau) \coloneqq & \sum_{r \in \mathbb{Z} + \frac{1}{2}} e^{2\pi i r \mathfrak{z}} q^{\frac{r^2}{2}} \ ,\\
	\vartheta_3(\mathfrak{z}|\tau) \coloneqq & \ \sum_{n \in \mathbb{Z}} e^{2\pi i n \mathfrak{z}} q^{\frac{n^2}{2}},
	& \vartheta_4(\mathfrak{z}|\tau) \coloneqq & \sum_{n \in \mathbb{Z}} (-1)^n e^{2\pi i n \mathfrak{z}} q^{\frac{n^2}{2}} \ .
\end{align}
We often omit $|\tau$ from the notation. The Jacobi theta function and the Dedekind $\eta$-function are related by $\vartheta'(0) = 2\pi \eta(\tau)^3$.

The $\vartheta$ functions enjoy simple shift properties,
\begin{align}
  \vartheta_{1,2}(\mathfrak{z} + 1) = & \ - \vartheta_{1,2}(\mathfrak{z}),  & \vartheta_{3,4}(\mathfrak{z} + 1) = & \ + \vartheta_{3,4}(\mathfrak{z}) , \\
	\vartheta_{1,4}(\mathfrak{z} + \tau) = & \ - \lambda \vartheta_{1,4}(\mathfrak{z}),
	& \vartheta_{2,3}(\mathfrak{z} + \tau) = & \ + \lambda \vartheta_{2,3}(\mathfrak{z}) \ ,
\end{align}
where $\lambda \coloneqq e^{- 2\pi i \mathfrak{z}} e^{- \pi i \tau}$.

\vspace{1em}
The (twisted) Eisenstein series $E_k\big[\substack{\phi \\ \theta} \big]$ is defined as a $q$-series
\begin{align}
  E_{k \ge 1}\left[\begin{matrix}
    \phi \\ \theta
  \end{matrix}\right] \coloneqq & \ - \frac{B_k(\lambda)}{k!}  \\
  & \ + \frac{1}{(k-1)!}\sum_{r \ge 0}' \frac{(r + \lambda)^{k - 1}\theta^{-1} q^{r + \lambda}}{1 - \theta^{-1}q^{r + \lambda}}
  + \frac{(-1)^k}{(k-1)!}\sum_{r \ge 1} \frac{(r - \lambda)^{k - 1}\theta q^{r - \lambda}}{1 - \theta q^{r - \lambda}} \ , \nonumber
\end{align}
where $\phi = e^{2\pi i \lambda}$ and $\theta$ are often referred to as the characteristics. In this paper we will call $\phi$ the twist parameter as later discussions of twisted modules relate to this parameter. We will also call $k$ the (modular) weight of the Eisenstein series, as it is tied to the transformation property of $E_k$ under $SL(2, \mathbb{Z})$. $B_k(x)$ denotes the $k$-th Bernoulli polynomial, and the prime $^\prime$ means that the term with $r = 0$ should be omitted when $\phi = \theta = 1$. We also define $E_0\big[\substack{\phi\\\theta}\big] = -1$. In the limit $\phi, \theta \to 1$, we recover the standard Eisenstein series $E_k$,
\begin{equation}
	E_{2n}\begin{bmatrix}
		+1 \\ +1
	\end{bmatrix} = E_{2n}(\tau), \quad
	E_{2n + 1 \ge 3} \begin{bmatrix}
		+1 \\ +1
	\end{bmatrix} = E_{2n + 1}(\tau) = 0, \quad
	E_1 \begin{bmatrix}
		+1 \\ z
	\end{bmatrix}
	= \frac{1}{2\pi i } \frac{\vartheta_1'(\mathfrak{z})}{\vartheta_1(\mathfrak{z})} \ . 
\end{equation}
Note that $E_1 \big[\substack{1 \\ z}\big]$ has a simple pole at $z = 1$ (or, $\mathfrak{z} = 0$), since $\vartheta_1(\mathfrak{z}\to 0) \sim 2 \pi \mathfrak{z}q^{1/8}$, but $\vartheta'_1(0)\ne 0$.

There are several useful properties. First of all, the Eisenstein series enjoy the symmetry property
\begin{align}\label{Eisenstein-symmetry}
  E_k\left[\begin{matrix}
    \pm 1 \\ z^{-1}
  \end{matrix}\right] = (-1)^k E_k\left[\begin{matrix}
    \pm 1 \\ z
  \end{matrix}\right] \ .
\end{align}
The twisted Eisenstein series of neighboring weights are related by
\begin{align}\label{EisensteinDerivative}
  q \partial_q E_k\left[\begin{matrix}
    \phi \\ b
  \end{matrix}
  \right] = (- k) b \partial_b E_{k + 1}\left[\begin{matrix}
    \phi \\ b
  \end{matrix}
  \right]\ .
\end{align}
Here we see again that $q\partial_q = D_q^{(1)}$ and $b \partial_b$ raises the modular weight by two and one unit, respectively. More explicitly,
\begin{align}
  \text{odd} ~ n: \quad 0 = z \partial_z E_n \begin{bmatrix}
    1 \\ z
  \end{bmatrix} +  & \ (n + 1)E_{n + 1}\begin{bmatrix}
    1 \\ z
  \end{bmatrix} + E_{n + 1}(\tau) \nonumber\\
  & \  + E_1 \begin{bmatrix}
    1 \\ z
  \end{bmatrix}E_n \begin{bmatrix}
    1 \\ z
  \end{bmatrix}
  - \sum_{k = 2}^{n - 1}E_k(\tau)E_{n + 1 - k} \begin{bmatrix}
    1 \\ z
  \end{bmatrix} \ , \\
  \text{even} ~ n: \quad 0 = z \partial_z E_n \begin{bmatrix}
    1 \\ z
  \end{bmatrix} +  & \ (n + 1)E_{n + 1}\begin{bmatrix}
    1 \\ z
  \end{bmatrix} - E_n(\tau) E_1\begin{bmatrix}
    1 \\ z
  \end{bmatrix} \nonumber\\
  & \  + E_1 \begin{bmatrix}
    1 \\ z
  \end{bmatrix}E_n \begin{bmatrix}
    1 \\ z
  \end{bmatrix}
  - \sum_{k = 2}^{n - 1}E_k(\tau)E_{n + 1 - k} \begin{bmatrix}
    1 \\ z
  \end{bmatrix} \ .
\end{align}

When shifting the argument $\mathfrak{z}$ of the Eisenstein series by half-integral or integral units of $\tau$, or equivalently, shifting $z$ by $q^{\frac{n}{2}}$, one has
\begin{align}\label{Eisenstein-half-shift}
  E_k\left[\begin{matrix}
    \pm 1\\ z q^{\frac{n}{2}}
  \end{matrix}\right]
  =
  \sum_{\ell = 0}^{k} \left(\frac{n}{2}\right)^\ell \frac{1}{\ell !}
  E_{k - \ell}\left[\begin{matrix}
    (-1)^n(\pm 1) \\ z
  \end{matrix}\right] \ , \qquad n \in \mathbb{Z} \ .
\end{align}
This property is crucial for solving the difference equation in the main text.

The Eisenstein series transforms non-trivially under $SL(2, \mathbb{Z})$ generated by
\begin{align}
  S: \tau \to - \frac{1}{\tau}, \ \mathfrak{z} \to \frac{\mathfrak{z}}{\tau}, \qquad
  \qquad
  T: \tau \to \tau + 1 , \ \mathfrak{z} \to \mathfrak{z} \ .
\end{align}
Concretely, $E_k \big[\substack{\pm 1 \\ \pm z}\big]$ transform under $S$,
\begin{align}
  E_n \begin{bmatrix}
    +1 \\ +z
  \end{bmatrix} \xrightarrow{S} &
  \left(\frac{1}{2\pi i}\right)^n\left[\bigg(\sum_{k \ge 0}\frac{1}{k!}(- \log z)^k y^k\bigg)
  \bigg(\sum_{\ell \ge 0}(\log q)^\ell y^\ell E_\ell \begin{bmatrix}
    + 1 \\ z
  \end{bmatrix}\bigg)\right]_n\ ,\\
  E_n \begin{bmatrix}
    -1 \\ +z
  \end{bmatrix} \xrightarrow{S} &
  \left(\frac{1}{2\pi i}\right)^n\left[\bigg(\sum_{k \ge 0}\frac{1}{k!}(- \log z)^k y^k\bigg)
  \bigg(\sum_{\ell \ge 0}(\log q)^\ell y^\ell E_\ell \begin{bmatrix}
    + 1 \\ -z
  \end{bmatrix}\bigg)\right]_n\ ,\\
  E_n \begin{bmatrix}
    1 \\ -z
  \end{bmatrix} \xrightarrow{S} &
  \left(\frac{1}{2\pi i}\right)^n\left[\bigg(\sum_{k \ge 0}\frac{1}{k!}(- \log z)^k y^k\bigg)
  \bigg(\sum_{\ell \ge 0}(\log q)^\ell y^\ell E_\ell \begin{bmatrix}
    -1 \\ +z
  \end{bmatrix}\bigg)\right]_n\ ,\\
  E_n \begin{bmatrix}
    -1 \\ -z 
  \end{bmatrix} \xrightarrow{S} &
  \left(\frac{1}{2\pi i}\right)^n\left[\bigg(\sum_{k \ge 0}\frac{1}{k!}(- \log z)^k y^k\bigg)
  \bigg(\sum_{\ell \ge 0}(\log q)^\ell y^\ell E_\ell \begin{bmatrix}
    -1 \\ -z
  \end{bmatrix}\bigg)\right]_n\ ,
\end{align}
where $[ \ldots ]_n$ extracts the coefficient of $y^n$. Under the $T$-action,
\begin{align}
  E_n \begin{bmatrix}
    + 1 \\ + z
  \end{bmatrix} \xrightarrow{T}& \ E_n \begin{bmatrix}
    + 1 \\ + z
  \end{bmatrix}, & 
  E_n \begin{bmatrix}
    - 1 \\ + z
  \end{bmatrix} \xrightarrow{T}& \
  E_n \begin{bmatrix}
    - 1 \\ - z
  \end{bmatrix} \\
  E_n \begin{bmatrix}
    + 1 \\ - z
  \end{bmatrix} \xrightarrow{T}& \ E_n \begin{bmatrix}
    + 1 \\ - z
  \end{bmatrix}, & 
  E_n \begin{bmatrix}
    - 1 \\ - z
  \end{bmatrix} \xrightarrow{T}& \ 
  E_n \begin{bmatrix}
    - 1 \\ + z
  \end{bmatrix} \ .
\end{align}
We may also combine the two and obtain that under $STS$,
\begin{align}
  E_n \begin{bmatrix}
    -1 \\ z
  \end{bmatrix} \xrightarrow{STS}
  \left(\frac{1}{2\pi i}\right)^n\left[\bigg(\sum_{k \ge 0}\frac{1}{k!}(- \log z)^k y^k\bigg)
  \bigg(\sum_{\ell \ge 0}(\log q - 2\pi i)^\ell y^\ell E_\ell \begin{bmatrix}
    -1 \\ +z
  \end{bmatrix}\bigg)\right]_n\ . \nonumber
\end{align}

\section{Useful identities}\label{app:useful-identities}

In this appendix we collect some useful identities of the theta function and Eisenstein series. The flavored twisted Eisenstein series can be written in terms of logarithmic derivatives of theta functions,
\begin{align}\label{EisensteinToTheta}
	E_k\left[\begin{matrix}
		+ 1 \\ z
	\end{matrix}\right] = - \left[e^{ - \frac{y}{2\pi i }\mathcal{D}_\mathfrak{z} - P_2(y) }\right]_k \vartheta_1(\mathfrak{z}) \ ,
\end{align}
where $P_2$ is a Weierstrass elliptic-$P$ function,
\begin{align}
	P_2(y) \coloneqq - \sum_{n = 1}^{\infty} \frac{1}{2n} E_{2n}(\tau)y^{2n} \ ,
\end{align}
$[f(y)]_k$ denotes the $k$-th coefficient of the Taylor series of $f(y)$ around $y=0$, and we define an abstract differential operator $\mathcal{D}_\mathfrak{z}^n$ by
\begin{align}
	\underbrace{\mathcal{D}_\mathfrak{z} \ldots \mathcal{D}_\mathfrak{z}}_{n \text{ copies}} \vartheta_i(\mathfrak{z}) = \mathcal{D}_\mathfrak{z}^n \vartheta_i(\mathfrak{z}) \equiv \frac{\vartheta^{(n)}_i(\mathfrak{z})}{\vartheta_i(\mathfrak{z})} \ .
\end{align}
More explicitly, we have $E_k[\substack{+1\\z}]$ in terms of $\vartheta_1^{(n)}(\mathfrak{z})$ and $E_{2n}(\tau)$,
\begin{align}\label{EisensteinToTheta-2}
	E_k\left[\begin{matrix}
		+ 1 \\ z
	\end{matrix}\right] = \sum_{\ell = 0}^{\floor{k/2}}  \frac{(-1)^{k + 1}}{(k - 2\ell)!}\left(\frac{1}{2\pi i}\right)^{k - 2\ell} \mathbb{E}_{2\ell}(\tau) \frac{\vartheta_1^{(k - 2\ell)}(\mathfrak{z})}{\vartheta_1(\mathfrak{z})} \ ,
\end{align}
where
\begin{align}
  \mathbb{E}_{2\ell}(\tau) \coloneqq \sum_{\substack{\{n_p\} \\ \sum_{p \ge 1} (2p)n_p = 2\ell}} \prod_{p\ge 1} \frac{1}{n_p !} \left(\frac{1}{2p}E_{2p}\right)^{n_p} \ .
\end{align}
The conversion from $E_k\left[\substack{- 1 \\ \pm z}\right]$ can be obtained by replacing $\vartheta_1$ with $\vartheta_{2,3,4}$ accordingly,
\begin{equation}
  \vartheta_4 \leftrightarrow \begin{bmatrix}
    -1 \\ z
  \end{bmatrix}, \quad
  \vartheta_2 \leftrightarrow \begin{bmatrix}
    +1 \\ -z
  \end{bmatrix}, \quad
  \vartheta_3 \leftrightarrow \begin{bmatrix}
    -1 \\ -z
  \end{bmatrix} \ .
\end{equation}

The unflavored twisted Eisenstein series can be written in terms of $\vartheta_i^4$.
\begin{align}
E_2 \begin{bsmallmatrix} -1 \\ 1 \end{bsmallmatrix} = & \  \frac{\vartheta_2(0)^4 + \vartheta_3(0)^4}{24}, \qquad
E_2 \begin{bsmallmatrix} -1 \\ -1 \end{bsmallmatrix} = \frac{-\vartheta_2(0)^4 + \vartheta_4(0)^4}{24}, \\
E_2 \begin{bsmallmatrix} 1 \\ -1 \end{bsmallmatrix} = & \ \frac{-\vartheta_4(0)^4 - \vartheta_3(0)^4}{24}, \\[6pt]
E_k \begin{bsmallmatrix} -1 \\ -1 \end{bsmallmatrix} = & \  0,\quad
E_k \begin{bsmallmatrix} 1 \\ -1 \end{bsmallmatrix} = 0,\quad
E_k \begin{bsmallmatrix} -1 \\ 1 \end{bsmallmatrix} = 0 \quad (k\ \text{odd}), \\[6pt]
E_4 \begin{bsmallmatrix} -1 \\ -1 \end{bsmallmatrix} = & \  E_4(\tau) - \frac{\vartheta_3(0)^8}{16 \cdot 24}, \qquad
E_4 \begin{bsmallmatrix} -1 \\ 1 \end{bsmallmatrix} = E_4(\tau) - \frac{\vartheta_4(0)^8}{16 \cdot 24}, \\
E_4 \begin{bsmallmatrix} 1 \\ -1 \end{bsmallmatrix} = & \  E_4(\tau) - \frac{\vartheta_2(0)^8}{16 \cdot 24}, \\[6pt]
E_6 \begin{bsmallmatrix} -1 \\ -1 \end{bsmallmatrix} = & \  E_6(\tau) + \frac{\vartheta_3(0)^{12}}{640 \cdot 24} - \frac{\vartheta_2(0)^4\,\vartheta_3(0)^8}{7680}, \\
E_6 \begin{bsmallmatrix} 1 \\ -1 \end{bsmallmatrix} = & \  E_6(\tau) + \frac{\vartheta_2(0)^{12}}{640 \cdot 24} - \frac{\vartheta_3(0)^4\,\vartheta_2(0)^8}{7680}, \\
E_6 \begin{bsmallmatrix} -1 \\ 1 \end{bsmallmatrix} = & \  E_6(\tau) - \frac{\vartheta_4(0)^{12}}{640 \cdot 24} + \frac{\vartheta_3(0)^4\,\vartheta_4(0)^8}{7680}, \\[6pt]
E_8 \begin{bsmallmatrix} -1 \\ -1 \end{bsmallmatrix} = & \  E_8(\tau) - \frac{\vartheta_2(0)^8\,\vartheta_3(0)^8}{322560} + \frac{\vartheta_2(0)^4\,\vartheta_3(0)^{12}}{322560} - \frac{17\,\vartheta_3(0)^{16}}{10321920}, \\
E_8 \begin{bsmallmatrix} -1 \\ 1 \end{bsmallmatrix} = & \  E_8(\tau) - \frac{17\,\vartheta_2(0)^{16}}{10321920} + \frac{\vartheta_2(0)^{12}\,\vartheta_3(0)^4}{286720} - \frac{19\,\vartheta_2(0)^8\,\vartheta_3(0)^8}{5160960} + \frac{\vartheta_2(0)^4\,\vartheta_3(0)^{12}}{286720} \nonumber\\
& \ - \frac{17\,\vartheta_3(0)^{16}}{10321920}, \\
E_8 \begin{bsmallmatrix} 1 \\ -1 \end{bsmallmatrix} = & \  E_8(\tau) - \frac{17\,\vartheta_2(0)^{16}}{10321920} + \frac{\vartheta_2(0)^{12}\,\vartheta_3(0)^4}{322560} - \frac{\vartheta_2(0)^8\,\vartheta_3(0)^8}{322560}, \\[6pt]
E_4(\tau) = & \  \frac{\vartheta_2(0)^8 + \vartheta_3(0)^8 + \vartheta_4(0)^8}{1440}, \\
E_6(\tau) = & \  -\frac{1}{60480}\left(\vartheta_2(0)^4 + \vartheta_3(0)^4\right)\left(\vartheta_3(0)^4 + \vartheta_4(0)^4\right)\left(\vartheta_4(0)^4 - \vartheta_2(0)^4\right), \\
E_8(\tau) = & \  \frac{\vartheta_2(0)^{16} + \vartheta_3(0)^{16} + \vartheta_4(0)^{16}}{2419200}.
\end{align}

\bibliographystyle{utphys2}

\bibliography{ref}

@article{Amariti:2024bsr,
	archiveprefix = {arXiv},
	author = {Amariti, Antonio and Glorioso, Pietro and Morgante, Davide and Zanetti, Andrea},
	doi = {10.1016/j.nuclphysb.2024.116773},
	eprint = {2403.17190},
	journal = {Nucl. Phys. B},
	pages = {116773},
	primaryclass = {hep-th},
	title = {{Cardy matches Bethe on the surface: A tale of a brane and a black hole}},
	volume = {1010},
	year = {2025}}

@article{David:2021qaa,
	archiveprefix = {arXiv},
	author = {David, Marina and Lezcano Gonz{\'a}lez, Alfredo and Nian, Jun and Pando Zayas, Leopoldo A.},
	doi = {10.1007/JHEP04(2022)160},
	eprint = {2106.09730},
	journal = {JHEP},
	pages = {160},
	primaryclass = {hep-th},
	reportnumber = {LCTP-21-14},
	title = {{Logarithmic corrections to the entropy of rotating black holes and black strings in AdS$_{5}$}},
	volume = {04},
	year = {2022}}

@article{GonzalezLezcano:2020yeb,
	archiveprefix = {arXiv},
	author = {Gonz{\'a}lez Lezcano, Alfredo and Hong, Junho and Liu, James T. and Pando Zayas, Leopoldo A.},
	doi = {10.1007/JHEP01(2021)001},
	eprint = {2007.12604},
	journal = {JHEP},
	pages = {001},
	primaryclass = {hep-th},
	reportnumber = {LCTP-20-16},
	title = {{Sub-leading Structures in Superconformal Indices: Subdominant Saddles and Logarithmic Contributions}},
	volume = {01},
	year = {2021}}

@article{Deddo:2025jrg,
	archiveprefix = {arXiv},
	author = {Deddo, Evan and Pando Zayas, Leopoldo A. and Zhou, Wenjie},
	doi = {10.1007/JHEP05(2025)170},
	eprint = {2502.01614},
	journal = {JHEP},
	pages = {170},
	primaryclass = {hep-th},
	title = {{The superconformal index and black hole instabilities}},
	volume = {05},
	year = {2025}}

@article{Eniceicu:2023uvd,
	archiveprefix = {arXiv},
	author = {Eniceicu, Dan Stefan},
	eprint = {2302.04887},
	month = {2},
	primaryclass = {hep-th},
	title = {{Comments on the Giant-Graviton Expansion of the Superconformal Index}},
	year = {2023}}

@article{Liu:2022olj,
	archiveprefix = {arXiv},
	author = {Liu, James T. and Rajappa, Neville Joshua},
	doi = {10.1007/JHEP04(2023)078},
	eprint = {2212.05408},
	journal = {JHEP},
	pages = {078},
	primaryclass = {hep-th},
	reportnumber = {LCTP-22-16},
	title = {{Finite N indices and the giant graviton expansion}},
	volume = {04},
	year = {2023}}

@article{Imamura:2021ytr,
	archiveprefix = {arXiv},
	author = {Imamura, Yosuke},
	doi = {10.1093/ptep/ptab141},
	eprint = {2108.12090},
	journal = {PTEP},
	number = {12},
	pages = {123B05},
	primaryclass = {hep-th},
	reportnumber = {TIT/HEP-686},
	title = {{Finite-N superconformal index via the AdS/CFT correspondence}},
	volume = {2021},
	year = {2021}}

@article{Colombo:2021kbb,
	archiveprefix = {arXiv},
	author = {Colombo, Edoardo},
	doi = {10.1007/JHEP12(2022)013},
	eprint = {2110.01911},
	journal = {JHEP},
	pages = {013},
	primaryclass = {hep-th},
	title = {{The large-N limit of 4d superconformal indices for general BPS charges}},
	volume = {12},
	year = {2022}}

@article{Cabo-Bizet:2024kfe,
	archiveprefix = {arXiv},
	author = {Cabo-Bizet, Alejandro and Li, Wei},
	doi = {10.1007/JHEP07(2025)206},
	eprint = {2411.12018},
	journal = {JHEP},
	pages = {206},
	primaryclass = {hep-th},
	title = {{Generalized Bethe expansions of superconformal indices}},
	volume = {07},
	year = {2025}}

@article{Eleftheriou:2023jxr,
	archiveprefix = {arXiv},
	author = {Eleftheriou, Giorgos and Murthy, Sameer and Rossell{\'o}, Mart{\'\i}},
	doi = {10.21468/SciPostPhys.17.4.098},
	eprint = {2312.14921},
	journal = {SciPost Phys.},
	number = {4},
	pages = {098},
	primaryclass = {hep-th},
	title = {{The giant graviton expansion in $AdS_5 \times S^5$}},
	volume = {17},
	year = {2024}}

@article{Deddo:2024liu,
	archiveprefix = {arXiv},
	author = {Deddo, Evan and Liu, James T. and Pando Zayas, Leopoldo A. and Saskowski, Robert J.},
	doi = {10.1103/PhysRevLett.132.261501},
	eprint = {2402.19452},
	journal = {Phys. Rev. Lett.},
	number = {26},
	pages = {261501},
	primaryclass = {hep-th},
	reportnumber = {LCTP-24-04},
	title = {{Giant Graviton Expansion from Bubbling Geometry: Discreteness from Quantized Geometry}},
	volume = {132},
	year = {2024}}

@article{Dolan:2008qi,
	archiveprefix = {arXiv},
	author = {Dolan, F. A. and Osborn, H.},
	doi = {10.1016/j.nuclphysb.2009.01.028},
	eprint = {0801.4947},
	journal = {Nucl. Phys. B},
	pages = {137--178},
	primaryclass = {hep-th},
	reportnumber = {DAMTP-08-07, DIAS-STP-08-02, SHEP-08-06},
	title = {{Applications of the Superconformal Index for Protected Operators and q-Hypergeometric Identities to N=1 Dual Theories}},
	volume = {818},
	year = {2009}}

@article{Spiridonov:2008zr,
	archiveprefix = {arXiv},
	author = {Spiridonov, V. P. and Vartanov, G. S.},
	doi = {10.1016/j.nuclphysb.2009.08.022},
	eprint = {0811.1909},
	journal = {Nucl. Phys. B},
	pages = {192--216},
	primaryclass = {hep-th},
	title = {{Superconformal indices for N = 1 theories with multiple duals}},
	volume = {824},
	year = {2010}}

@article{Li:2020ljm,
	archiveprefix = {arXiv},
	author = {Li, Si and Zhou, Jie},
	doi = {10.1007/s00220-021-04232-6},
	eprint = {2008.07503},
	journal = {Commun. Math. Phys.},
	number = {3},
	pages = {1403--1474},
	primaryclass = {math.DG},
	title = {{Regularized Integrals on Riemann Surfaces and Modular Forms}},
	volume = {388},
	year = {2021}}

@article{ArabiArdehali:2019orz,
	archiveprefix = {arXiv},
	author = {Arabi Ardehali, Arash and Hong, Junho and Liu, James T.},
	doi = {10.1007/JHEP07(2020)073},
	eprint = {1912.04169},
	journal = {JHEP},
	pages = {073},
	primaryclass = {hep-th},
	reportnumber = {LCTP 19-32},
	title = {{Asymptotic growth of the 4d $ \mathcal{N} $ = 4 index and partially deconfined phases}},
	volume = {07},
	year = {2020}}

@article{Tachikawa:2009rb,
	archiveprefix = {arXiv},
	author = {Tachikawa, Yuji},
	doi = {10.1088/1126-6708/2009/07/067},
	eprint = {0905.4074},
	journal = {JHEP},
	pages = {067},
	primaryclass = {hep-th},
	title = {{Six-dimensional D(N) theory and four-dimensional SO-USp quivers}},
	volume = {07},
	year = {2009}}

@article{Beccaria:2024vfx,
	archiveprefix = {arXiv},
	author = {Beccaria, Matteo and Cabo-Bizet, Alejandro},
	doi = {10.1007/JHEP04(2024)110},
	eprint = {2402.12172},
	journal = {JHEP},
	pages = {110},
	primaryclass = {hep-th},
	title = {{Large N Schur index of $ \mathcal{N} $ = 4 SYM from semiclassical D3 brane}},
	volume = {04},
	year = {2024}}

@article{Xie:2026xxg,
	archiveprefix = {arXiv},
	author = {Xie, Dan},
	eprint = {2606.16714},
	month = {6},
	primaryclass = {hep-th},
	title = {{On the Representation Theory of Non-Admissible $W$-Algebras: Part I}},
	year = {2026}}

@article{Pan:2024hcz,
	archiveprefix = {arXiv},
	author = {Pan, Yiwen and Yan, Wenbin},
	eprint = {2410.15695},
	month = {10},
	primaryclass = {hep-th},
	title = {{Mirror symmetry for circle compactified 4d $A_1$ class-$S$ theories}},
	year = {2024}}

@article{Li:2025nhc,
	archiveprefix = {arXiv},
	author = {Li, Yutong and Pan, Yiwen and Yan, Wenbin},
	eprint = {2510.03888},
	month = {10},
	primaryclass = {hep-th},
	title = {{Chiral algebra, Wilson lines, and mixed Hodge structure of Coulomb branch}},
	year = {2025}}

@article{Chandra:2025qpv,
	archiveprefix = {arXiv},
	author = {Chandra, A. Ramesh and Mukhi, Sunil and Singh, Palash},
	doi = {10.1007/JHEP06(2026)211},
	eprint = {2512.02107},
	journal = {JHEP},
	pages = {211},
	primaryclass = {hep-th},
	title = {{Generalised 4d partition functions and modular differential equations}},
	volume = {06},
	year = {2026}}

@article{Deb:2025ddc,
	archiveprefix = {arXiv},
	author = {Deb, Anirudh},
	eprint = {2512.02102},
	month = {12},
	primaryclass = {hep-th},
	reportnumber = {YITP-SB-2025-20},
	title = {{Generalized Schur limit, modular differential equations and quantum monodromy traces}},
	year = {2025}}

@article{Deb:2025ypl,
	archiveprefix = {arXiv},
	author = {Deb, Anirudh and Razamat, Shlomo S.},
	doi = {10.1103/ldxd-2jm5},
	eprint = {2506.13764},
	journal = {Phys. Rev. D},
	number = {4},
	pages = {045011},
	primaryclass = {hep-th},
	reportnumber = {YITP-SB-2025-12},
	title = {{Generalized Schur partition functions and RG flows}},
	volume = {113},
	year = {2026}}

@article{Benini:2018ywd,
	archiveprefix = {arXiv},
	author = {Benini, Francesco and Milan, Elisa},
	doi = {10.1103/PhysRevX.10.021037},
	eprint = {1812.09613},
	journal = {Phys. Rev. X},
	number = {2},
	pages = {021037},
	primaryclass = {hep-th},
	reportnumber = {SISSA 56/2018/FISI},
	title = {{Black Holes in 4D $\mathcal{N}$=4 Super-Yang-Mills Field Theory}},
	volume = {10},
	year = {2020}}

@article{Amariti:2025vjd,
	archiveprefix = {arXiv},
	author = {Amariti, Antonio and Glorioso, Pietro},
	doi = {10.1016/j.nuclphysb.2025.117215},
	eprint = {2506.15296},
	journal = {Nucl. Phys. B},
	pages = {117215},
	primaryclass = {hep-th},
	title = {{Exact results on the Bethe Ansatz evaluation of the SCI}},
	volume = {1022},
	year = {2026}}

@article{Aharony:2021zkr,
	archiveprefix = {arXiv},
	author = {Aharony, Ofer and Benini, Francesco and Mamroud, Ohad and Milan, Elisa},
	doi = {10.1103/PhysRevD.104.086026},
	eprint = {2104.13932},
	journal = {Phys. Rev. D},
	pages = {086026},
	primaryclass = {hep-th},
	reportnumber = {SISSA 01/2021/FISI},
	title = {{A gravity interpretation for the Bethe Ansatz expansion of the $\mathcal{N}=4$ SYM index}},
	volume = {104},
	year = {2021}}

@article{Lanir:2019abx,
	archiveprefix = {arXiv},
	author = {Lanir, Assaf and Nedelin, Anton and Sela, Orr},
	doi = {10.1007/JHEP04(2020)091},
	eprint = {1908.01737},
	journal = {JHEP},
	pages = {091},
	primaryclass = {hep-th},
	title = {{Black hole entropy function for toric theories via Bethe Ansatz}},
	volume = {04},
	year = {2020}}

@article{Benini:2020gjh,
	archiveprefix = {arXiv},
	author = {Benini, Francesco and Colombo, Edoardo and Soltani, Saman and Zaffaroni, Alberto and Zhang, Ziruo},
	doi = {10.1088/1361-6382/abb39b},
	eprint = {2005.12308},
	journal = {Class. Quant. Grav.},
	number = {21},
	pages = {215021},
	primaryclass = {hep-th},
	reportnumber = {SISSA 11/2020/FISI},
	title = {{Superconformal indices at large $N$ and the entropy of AdS$_5$ $\times$ SE$_5$ black holes}},
	volume = {37},
	year = {2020}}

@article{Hatsuda:2023imp,
	archiveprefix = {arXiv},
	author = {Hatsuda, Yasuyuki and Okazaki, Tadashi},
	doi = {10.1007/JHEP01(2024)096},
	eprint = {2309.11712},
	journal = {JHEP},
	pages = {096},
	primaryclass = {hep-th},
	reportnumber = {RUP-23-17},
	title = {{Large N and large representations of Schur line defect correlators}},
	volume = {01},
	year = {2024}}

@article{Kim:2012ava,
	archiveprefix = {arXiv},
	author = {Kim, Hee-Cheol and Kim, Seok},
	doi = {10.1007/JHEP05(2013)144},
	eprint = {1206.6339},
	journal = {JHEP},
	pages = {144},
	primaryclass = {hep-th},
	reportnumber = {SNUTP12-002, KIAS-P12038},
	title = {{M5-branes from gauge theories on the 5-sphere}},
	volume = {05},
	year = {2013}}

@article{Imamura:2024zvw,
	archiveprefix = {arXiv},
	author = {Imamura, Yosuke and Sei, Akihiro and Yokoyama, Daisuke},
	doi = {10.1007/JHEP09(2024)202},
	eprint = {2406.19777},
	journal = {JHEP},
	pages = {202},
	primaryclass = {hep-th},
	reportnumber = {TIT/HEP-702},
	title = {{Giant graviton expansion for general Wilson line operator indices}},
	volume = {09},
	year = {2024}}

@article{Beccaria:2023zjw,
	archiveprefix = {arXiv},
	author = {Beccaria, Matteo and Cabo-Bizet, Alejandro},
	doi = {10.1007/JHEP08(2023)073},
	eprint = {2305.17730},
	journal = {JHEP},
	pages = {073},
	primaryclass = {hep-th},
	title = {{On the brane expansion of the Schur index}},
	volume = {08},
	year = {2023}}

@article{Ezroura:2024wmp,
	archiveprefix = {arXiv},
	author = {Ezroura, Nizar and Liu, James T. and Rajappa, Neville Joshua},
	doi = {10.1007/JHEP01(2025)028},
	eprint = {2408.02759},
	journal = {JHEP},
	pages = {028},
	primaryclass = {hep-th},
	reportnumber = {LCTP-24-12},
	title = {{Analytic continuation and the giant graviton expansion}},
	volume = {01},
	year = {2025}}

@article{Beem:2023ofp,
	archiveprefix = {arXiv},
	author = {Beem, Christopher and Martone, Mario and Sacchi, Matteo and Singh, Palash and Stedman, Jake},
	eprint = {2311.12123},
	month = {11},
	primaryclass = {hep-th},
	title = {{Simplifying the Type $A$ Argyres-Douglas Landscape}},
	year = {2023}}

@article{Arakawa:2018egx,
	archiveprefix = {arXiv},
	author = {Arakawa, Tomoyuki},
	eprint = {1811.01577},
	month = {11},
	primaryclass = {math.RT},
	title = {{Chiral algebras of class $\mathcal{S}$ and Moore-Tachikawa symplectic varieties}},
	year = {2018}}

@article{Arakawa:2023cki,
	archiveprefix = {arXiv},
	author = {Arakawa, Tomoyuki and Kuwabara, Toshiro and M\"oller, Sven},
	eprint = {2309.17308},
	month = {9},
	primaryclass = {math.RT},
	title = {{Hilbert Schemes of Points in the Plane and Quasi-Lisse Vertex Algebras with $\mathcal{N}=4$ Symmetry}},
	year = {2023}}

@article{Beem:2013sza,
	archiveprefix = {arXiv},
	author = {Beem, Christopher and Lemos, Madalena and Liendo, Pedro and Peelaers, Wolfger and Rastelli, Leonardo and van Rees, Balt C.},
	doi = {10.1007/s00220-014-2272-x},
	eprint = {1312.5344},
	journal = {Commun. Math. Phys.},
	number = {3},
	pages = {1359--1433},
	primaryclass = {hep-th},
	reportnumber = {YITP-SB-13-45, CERN-PH-TH-2013-311, HU-EP-13-78},
	title = {{Infinite Chiral Symmetry in Four Dimensions}},
	volume = {336},
	year = {2015}}

@article{Beem:2014rza,
	archiveprefix = {arXiv},
	author = {Beem, Christopher and Peelaers, Wolfger and Rastelli, Leonardo and van Rees, Balt C.},
	doi = {10.1007/JHEP05(2015)020},
	eprint = {1408.6522},
	journal = {JHEP},
	pages = {020},
	primaryclass = {hep-th},
	reportnumber = {YITP-SB-14-30, CERN-PH-TH-2014-165, YITP-SB-14-30, CERN-PH-TH-2014-165},
	title = {{Chiral algebras of class S}},
	volume = {05},
	year = {2015}}

@article{Benini:2013xpa,
	archiveprefix = {arXiv},
	author = {Benini, Francesco and Eager, Richard and Hori, Kentaro and Tachikawa, Yuji},
	doi = {10.1007/s00220-014-2210-y},
	eprint = {1308.4896},
	journal = {Commun. Math. Phys.},
	number = {3},
	pages = {1241--1286},
	primaryclass = {hep-th},
	reportnumber = {IPMU-13-0146, UT-13-29},
	title = {{Elliptic Genera of 2d ${\mathcal{N}}$ = 2 Gauge Theories}},
	volume = {333},
	year = {2015}}

@article{Bianchi:2019sxz,
	archiveprefix = {arXiv},
	author = {Bianchi, Lorenzo and Lemos, Madalena},
	doi = {10.1007/JHEP06(2020)056},
	eprint = {1911.05082},
	journal = {JHEP},
	pages = {056},
	primaryclass = {hep-th},
	reportnumber = {CERN-TH-2019-190},
	title = {{Superconformal surfaces in four dimensions}},
	volume = {06},
	year = {2020}}

@article{Gadde:2011ik,
	archiveprefix = {arXiv},
	author = {Gadde, Abhijit and Rastelli, Leonardo and Razamat, Shlomo S. and Yan, Wenbin},
	doi = {10.1103/PhysRevLett.106.241602},
	eprint = {1104.3850},
	journal = {Phys.Rev.Lett.},
	pages = {241602},
	primaryclass = {hep-th},
	reportnumber = {YITP-SB-11-13},
	slaccitation = {%%CITATION = ARXIV:1104.3850;%%},
	title = {{The 4d Superconformal Index from $q$-deformed 2d Yang-Mills}},
	volume = {106},
	year = {2011}}

@article{Gadde:2011uv,
	archiveprefix = {arXiv},
	author = {Gadde, Abhijit and Rastelli, Leonardo and Razamat, Shlomo S. and Yan, Wenbin},
	doi = {10.1007/s00220-012-1607-8},
	eprint = {1110.3740},
	journal = {Commun. Math. Phys.},
	pages = {147-193},
	primaryclass = {hep-th},
	reportnumber = {YITP-SB-11-30},
	slaccitation = {%%CITATION = ARXIV:1110.3740;%%},
	title = {{Gauge Theories and Macdonald Polynomials}},
	volume = {319},
	year = {2013}}

@article{Imamura:2011su,
	archiveprefix = {arXiv},
	author = {Imamura, Yosuke and Yokoyama, Shuichi},
	doi = {10.1007/JHEP04(2011)007},
	eprint = {1101.0557},
	journal = {JHEP},
	pages = {007},
	primaryclass = {hep-th},
	reportnumber = {UT-11-01, TIT-HEP-607},
	slaccitation = {%%CITATION = ARXIV:1101.0557;%%},
	title = {{Index for three dimensional superconformal field theories with general R-charge assignments}},
	volume = {1104},
	year = {2011}}

@article{Kapustin:2011jm,
	archiveprefix = {arXiv},
	author = {Kapustin, Anton and Willett, Brian},
	eprint = {1106.2484},
	primaryclass = {hep-th},
	reportnumber = {68-2840},
	slaccitation = {%%CITATION = ARXIV:1106.2484;%%},
	title = {{Generalized Superconformal Index for Three Dimensional Field Theories}},
	year = {2011}}

@article{Lemos:2014lua,
	archiveprefix = {arXiv},
	author = {Lemos, Madalena and Peelaers, Wolfger},
	doi = {10.1007/JHEP02(2015)113},
	eprint = {1411.3252},
	journal = {JHEP},
	pages = {113},
	primaryclass = {hep-th},
	reportnumber = {YITP-SB-14-41},
	title = {{Chiral Algebras for Trinion Theories}},
	volume = {02},
	year = {2015}}

@article{Nekrasov:2009uh,
	archiveprefix = {arXiv},
	author = {Nekrasov, Nikita A. and Shatashvili, Samson L.},
	doi = {10.1016/j.nuclphysbps.2009.07.047},
	eprint = {0901.4744},
	journal = {Nucl.Phys.Proc.Suppl.},
	pages = {91-112},
	primaryclass = {hep-th},
	reportnumber = {IHES-P-09-09, TCD-MATH-09-04, HMI-09-01, NSF-KITP-09-11},
	slaccitation = {%%CITATION = ARXIV:0901.4744;%%},
	title = {{Supersymmetric vacua and Bethe ansatz}},
	volume = {192-193},
	year = {2009}}

@article{Nekrasov:2009ui,
	archiveprefix = {arXiv},
	author = {Nekrasov, Nikita A. and Shatashvili, Samson L.},
	doi = {10.1143/PTPS.177.105},
	eprint = {0901.4748},
	journal = {Prog.Theor.Phys.Suppl.},
	note = {21 pp., short version II, conference in honour of T.Eguchi's 60th anniversary},
	pages = {105-119},
	primaryclass = {hep-th},
	reportnumber = {IHES-P-08-59, TCD-MATH-09-05, HMI-09-02, NSF-KITP-09-12},
	slaccitation = {%%CITATION = ARXIV:0901.4748;%%},
	title = {{Quantum integrability and supersymmetric vacua}},
	volume = {177},
	year = {2009}}

@article{Orlando:2010uu,
	archiveprefix = {arXiv},
	author = {Orlando, Domenico and Reffert, Susanne},
	doi = {10.1007/JHEP10(2010)071},
	eprint = {1005.4445},
	journal = {JHEP},
	pages = {071},
	primaryclass = {hep-th},
	reportnumber = {IPMU10-0088},
	slaccitation = {%%CITATION = ARXIV:1005.4445;%%},
	title = {{Relating Gauge Theories via Gauge/Bethe Correspondence}},
	volume = {1010},
	year = {2010}}

@article{Pan:2021mrw,
	archiveprefix = {arXiv},
	author = {Pan, Yiwen and Peelaers, Wolfger},
	doi = {10.1103/PhysRevD.106.045017},
	eprint = {2112.09705},
	journal = {Phys. Rev. D},
	number = {4},
	pages = {045017},
	primaryclass = {hep-th},
	title = {{Exact Schur index in closed form}},
	volume = {106},
	year = {2022}}

@article{Romelsberger:2005eg,
	archiveprefix = {arXiv},
	author = {Romelsberger, Christian},
	doi = {10.1016/j.nuclphysb.2006.03.037},
	eprint = {hep-th/0510060},
	journal = {Nucl. Phys.},
	pages = {329-353},
	slaccitation = {%%CITATION = HEP-TH/0510060;%%},
	title = {{Counting Chiral Primaries in ${\mathcal{N}}\!=1$, $d=4$ Superconformal Field Theories}},
	volume = {B747},
	year = {2006}}

@article{Seiberg:1994pq,
	archiveprefix = {arXiv},
	author = {Seiberg, N.},
	doi = {10.1016/0550-3213(94)00023-8},
	eprint = {hep-th/9411149},
	journal = {Nucl.Phys.},
	pages = {129-146},
	primaryclass = {hep-th},
	reportnumber = {RU-94-82, IASSNS-HEP-94-98},
	slaccitation = {%%CITATION = HEP-TH/9411149;%%},
	title = {{Electric - magnetic duality in supersymmetric nonAbelian gauge theories}},
	volume = {B435},
	year = {1995}}

@article{Witten:1986bf,
	author = {Witten, Edward},
	doi = {10.1007/BF01208956},
	journal = {Commun. Math. Phys.},
	pages = {525},
	reportnumber = {PUPT-1024},
	slaccitation = {%%CITATION = CMPHA,109,525;%%},
	title = {{Elliptic Genera and Quantum Field Theory}},
	volume = {109},
	year = {1987}}

@article{2004math.....11044S,
	archiveprefix = {arXiv},
	author = {Spiridonov, Vyacheslav P and Warnaar, S Ole},
	eprint = {math/0411044},
	journal = {Advances in Mathematics},
	number = {1},
	pages = {91--132},
	publisher = {Elsevier},
	title = {Inversions of integral operators and elliptic beta integrals on root systems},
	volume = {207},
	year = {2006}}

@article{2017arXiv170906142F,
	adsurl = {https://ui.adsabs.harvard.edu/abs/2017arXiv170906142F},
	archiveprefix = {arXiv},
	author = {{Fredrickson}, Laura and {Neitzke}, Andrew},
	doi = {10.48550/arXiv.1709.06142},
	eid = {arXiv:1709.06142},
	eprint = {1709.06142},
	journal = {arXiv e-prints},
	month = sep,
	pages = {arXiv:1709.06142},
	primaryclass = {math.DG},
	title = {{From $S^1$-fixed points to $\mathcal{W}$-algebra representations}},
	year = 2017}

@article{Arai:2020qaj,
	archiveprefix = {arXiv},
	author = {Arai, Reona and Fujiwara, Shota and Imamura, Yosuke and Mori, Tatsuya},
	doi = {10.1103/PhysRevD.101.086017},
	eprint = {2001.11667},
	journal = {Phys. Rev. D},
	number = {8},
	pages = {086017},
	primaryclass = {hep-th},
	reportnumber = {TIT/HEP-677},
	title = {{Schur index of the ${\cal N}=4$ $U(N)$ supersymmetric Yang-Mills theory via the AdS/CFT correspondence}},
	volume = {101},
	year = {2020}}

@article{Arakawa:2016hkg,
	archiveprefix = {arXiv},
	author = {Arakawa, Tomoyuki and Kawasetsu, Kazuya},
	eprint = {1610.05865},
	primaryclass = {math.QA},
	slaccitation = {%%CITATION = ARXIV:1610.05865;%%},
	title = {{Quasi-lisse vertex algebras and modular linear differential equations}},
	year = {2016}}

@article{Beccaria:2024szi,
	archiveprefix = {arXiv},
	author = {Beccaria, Matteo and Cabo-Bizet, Alejandro},
	eprint = {2403.06509},
	month = {3},
	primaryclass = {hep-th},
	title = {{Giant graviton expansion of Schur index and quasimodular forms}},
	year = {2024}}

@article{Beemetal,
	archiveprefix = {arXiv},
	author = {Beem, Christopher and Razamat, Shlomo S. and Singh, Palash},
	eprint = {2112.10715},
	month = {12},
	primaryclass = {hep-th},
	title = {{Schur Indices of Class $\mathcal{S}$ and Quasimodular Forms}},
	year = {2021}}

@article{Benini:2018mlo,
	archiveprefix = {arXiv},
	author = {Benini, Francesco and Milan, Paolo},
	doi = {10.1007/s00220-019-03679-y},
	eprint = {1811.04107},
	journal = {Commun. Math. Phys.},
	number = {2},
	pages = {1413--1440},
	primaryclass = {hep-th},
	reportnumber = {SISSA 46/2018/FISI},
	title = {{A Bethe Ansatz type formula for the superconformal index}},
	volume = {376},
	year = {2020}}

@article{Benini:2021ano,
	archiveprefix = {arXiv},
	author = {Benini, Francesco and Rizi, Giovanni},
	doi = {10.1007/JHEP05(2021)061},
	eprint = {2102.03638},
	journal = {JHEP},
	pages = {061},
	primaryclass = {hep-th},
	reportnumber = {SISSA 07/2021/FISI},
	title = {{Superconformal index of low-rank gauge theories via the Bethe Ansatz}},
	volume = {05},
	year = {2021}}

@article{Bourdier:2015sga,
	archiveprefix = {arXiv},
	author = {Bourdier, Jun and Drukker, Nadav and Felix, Jan},
	doi = {10.1007/JHEP01(2016)167},
	eprint = {1510.07041},
	journal = {JHEP},
	pages = {167},
	primaryclass = {hep-th},
	title = {{The $\mathcal{N}=2$ Schur index from free fermions}},
	volume = {01},
	year = {2016}}

@article{Bourdier:2015wda,
	archiveprefix = {arXiv},
	author = {Bourdier, Jun and Drukker, Nadav and Felix, Jan},
	doi = {10.1007/JHEP11(2015)210},
	eprint = {1507.08659},
	journal = {JHEP},
	pages = {210},
	primaryclass = {hep-th},
	title = {{The exact Schur index of $\mathcal{N}=4$ SYM}},
	volume = {11},
	year = {2015}}

@article{Cordova:2016uwk,
	archiveprefix = {arXiv},
	author = {Cordova, Clay and Gaiotto, Davide and Shao, Shu-Heng},
	doi = {10.1007/JHEP11(2016)106},
	eprint = {1606.08429},
	journal = {JHEP},
	pages = {106},
	primaryclass = {hep-th},
	title = {{Infrared Computations of Defect Schur Indices}},
	volume = {11},
	year = {2016}}

@article{Cordova:2017mhb,
	archiveprefix = {arXiv},
	author = {Cordova, Clay and Gaiotto, Davide and Shao, Shu-Heng},
	doi = {10.1007/JHEP05(2017)140},
	eprint = {1704.01955},
	journal = {JHEP},
	pages = {140},
	primaryclass = {hep-th},
	title = {{Surface Defects and Chiral Algebras}},
	volume = {05},
	year = {2017}}

@article{Creutzig:2017qyf,
	archiveprefix = {arXiv},
	author = {Creutzig, Thomas},
	eprint = {1701.05926},
	primaryclass = {hep-th},
	slaccitation = {%%CITATION = ARXIV:1701.05926;%%},
	title = {{W-algebras for Argyres-Douglas theories}},
	year = {2017}}

@article{Creutzig:2018lbc,
	archiveprefix = {arXiv},
	author = {Creutzig, Thomas},
	doi = {10.1007/JHEP11(2018)188},
	eprint = {1809.01725},
	journal = {JHEP},
	pages = {188},
	primaryclass = {hep-th},
	slaccitation = {%%CITATION = ARXIV:1809.01725;%%},
	title = {{Logarithmic W-algebras and Argyres-Douglas theories at higher rank}},
	volume = {11},
	year = {2018}}

@article{Gaiotto:2021xce,
	archiveprefix = {arXiv},
	author = {Gaiotto, Davide and Lee, Ji Hoon},
	eprint = {2109.02545},
	month = {9},
	primaryclass = {hep-th},
	title = {{The Giant Graviton Expansion}},
	year = {2021}}

@article{Guo:2023mkn,
	archiveprefix = {arXiv},
	author = {Guo, Zhaoting and Li, Yutong and Pan, Yiwen and Wang, Yufan},
	doi = {10.1103/PhysRevD.108.106002},
	eprint = {2307.15650},
	journal = {Phys. Rev. D},
	number = {10},
	pages = {106002},
	primaryclass = {hep-th},
	title = {{$\mathcal{N}=2$ Schur index and line operators}},
	volume = {108},
	year = {2023}}

@article{Hatsuda:2022xdv,
	archiveprefix = {arXiv},
	author = {Hatsuda, Yasuyuki and Okazaki, Tadashi},
	doi = {10.1007/JHEP01(2023)029},
	eprint = {2208.01426},
	journal = {JHEP},
	pages = {029},
	primaryclass = {hep-th},
	reportnumber = {RUP-22-17, KIAS-P22059},
	title = {{$ \mathcal{N} $ = 2$^{*}$ Schur indices}},
	volume = {01},
	year = {2023}}

@article{Hatsuda:2023iwi,
	archiveprefix = {arXiv},
	author = {Hatsuda, Yasuyuki and Okazaki, Tadashi},
	doi = {10.1007/JHEP06(2023)169},
	eprint = {2303.14887},
	journal = {JHEP},
	pages = {169},
	primaryclass = {hep-th},
	reportnumber = {RUP-23-7},
	title = {{Exact $ \mathcal{N} $ = 2$^{*}$ Schur line defect correlators}},
	volume = {06},
	year = {2023}}

@article{Hatsuda:2025mvj,
	archiveprefix = {arXiv},
	author = {Hatsuda, Yasuyuki},
	eprint = {2503.03952},
	month = {3},
	primaryclass = {hep-th},
	reportnumber = {RUP-25-6},
	title = {{Deformed Schur indices and Macdonald polynomials}},
	year = {2025}}

@article{Kang:2021lic,
	archiveprefix = {arXiv},
	author = {Kang, Monica Jinwoo and Lawrie, Craig and Song, Jaewon},
	eprint = {2106.12579},
	month = {6},
	primaryclass = {hep-th},
	reportnumber = {CALT-TH-2021-026},
	title = {{Infinitely many 4d $\mathcal{N}=2$ SCFTs with $a=c$ and beyond}},
	year = {2021}}

@article{Lemos:2012ph,
	archiveprefix = {arXiv},
	author = {Lemos, Madalena and Peelaers, Wolfger and Rastelli, Leonardo},
	doi = {10.1007/JHEP05(2014)120},
	eprint = {1212.1271},
	journal = {JHEP},
	pages = {120},
	primaryclass = {hep-th},
	reportnumber = {YITP-SB-12-45},
	title = {{The superconformal index of class $S$ theories of type $D$}},
	volume = {05},
	year = {2014}}

@article{Lezcano:2021qbj,
	archiveprefix = {arXiv},
	author = {Lezcano, Alfredo Gonz\'alez and Hong, Junho and Liu, James T. and Zayas, Leopoldo A. Pando},
	doi = {10.1007/JHEP06(2021)126},
	eprint = {2101.12233},
	journal = {JHEP},
	pages = {126},
	primaryclass = {hep-th},
	reportnumber = {LCTP-21-02},
	title = {{The Bethe-Ansatz approach to the $ \mathcal{N} $ = 4 superconformal index at finite rank}},
	volume = {06},
	year = {2021}}

@article{Nishinaka:2018zwq,
	archiveprefix = {arXiv},
	author = {Nishinaka, Takahiro and Sasa, Shinya and Zhu, Rui-Dong},
	doi = {10.1007/JHEP03(2019)091},
	eprint = {1811.11772},
	journal = {JHEP},
	pages = {091},
	primaryclass = {hep-th},
	reportnumber = {UT-18-27},
	title = {{On the Correspondence between Surface Operators in Argyres-Douglas Theories and Modules of Chiral Algebra}},
	volume = {03},
	year = {2019}}

@article{Pan:2017zie,
	archiveprefix = {arXiv},
	author = {Pan, Yiwen and Peelaers, Wolfger},
	doi = {10.1007/JHEP02(2018)138},
	eprint = {1710.04306},
	journal = {JHEP},
	pages = {138},
	primaryclass = {hep-th},
	reportnumber = {UUITP-34-17},
	title = {{Chiral Algebras, Localization and Surface Defects}},
	volume = {02},
	year = {2018}}

@article{Pan:2024epf,
	archiveprefix = {arXiv},
	author = {Pan, Yiwen and Yan, Wenbin},
	eprint = {2412.03155},
	month = {12},
	primaryclass = {hep-th},
	title = {{Mirror symmetry for 4d $A_1$ class-$\mathcal{S}$ theories: modularity, defects and Coulomb branch}},
	year = {2024}}

@inbook{Rastelli:2014jja,
	archiveprefix = {arXiv},
	author = {Rastelli, Leonardo and Razamat, Shlomo S.},
	booktitle = {New Dualities of Supersymmetric Gauge Theories},
	doi = {10.1007/978-3-319-18769-3_9},
	eprint = {1412.7131},
	pages = {261--305},
	primaryclass = {hep-th},
	publisher = {Springer},
	title = {{The Superconformal Index of Theories of Class $\mathcal {S}$}},
	year = {2016}}

@article{Razamat:2012uv,
	archiveprefix = {arXiv},
	author = {Razamat, Shlomo S.},
	doi = {10.1007/JHEP10(2012)191},
	eprint = {1208.5056},
	journal = {JHEP},
	pages = {191},
	primaryclass = {hep-th},
	title = {{On a modular property of N=2 superconformal theories in four dimensions}},
	volume = {10},
	year = {2012}}

@article{Shan:2023xtw,
	archiveprefix = {arXiv},
	author = {Shan, Peng and Xie, Dan and Yan, Wenbin},
	eprint = {2306.15214},
	month = {6},
	primaryclass = {hep-th},
	title = {{Mirror symmetry for circle compactified 4d $\mathcal{N}=2$ SCFTs}},
	year = {2023}}

@article{Xie:2019zlb,
	archiveprefix = {arXiv},
	author = {Xie, Dan and Yan, Wenbin},
	eprint = {1904.09094},
	primaryclass = {hep-th},
	slaccitation = {%%CITATION = ARXIV:1904.09094;%%},
	title = {{Schur sector of Argyres-Douglas theory and $W$-algebra}},
	year = {2019}}

@article{Du:2023kfu,
	archiveprefix = {arXiv},
	author = {Du, Bao-ning and Huang, Min-xin and Wang, Xin},
	doi = {10.1007/JHEP03(2024)009},
	eprint = {2311.08714},
	journal = {JHEP},
	pages = {009},
	primaryclass = {hep-th},
	reportnumber = {USTC-ICTS/PCFT-23-33, KIAS-Q23022},
	title = {{Schur indices for $ \mathcal{N} $ = 4 super-Yang-Mills with more general gauge groups}},
	volume = {03},
	year = {2024}}

@article{Closset:2017bse,
	archiveprefix = {arXiv},
	author = {Closset, Cyril and Kim, Heeyeon and Willett, Brian},
	doi = {10.1007/JHEP08(2017)090},
	eprint = {1707.05774},
	journal = {JHEP},
	pages = {090},
	primaryclass = {hep-th},
	reportnumber = {CERN-TH-2017-180},
	title = {{$ \mathcal{N} $ = 1 supersymmetric indices and the four-dimensional A-model}},
	volume = {08},
	year = {2017}}

@article{Huang:2022bry,
	archiveprefix = {arXiv},
	author = {Huang, Min-xin},
	eprint = {2205.00818},
	month = {5},
	primaryclass = {hep-th},
	reportnumber = {USTC-ICTS/PCFT-22-14},
	title = {{Modular Anomaly Equation for Schur Index of $\mathcal{N}=4$ Super-Yang-Mills}},
	year = {2022}}

@article{Gadde:2010te,
	archiveprefix = {arXiv},
	author = {Gadde, Abhijit and Rastelli, Leonardo and Razamat, Shlomo S. and Yan, Wenbin},
	doi = {10.1007/JHEP08(2010)107},
	eprint = {1003.4244},
	journal = {JHEP},
	pages = {107},
	primaryclass = {hep-th},
	reportnumber = {YITP-SB-10-7},
	title = {{The Superconformal Index of the $E_{6}$ SCFT}},
	volume = {08},
	year = {2010}}

@article{Fredrickson:2017yka,
	archiveprefix = {arXiv},
	author = {Fredrickson, Laura and Pei, Du and Yan, Wenbin and Ye, Ke},
	doi = {10.1007/JHEP01(2018)150},
	eprint = {1701.08782},
	journal = {JHEP},
	pages = {150},
	primaryclass = {hep-th},
	reportnumber = {CALT-TH-2016-038},
	title = {{Argyres-Douglas Theories, Chiral Algebras and Wild Hitchin Characters}},
	volume = {01},
	year = {2018}}

@article{Dedushenko:2019yiw,
	archiveprefix = {arXiv},
	author = {Dedushenko, Mykola and Fluder, Martin},
	eprint = {1904.02704},
	primaryclass = {hep-th},
	reportnumber = {CALT-TH 2019-011, IPMU19-0045},
	slaccitation = {%%CITATION = ARXIV:1904.02704;%%},
	title = {{Chiral Algebra, Localization, Modularity, Surface defects, And All That}},
	year = {2019}}

@article{Creutzig:2017uxh,
	archiveprefix = {arXiv},
	author = {Creutzig, Thomas and Gaiotto, Davide},
	eprint = {1708.00875},
	primaryclass = {hep-th},
	slaccitation = {%%CITATION = ARXIV:1708.00875;%%},
	title = {{Vertex Algebras for S-duality}},
	year = {2017}}

@article{Choi:2017nur,
	archiveprefix = {arXiv},
	author = {Choi, Jaewang and Nishinaka, Takahiro},
	doi = {10.1007/JHEP04(2018)004},
	eprint = {1711.07941},
	journal = {JHEP},
	pages = {004},
	primaryclass = {hep-th},
	slaccitation = {%%CITATION = ARXIV:1711.07941;%%},
	title = {{On the chiral algebra of Argyres-Douglas theories and S-duality}},
	volume = {04},
	year = {2018}}

@article{Xie:2016evu,
	archiveprefix = {arXiv},
	author = {Xie, Dan and Yan, Wenbin and Yau, Shing-Tung},
	doi = {10.1103/PhysRevD.103.065003},
	eprint = {1604.02155},
	journal = {Phys. Rev. D},
	number = {6},
	pages = {065003},
	primaryclass = {hep-th},
	title = {{Chiral algebra of the Argyres-Douglas theory from M5 branes}},
	volume = {103},
	year = {2021}}

@article{Kinney:2005ej,
	archiveprefix = {arXiv},
	author = {Kinney, Justin and Maldacena, Juan Martin and Minwalla, Shiraz and Raju, Suvrat},
	doi = {10.1007/s00220-007-0258-7},
	eprint = {hep-th/0510251},
	journal = {Commun. Math. Phys.},
	pages = {209-254},
	primaryclass = {hep-th},
	slaccitation = {%%CITATION = HEP-TH/0510251;%%},
	title = {{An Index for 4 dimensional super conformal theories}},
	volume = {275},
	year = {2007}}

@article{Dedushenko:2018bpp,
	archiveprefix = {arXiv},
	author = {Dedushenko, Mykola and Gukov, Sergei and Nakajima, Hiraku and Pei, Du and Ye, Ke},
	doi = {10.1088/1751-8121/abb481},
	eprint = {1809.04638},
	journal = {J. Phys. A},
	number = {43},
	pages = {43LT01},
	primaryclass = {hep-th},
	reportnumber = {CALT-TH-2018-033},
	title = {{3d TQFTs from Argyres-Douglas theories}},
	volume = {53},
	year = {2020}}

@article{Pan:2025vyu,
	archiveprefix = {arXiv},
	author = {Pan, Yiwen and Yang, Peihe},
	doi = {10.1103/4n8q-cpmb},
	eprint = {2509.20439},
	journal = {Phys. Rev. D},
	number = {4},
	pages = {045007},
	primaryclass = {hep-th},
	title = {{Exact Schur index in closed form for non-Lagrangian theories}},
	volume = {113},
	year = {2026}}

@article{GonzalezLezcano:2019nca,
	archiveprefix = {arXiv},
	author = {Gonz{\'a}lez Lezcano, Alfredo and Pando Zayas, Leopoldo A.},
	doi = {10.1007/JHEP03(2020)088},
	eprint = {1907.12841},
	journal = {JHEP},
	pages = {088},
	primaryclass = {hep-th},
	reportnumber = {LCTP-19-17},
	title = {{Microstate counting via Bethe Ans{\"a}tze in the 4d $ \mathcal{N} $ = 1 superconformal index}},
	volume = {03},
	year = {2020}}

@article{vanLeuven:2025gwr,
	archiveprefix = {arXiv},
	author = {van Leuven, Sam and Mathieson, Kayleigh and Roy, Pratik},
	doi = {10.1007/JHEP05(2026)204},
	eprint = {2511.10732},
	journal = {JHEP},
	pages = {204},
	primaryclass = {hep-th},
	title = {{Residue sums for superconformal indices}},
	volume = {05},
	year = {2026}}

\end{document}